\documentclass[usenatbib]{mn2e}

\usepackage{epsfig,times}
\usepackage[]{natbib}

\newcommand{\mklt}[0]{$M_K^* - 5 \log h = (-23.79 \pm 0.24)$ mag}
\newcommand{\mkl}[0]{$-$23.79 $\pm$ 0.24}

\newcommand{\pklt}[0]{$\Phi_K^* = (1.11 \pm 0.12) \times 10^{-2} \, h^3 \,
\mathrm{Mpc}^{-3}$}
\newcommand{\pkl}[0]{1.11 $\pm$ 0.12}

\newcommand{\mkmt}[0]{$M_K^* - 5 \log h = (-24.04 \pm 0.26)$ mag}
\newcommand{\mkm}[0]{$-$24.04 $\pm$ 0.26}

\newcommand{\pkmt}[0]{$\Phi_K^* = (0.71 \pm 0.25) \times 10^{-2} \, h^3 \,
\mathrm{Mpc}^{-3}$}
\newcommand{\pkm}[0]{0.71 $\pm$ 0.25}

\newcommand{\mjlt}[0]{$M_J^* - 5 \log h = (-22.45 \pm 0.24)$ mag}
\newcommand{\mjl}[0]{$-$22.45 $\pm$ 0.24}

\newcommand{\pjlt}[0]{$\Phi_J^* = (1.49 \pm 0.22) \times 10^{-2} \, h^3 \,
\mathrm{Mpc}^{-3}$}
\newcommand{\pjl}[0]{1.49 $\pm$ 0.22}

\newcommand{\mjmt}[0]{$M_J^* - 5 \log h = (-23.06 \pm 0.24)$ mag}
\newcommand{\mjm}[0]{$-$23.06 $\pm$ 0.24}

\newcommand{\pjmt}[0]{$\Phi_J^* = (0.76 \pm 0.25) \, h^3 \,
\mathrm{Mpc}^{-3}$}
\newcommand{\pjm}[0]{0.76 $\pm$ 0.25}

\title[MUNICS -- V.\ Evolution of near-infrared galaxy luminosity
functions]{The Munich Near-Infrared Cluster Survey (MUNICS) -- V.\ The
evolution of the rest-frame $K$-band and $J$-band galaxy luminosity
functions to $z \sim 0.7$}

\author[Georg Feulner et al.]
 {Georg~Feulner,$^1$\footnotemark[1]\footnotemark[2] 
  Ralf~Bender,$^{1,2}$\footnotemark[2]
  Niv~Drory,$^{3}$\footnotemark[2]
  Ulrich~Hopp,$^1$\footnotemark[2]\footnotemark[3]
  Jan~Snigula,$^1$\footnotemark[2]\newauthor\vspace*{.25em}
  Gary~J.~Hill$^3$\\
  $^1$Universit\"ats-Sternwarte M\"unchen, Scheinerstra\ss e 1, 
  D-81679 M\"unchen, Germany\\
  $^2$Max-Planck-Institut f\"ur Extraterrestrische Physik,
 Giessenbachstra\ss e, D-85748 Garching bei M\"unchen, Germany\\
  $^3$McDonald Observatory, University of Texas at Austin, Austin, Texas 78712}

\date{Accepted --; Received --; }

\pagerange{\pageref{firstpage}--\pageref{lastpage}} \pubyear{2002}

\begin{document}

\label{firstpage}

\maketitle

%
%
\begin{abstract}
  We present spectroscopic follow-up observations of galaxies from the
  Munich Near-Infrared Cluster Survey (MUNICS). MUNICS is a wide-field
  medium-deep $K'$-band selected survey covering 1 square degree in
  the near-infrared $K'$ and $J$ pass-bands, and 0.35 square degrees
  in $I$, $R$, $V$, and -- recently completed -- $B$. The
  spectroscopic sample comprises observations of objects down to a
  limit of $K' \le 17.5$ in five survey fields (0.17 square degrees in
  total), and a sparsely selected deeper sample ($K' \le 19.0$)
  constructed in one of the survey patches (0.03 square degrees). Here
  we describe the selection procedure of objects for spectroscopic
  observations, the observations themselves, the data reduction, and
  the construction of the spectroscopic catalogue containing roughly
  500 galaxies with secure redshifts. Furthermore we discuss global
  properties of the sample like its distribution in colour-redshift
  space, the accuracy of redshift determination, and the completeness
  function of the data. We derive the rest-frame $K'$-band luminosity
  function of galaxies at median redshifts of $z = 0.2$, $z = 0.4$,
  and $z = 0.7$. We find evidence for mild evolution of magnitudes
  ($\Delta M_K^* \simeq -0.70$~mag) and number densities ($\Delta
  \Phi_K^*/\Phi_K^* \simeq - 0.35$) to redshift one.  Furthermore, we
  present the rest-frame $J$-band luminosity function of galaxies at
  these redshifts, the first determination of this quantity at higher
  redshifts, with a behaviour similar to the $K$-band luminosity
  function.
\end{abstract}


\footnotetext[1]{E-mail: feulner@usm.uni-muenchen.de}

\footnotetext[2]{Visiting astronomer at the German-Spanish
  Astronomical Center, Calar Alto, operated by the Max-Planck-Institut
  f\"ur Astronomie, Heidelberg, jointly with the Spanish National
  Commission for Astronomy.}

\footnotetext[3]{Visiting astronomer at the European Southern
  Observatory, Chile, proposal number N 66.A-0129 and 66.A-0123.}

%
%

\section{Introduction}
\label{s:introduction}

The study of the luminosity function of galaxies at different
wavelengths and at different cosmic epochs is one of the most
important methods to address the problem of the formation and
evolution of galaxies within observational cosmology.

During the last decades, most field galaxy surveys have been optically
selected, mostly in the $I$ or $B$ band (see e.g.\ \citealt{Ellis1997}
for a review). In recent years however, there has been growing
interest in the study of near-infrared-selected samples of
galaxies. Especially the $K$-band at wavelengths of roughly $\lambda
\simeq 2 \mu\mathrm{m}$ offers the unique opportunity to detect
evolved galaxies, since the $K$-band light of galaxies is largely
dominated by radiation from the evolved stellar population
(\citealt{RR93,KC98a,BE00}). Moreover, in the case of the $K$-band,
the transformation which has to be applied to convert the observed
flux of an object's redshifted spectral energy distribution through a
given filter to the corresponding \textit{rest-frame} flux, the $k(z)$
correction, is much smaller -- and indeed negative -- than in the
optical wavebands, and the variation of $k(z)$ corrections between
different galaxy types is small. These effects reduce systematic
errors introduced by applying $k(z)$ corrections during the
construction of the luminosity function considerably. As another
advantage of near-infrared selected surveys, the rest-frame $K$-band
luminosity can be used to derive \textit{stellar masses} for the
galaxies, as has been done by \citeauthor{munics3} (2001b, hereafter
MUNICS~III), for example. Obviously, this allows the direct study of
the history of the assembly of stellar mass in the galaxies and thus
helps to understand the mechanism of their formation.

The past ten years have seen a number of measurements of
near-infrared luminosity functions, all of which are summarised in
Tables~\ref{t:litk} and \ref{t:litj}. The easiest way to determine the
$K$-band luminosity function of galaxies is by follow-up near-infrared
photometry of existing, typically \textit{optically} selected, galaxy
redshift catalogues. \citet*{MSE93} and \citet{Szokolyetal98} have
chosen this approach, and later on \citet{Loveday00} has presented the
luminosity function at a median redshift $z = 0.051$ from $K$-band
imaging of $b_J$-selected galaxies from the Stromlo-APM Redshift
Survey \citep{APM90,APM96}. Note that the use of optically-selected
samples can lead to a bias against red (for example old or dusty)
galaxies.

For \textit{near-infrared} selected galaxy surveys, most measurements
of the $K$-band luminosity function so far were based on local ($z \la
0.15$) samples. \citet{GSFC97} have calculated the $K$-band luminosity
function at a median redshift $z =0.14$. \citet{Kochaneketal01} have
determined type-dependent luminosity functions at median redshift $z =
0.023$ from 2MASS \citep{2MASS} and the CfA2 \citep{CfA2} and UZC
\citep{UZC} catalogues. A similar strategy was adopted by
\citet{Coleetal2001_2} who combined the photometric 2MASS data with
the 2dF Galaxy Redshift Survey \citep{2dFGRS_2}, and
\citet{Baloghetal2001} who cross-correlate the 2MASS data with the Las
Campanas Redshift Survey \citep{LCRS96}. As will be shown below, there
is good agreement between these determinations of the rest-frame
$K$-band luminosity function in the local universe. However,
\citet{Huangetal2002} have presented a measurement of the local
$K$-band luminosity function from the Hawaii--AAO $K$-band Galaxy
Redshift Survey, finding a slightly brighter $M^*$ and a steeper
faint-end slope, which they attribute to the influence of different
redshift ranges of the local samples. They also note that their value
for the faint-end slope is in better agreement with optical luminosity
functions.

At higher redshifts, \citet{GPMC95} derived the $K$-band luminosity
function out to redshifts $z \le 0.8$ and find evidence for a
brightening of the characteristic luminosity at $z >
0.5$. \citet{CSHC96} present the evolution of the $K$-band luminosity
function of galaxies in four redshift bins $z \in [0,0.1]$, $z \in
[0.1,0.2]$, $z \in [0.2,0.6]$, and $z \in [0.6,1.0]$ based on a deep,
but rather small spectroscopic sample, complemented by the shallower
samples from \citet{SCHG94}. They find no evolution with
redshift. Using our own, much larger sample of spectroscopically
calibrated photometric redshifts from the MUNICS survey, we find mild
evolution to redshift one, with a brightening of 0.5 to 0.7 mag, and a
decrease in number density of about 25 per cent (\citealt{munics2};
hereafter MUNICS~II).

The spectroscopic sample of $K$-band selected galaxies described
in this paper enables us to derive the rest-frame $K$-band luminosity
function of galaxies at redshifts $z = 0.2$, $z = 0.4$, and $z =
0.7$ from data based on a survey much larger in area than the Hawaii
Deep Fields \citep{CSHC96}, making the luminosity function presented
in this paper much less affected by cosmic variance.

Furthermore, the rest-frame $J$-band luminosity function at these
redshifts is presented for the first time. So far, there are only two
local measurements of the luminosity function in this band
\citep{Baloghetal2001,Coleetal2001_2}.

The paper is organised as follows. Section~\ref{s:observations}
briefly describes the Munich Near-Infrared Cluster Survey (MUNICS),
the selection of objects for spectroscopy, the observations, as well
as the data reduction and the construction of the redshift
catalogue. The properties of the spectroscopic sample are discussed in
Section~\ref{s:prop}. Section~\ref{s:lf} describes the $K$-band and
$J$-band luminosity functions of galaxies as derived from the
spectroscopic data, with a detailed discussion of the results in
Section~\ref{s:discussion}. Finally, Section~\ref{s:conclusions}
concludes this work with a summary of our results.

We assume $\Omega_M = 0.3$, $\Omega_{\Lambda} = 0.7$ throughout this
paper. We write Hubble's Constant as $H_0 = 100\ h\ \mathrm{km\
s^{-1}\ Mpc^{-1}}$, using $h = 0.60$ unless explicit dependence on $h$
is given. For convenience, we write the apparent $K'$-band magnitude
as $m_K$ in some places.

\begin{table}
\caption{The five MUNICS survey fields for which spectroscopic data
are available. The table gives the field name, the field coordinates
for the equinox 2000, the limiting magnitudes for the spectroscopic
observations, and the number of objects observed spectroscopically in
each field. The field named S2F1 contains the sparse sample
observed with the ESO VLT, the data in the field S2F5 are rather
incomplete and are excluded from any further analysis.}
\label{t:fields}
\begin{center}
\begin{tabular}{llllr}
\hline
Field & $\alpha$ (2000.0) & $\delta$ (2000.0) & Lim.\ mag.\ & Spectra \\
\hline
S2F1 & 03:06:41 & $+$00:01:12 & $K \le 19.0$ & 347 \\
S2F5 & 03:06:41 & $-$00:13:30 & $K \le 17.5$ &  29 \\
S5F1 & 10:24:01 & $+$39:46:37 & $K \le 17.5$ & 121 \\
S6F5 & 11:57:56 & $+$65:35:55 & $K \le 17.5$ & 193 \\
S7F5 & 13:34:44 & $+$16:51:44 & $K \le 17.5$ & 140 \\
\hline
\end{tabular}
\end{center}
\end{table}

%
%

\section{Spectroscopic observations}
\label{s:observations}

\subsection{The Munich Near-Infrared Cluster Survey}
\label{s:munics}

The Munich Near-Infrared Cluster Survey (MUNICS) is a wide-field
medium-deep survey in the near-infrared $K'$ and $J$ pass-bands which
is fully described in \citeauthor{munics1} (2001a; hereafter
MUNICS~I). In brief, the survey consists of a $K'$-selected catalogue
down to $K' \le 19.5$ (50 per cent completeness limit for point
sources) covering an area of 1 square degree. Additionally, 0.35
square degrees have been observed in $I$, $R$, and $V$. $B$-band
imaging has been recently completed for the same area and will be
described elsewhere. The limiting magnitudes of this main part of the
survey are 23.5 in $V$, 23.5 in $R$, 22.5 in $I$, and 21.5 in $J$. The
magnitudes are in the Vega system and refer to 50\% completeness for
point sources. The fields used in this work are five randomly selected
patches of sky (see Table~\ref{t:fields} for details). The layout of
the survey, the observations and data reduction are described in
MUNICS~I.

\subsection{Selection of the spectroscopic sample}
\label{s:selection}

Objects for spectroscopic observations were chosen from the $K$-band
selected photometric catalogue of MUNICS in five survey fields, the
details of which can be found in Table~\ref{t:fields} (see MUNICS~I
for a description of the field nomenclature).

Object selection for spectroscopy was based on two criteria. Firstly,
we aimed at a $K$-band magnitude-limited sample. Due to the use of
optical spectrographs, the appropriate $K$-band limit is determined by
the typical colours of red galaxies (roughly $R\!-\!K \simeq 4$, see
Fig.~\ref{f:colours}) and the limits of the optical spectrographs at the
telescopes we used. Trying to keep the $K$-band completeness of the
spectroscopic sample as high as possible yields sample limits of $K
\le 17.5$ for spectroscopic observations at the Calar Alto 3.5-m
telescope and $K \la 19.0$ for observations at the Very Large
Telescope (VLT). Obviously, a small fraction of very red objects will
be lost, but their number density is comparatively small anyway (see,
for instance, \citet{Martini2001} and references therein). Results on
the few very red objects identified spectroscopically can be found in
Section~\ref{s:eros}.

Secondly, in selecting objects for spectroscopy, we have tried to
exclude stars. This was done using the image-based classification of
objects into point-like objects and extended sources as described in
MUNICS~I.

To test this classification procedure, we have built up a small
spectroscopic test sample of bright objects which was purely magnitude
selected and therefore contains both stars and galaxies. The results
of the comparison between image-based and spectroscopic classification
show that our morphological approach is able to distinguish stars and
galaxies with reasonable reliability. The results of this test are
summarised in Table~\ref{t:classification}.

On a purely magnitude-limited sample containing all objects with $K'
\le 16.5$, all objects classified as point-like are indeed stars, and
the stellar contamination is 20 per cent only. On the other hand, we
can also use the complete spectroscopic catalogue to investigate the
reliability of our classification method, yielding also very good
agreement between spectral and morphological classification.

\begin{table}
  \caption{Summary of results of the test of the image-based
  classification method for a sample of objects with $K' \le 16.5$ in
  two survey fields (upper part of table) and for the whole
  spectroscopic sample (lower part of table). The numbers quoted in
  the table give the number and fraction of objects with stellar or
  galactic spectrum among objects classified as point-like or
  extended, respectively.}

  \label{t:classification}
  \begin{center}
  \begin{tabular}{lrrrr}
  \hline
  Morphological classification & \multicolumn{4}{c}{Spectral
  classification} \\
& \multicolumn{2}{c}{Star} & \multicolumn{2}{c}{Galaxy} \\
  \hline
  \multicolumn{5}{l}{Sparse sample with $K' \le 16.5$:} \\
  Point-like &  39 & 100\% &   0 &   0\% \\
  Extended   &  11 &  20\% &  44 &  80\% \\
  \hline
  \multicolumn{5}{l}{Complete spectroscopic sample:} \\
  Point-like &  64 &  79\% &  17 &  21\% \\
  Extended   &  28 &   7\% & 397 &  93\%\\
  \hline
  \end{tabular}
  \end{center}
\end{table}

Thus our method to classify objects into point-like or extended
sources can reliably reduce stellar contamination of the spectroscopic
catalogue as long as the objects are not too faint. Pre-selecting our
spectroscopic sample by image morphology results in a loss of less
than 4 per cent of all galaxies which are classified as point-like,
and in a 7 per cent contamination by stars.

\subsection{Spectroscopic observations and data reduction}

The largest part of the spectroscopic observations was carried out
with the Multi-Object Spectrograph for Calar Alto (MOSCA) at the 3.5-m
telescope at Calar Alto Observatory (Spain), and with FOcal Reducer
and low-dispersion Spectrograph (FORS) 1 and 2 \citep{fors} at the
European Southern Observatory's Very Large Telescope (VLT). Part of
the sample was observed with the Low Resolution Spectrograph (LRS;
\citealt{lrs}) at the Hobby-Eberly Telescope (HET) at McDonald
Observatory, Texas, and with ESO Faint Object Spectrograph and Camera
(EFOSC) 2 at the ESO 3.6-m telescope on La Silla (Chile). All
observing runs are listed in Table~\ref{t:runs}.

\begin{table}
\caption{Observing runs of spectroscopic follow-up observations
for the MUNICS project. The runs in 2002 were mostly carried out in
service mode.}
\label{t:runs}
\begin{center}
\begin{tabular}{lll}
\hline
Date & Telescope & Instrument \\
\hline
15.12.1999       & ESO 3.6                & EFOSC2 \\
26.--31.5.2000   & Calar Alto 3.5         & MOSCA \\
27.5.2000        & Hobby-Eberly Telescope & LRS   \\
21.11.2000       & ESO VLT UT1 (Antu)     & FORS1 \\
21.--22.11.2000  & ESO VLT UT2 (Kueyen)   & FORS2 \\
24.--28.11.2000  & Calar Alto 3.5         & MOSCA \\
17.--20.1.2001   & Calar Alto 3.5         & MOSCA \\ 
26.3.--1.4..2001 & Calar Alto 3.5         & MOSCA \\ 
18.--21.5.2001   & Calar Alto 3.5         & MOSCA \\ 
15.--20.12.2001  & Calar Alto 3.5         & MOSCA \\ 
9.--12.4.2002    & Calar Alto 3.5         & MOSCA \\
9.--17.5.2002    & Calar Alto 3.5         & MOSCA \\
7.10.2002	 & Calar Alto 3.5         & MOSCA \\
9.--10.11.2002	 & Calar Alto 3.5         & MOSCA \\
\hline
\end{tabular}
\end{center}
\end{table}

MOSCA was used with the Green~500 grism and without any
filter. MOSCA's multi-object spectroscopy mode uses slit masks
containing typically 20$\dots$25 slits of 1.5\arcsec\ width. MOSCA is
equipped with a 2048$\times$4096 pixel CCD with 15$\mu$m pixel size,
yielding an effective area of 10\arcmin $\times$ 10\arcmin\ usable for
spectroscopy. 

At the VLT, FORS1 offers 19 movable slit-lets, whereas FORS2 is
equipped with a Mask Exchange Unit (MXU), which allows spectroscopy of
a larger number of objects. For both instruments a slit width of
1\arcsec\ was chosen. The CCD detectors are 2048$\times$2048 pixel in
size with 24$\mu$m pixels, thus allowing a field of roughly 3\arcmin
$\times$ 7\arcmin to be used for spectroscopy. Grism 300~I and filter
OG590 was used for the observations

In multi-object spectroscopy mode, the LRS at HET provides 13 slit-lets
of 1.5\arcsec\ width. The spectra were obtained through grism 300 and
filter GG385.

Finally, EFOSC2 at the ESO 3.6-m telescope was equipped with grism 11
and a slit-width of 1.0 \arcsec . The technical characteristics of all
instruments used for the spectroscopic observations are summarised in
Table~\ref{t:instr}.

\begin{table}
\caption{Technical parameters of the spectrographs used for the
observations. The effective resolution of the resulting spectra is
given for the grisms listed in the table and the appropriate
slit-widths mentioned in the text (the seeing during the spectroscopic
observations always was of the order of the slit-width).}
\label{t:instr}
\begin{center}
\begin{tabular}{llcc}
\hline
Instrument & Grism \& Filter & Spectral range & Resolution \\
\hline
MOSCA  & Green 500       & 4300 \AA\ -- 8000 \AA & 13.6 \AA \\
EFOSC2 & G11             & 3380 \AA\ -- 7520 \AA & 13.2 \AA \\
LRS    & G300 $+$ GG385  & 4000 \AA\ -- 8000 \AA & 13.9 \AA \\
FORS1  & G300I $+$ OG590 & 6000 \AA\ -- 9500 \AA & 13.0 \AA \\
FORS2  & G300I $+$ OG590 & 6000 \AA\ -- 9500 \AA & 13.0 \AA \\
\hline
\end{tabular}
\end{center}
\end{table}

The objects selected for spectroscopy were sorted into bins according
to their apparent $R$-band magnitude. The magnitude ranges, the
typical number of mask setups for each MUNICS field ($13\arcmin \times
13\arcmin$), and the exposure times are given in
Table~\ref{t:obs}. All objects with $R \le 21.5$ (roughly
corresponding to $K' \le 17.5$) were observed at the Calar Alto 3.5-m
telescope, the ESO 3.6-m telescope, and the HET, whereas spectroscopy
of the fainter sample was carried out at the ESO VLT.

\begin{table}
\caption{$R$-band magnitude bins, typical number of masks per MUNICS
field, and exposure times for the spectroscopic observations. Note
that the faintest objects with $R > 21.5$ were observed at the ESO
VLT, while all brighter objects were mainly observed at the 3.5-m
telescope at Calar Alto, a few also at the ESO 3.6-m telescope, or at
the HET.}
\label{t:obs}
\begin{center}
\begin{tabular}{ccc}
\hline
Magnitude & Number of masks & Exposure time \\
\hline
$R \le 18.5$        & 1 & $1 \times 1800$ s \\
$18.5 < R \le 19.5$ & 2 & $2 \times 2100$ s \\
$19.5 < R \le 20.5$ & 3 & $5 \times 2400$ s \\
$19.5 < R \le 21.5$ & 4 & $5 \times 5400$ s \\
$R > 21.5$          & 9 & $3 \times 3000$ s \\
\hline
\end{tabular}
\end{center}
\end{table}

Data reduction was performed using
\textsc{iraf}\footnote{\textsc{iraf}, the \textsc{image reduction and
analysis facility}, is distributed by the National Optical Astronomy
Observatories, which are operated by the Association of Universities
for Research in Astronomy, Inc., under cooperative agreement with the
National Science Foundation.}, except for cosmic ray
filtering. Firstly, a standard CCD reduction was performed, including
over-scan and bias correction, as well as interpolation of bad
pixels. Cosmic rays were identified by searching for narrow local
maxima in the image and fitting a bivariate rotated Gaussian to each
maximum.  A locally deviant pixel is then replaced by the mean value
of the surrounding pixels if the Gaussian obeys appropriate flux ratio
and sharpness criteria \citep{GoesslRiffeser2002}.

The following reduction of the spectra was performed within the
\textsc{iraf specred} package. The locations of the spectra on the
two-dimensional image were defined, and the trace of the spectra
produced by the distortion of the spectrograph was approximated by a
polynomial of fourth order. Flat-fields were constructed from
typically 21 images of an internal continuum lamp, and normalised by
fitting a set of low-order splines to the illumination function of
each slit. After dividing the frames by the appropriate flat-field,
the apertures for extracting one-dimensional spectra from the
two-dimensional images as well as the apertures for subtraction of the
night sky were defined. This was done manually for each object to
ensure maximum signal-to-noise ratio and accurate sky subtraction. If
possible, sky apertures on both sides of each object were used, and
the variation of the night sky perpendicular to the spectrum of each
object was approximated by a polynomial of second order. Note that we
used lower-order polynomials in cases where there were only few pixels
to define the sky region in order to prevent artefacts due to
ill-defined approximations. After that the one-dimensional spectra
were extracted by summing the pixel values for each wavelength bin
within the manually defined apertures and subtracting the
appropriately scaled sky value from the fit perpendicular to the
spectrum. Additionally we produced two-dimensional sky-subtracted
images to be able to check whether a feature seen in the
one-dimensional spectrum is a real spectral feature or a residual, for
example from the replacement of cosmic ray events or the subtraction
of sky-lines.

In the case of the MOSCA, EFOSC2 and LRS observations, spectra of
internal lamps were used for wavelength calibration. The wavelength
transformation was computed in form of a fourth order polynomial. Then
the one-dimensional spectra were transformed to wavelength coordinates
and re-binned using the appropriate resolution. Note that we usually
took only one wavelength calibration image per mask, either between
two mask exposures or after finishing all mask
exposures. Nevertheless, this does not severely affect the wavelength
calibration, since the maximum shift of the spectral pattern of a
calibration lamp, caused by the flexure of the instrument, was
measured to be 1 pixel for MOSCA between two extreme orientations of
the spectrograph. According to the manuals, the flexure of the other
instruments should have a similar order of magnitude. Also, the
accuracy of the wavelength calibration can be efficiently examined by
measuring the positions of night-sky emission lines. The comparison
showed negligible deviations. The influence on the redshift
determination will be discussed in Section~\ref{s:error}.  For the
FORS data we used the lines of the night sky for the wavelength
calibration, with the line data taken from
\citet{Osterbrock1996,Osterbrock1997}.

\begin{figure*}
  \centerline{\epsfig{figure=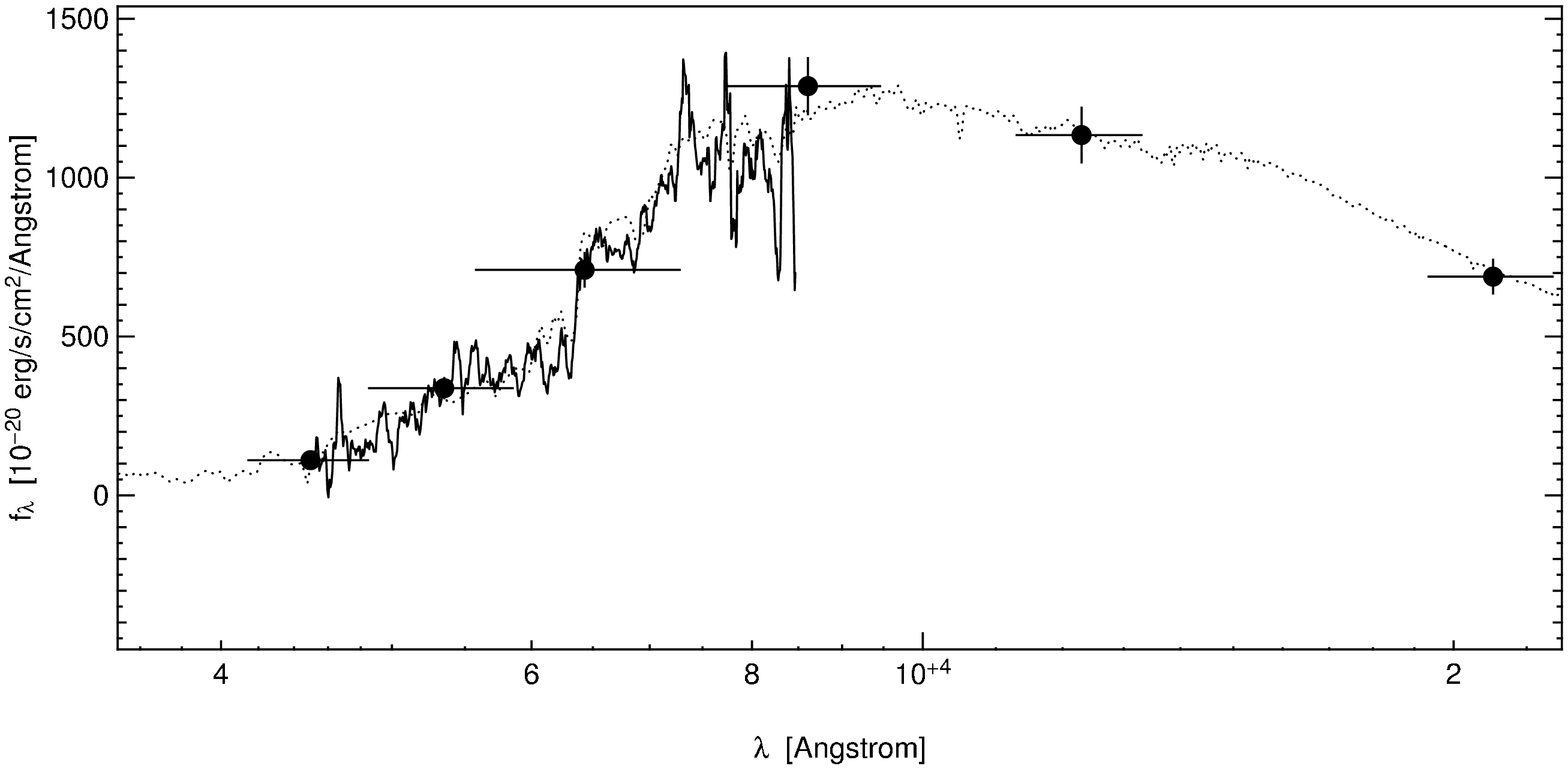,width=5.cm,angle=270}}
  \vspace*{.25cm}
  \centerline{\epsfig{figure=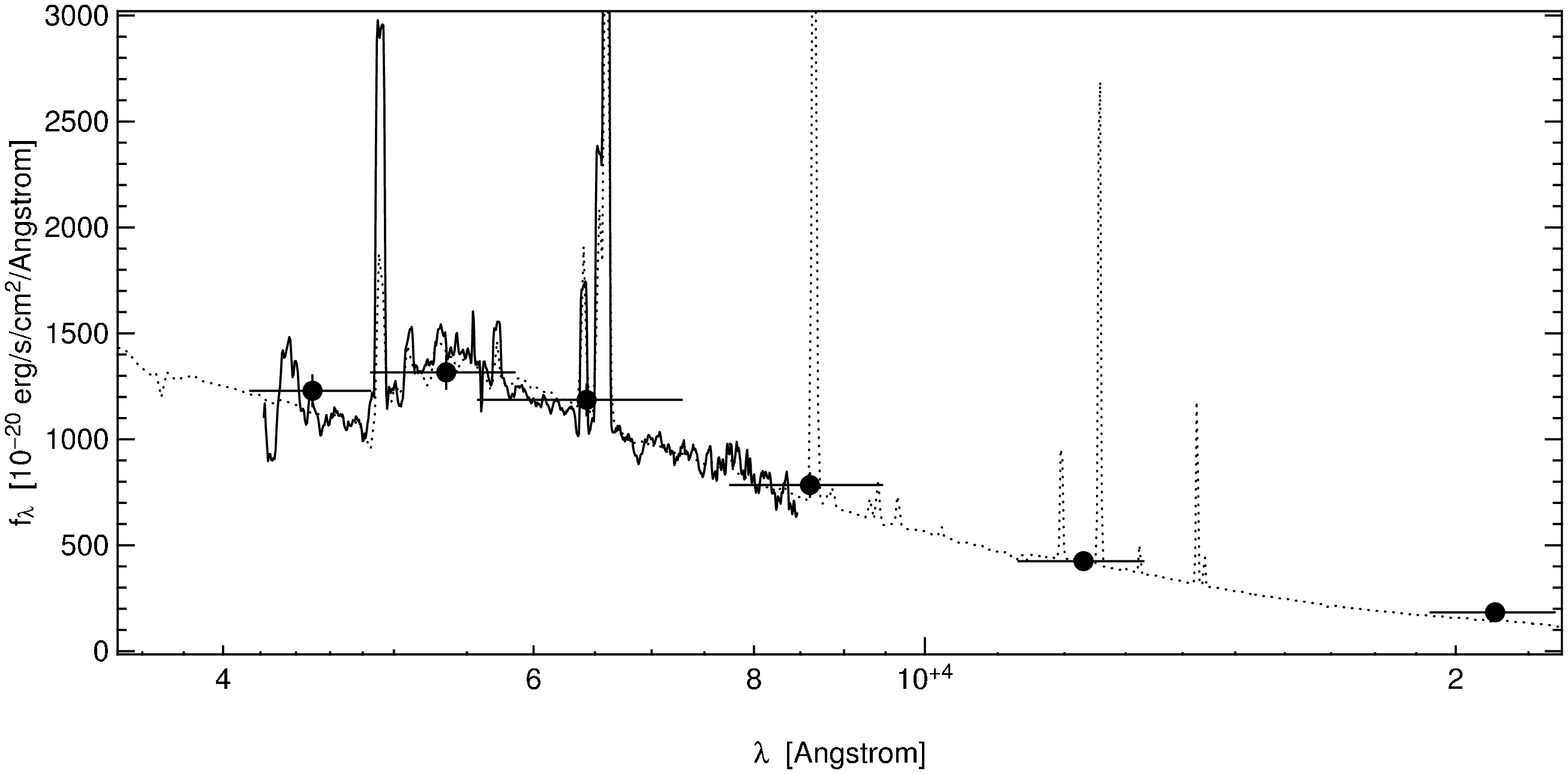,width=5.cm,angle=270}}
  \caption{Comparison of the flux calibrated spectrum (solid line),
  the flux derived from MUNICS broad-band photometry in the filters
  $B$, $V$, $R$, $I$, $J$, and $K$ (filled circles with error bars),
  and a fit of a stellar population synthesis model to the photometric
  data (dotted line) for two objects in the spectroscopic MUNICS
  catalogue.}
\label{f:flux}
\end{figure*}

Finally, the spectra were flux calibrated using spectro-photometric
standard stars observed during each night. Extinction correction was
based on average extinction curves either as published by the
observatories (ESO for La Silla and Paranal Observatory, and McDonald
Observatory) or by \citet{caext} for the Calar Alto observations. The
accuracy of the flux calibration was tested by comparing the flux in
individual broad-band photometry filters to the flux in the
flux-calibrated spectrum. Fig.~\ref{f:flux} shows an example for
this method. Clearly, the flux calibration is satisfying, with
deviations visible only at the ends of the spectral range of the
instrument, where the corrections due to the response function of the
grism become large.

Redshifts were determined using the \textsc{rv} package within
\textsc{iraf}. Prominent absorption and emission features were
identified in the spectra to obtain the corresponding redshift of a
galaxy.\footnote{So far we have not applied cross-correlation
techniques to determine redshifts, however we plan to do so in the
future to reduce the fraction of galaxies without secure redshift due
to low signal-to-noise ratios, although first tests show that we
cannot significantly increase the number of securely measured
redshifts by using correlation methods.} We compared the
one-dimensional spectrum with the two-dimensional sky-subtracted image
to exclude features possibly affected by residuals from night-sky
subtraction or cosmic filtering. Furthermore we ensured that the
radial velocity displacement between different spectral features is
not larger than the typical error expected from the measurement of the
line centre and the wavelength calibration, i.e.\ typically $\Delta z
\la 5 \times 10^{-4}$.

\subsection{Catalogue construction}
\label{s:cat}

\begin{figure*}
  \epsfig{figure=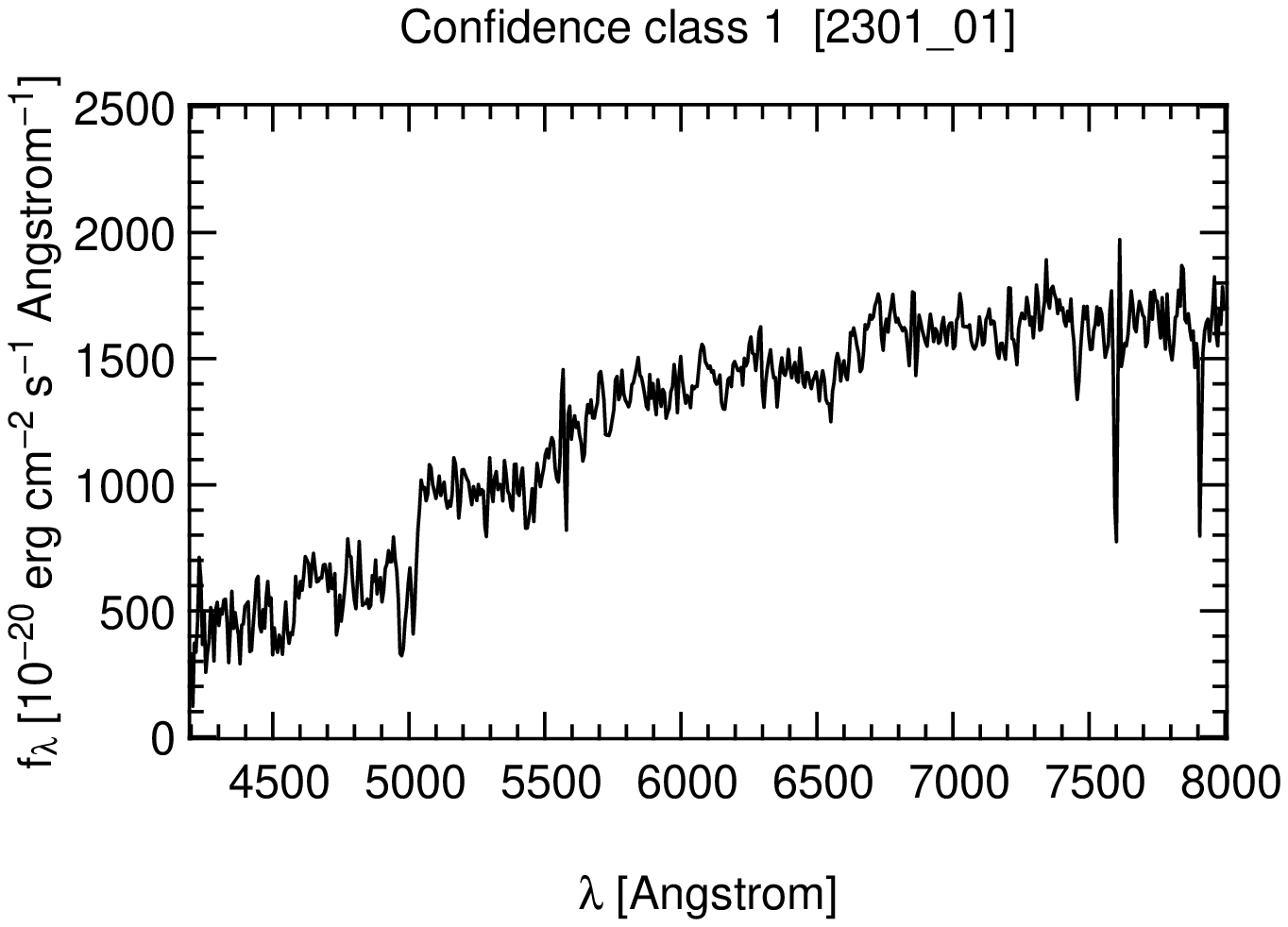,width=5.25cm}
  \hfill
  \epsfig{figure=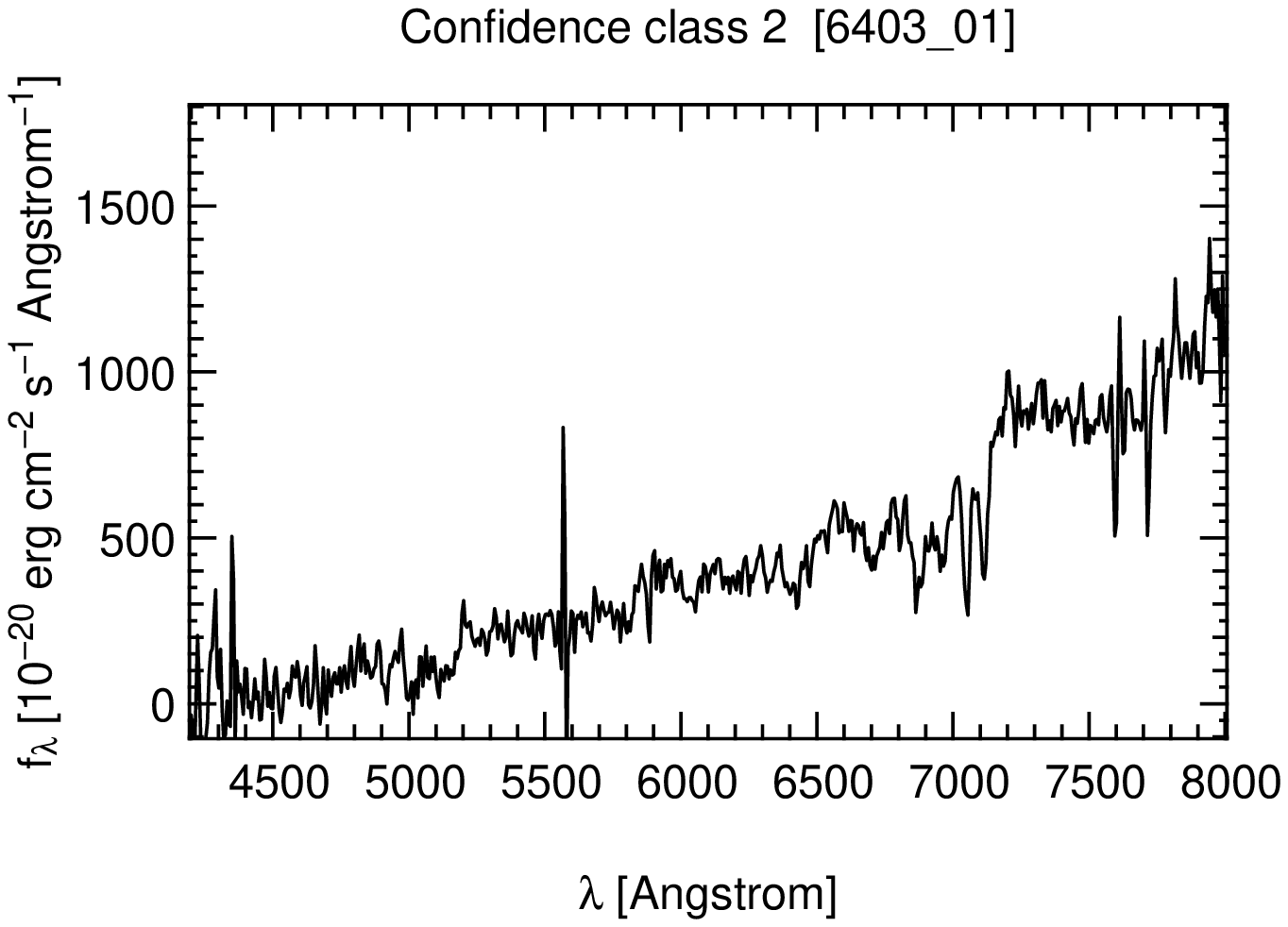,width=5.25cm}
  \hfill
  \epsfig{figure=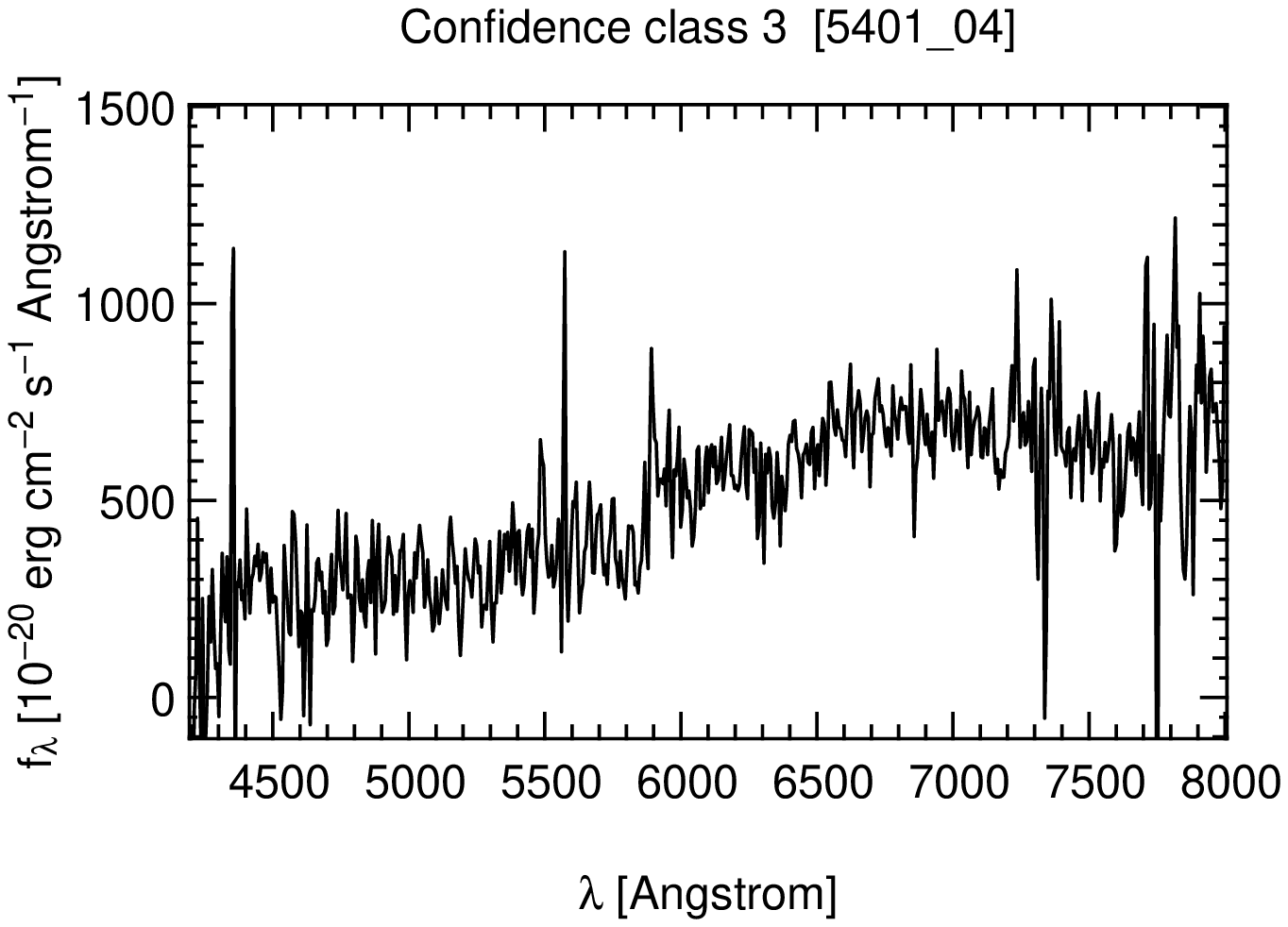,width=5.25cm}
 
  \vspace*{.5cm}

  \epsfig{figure=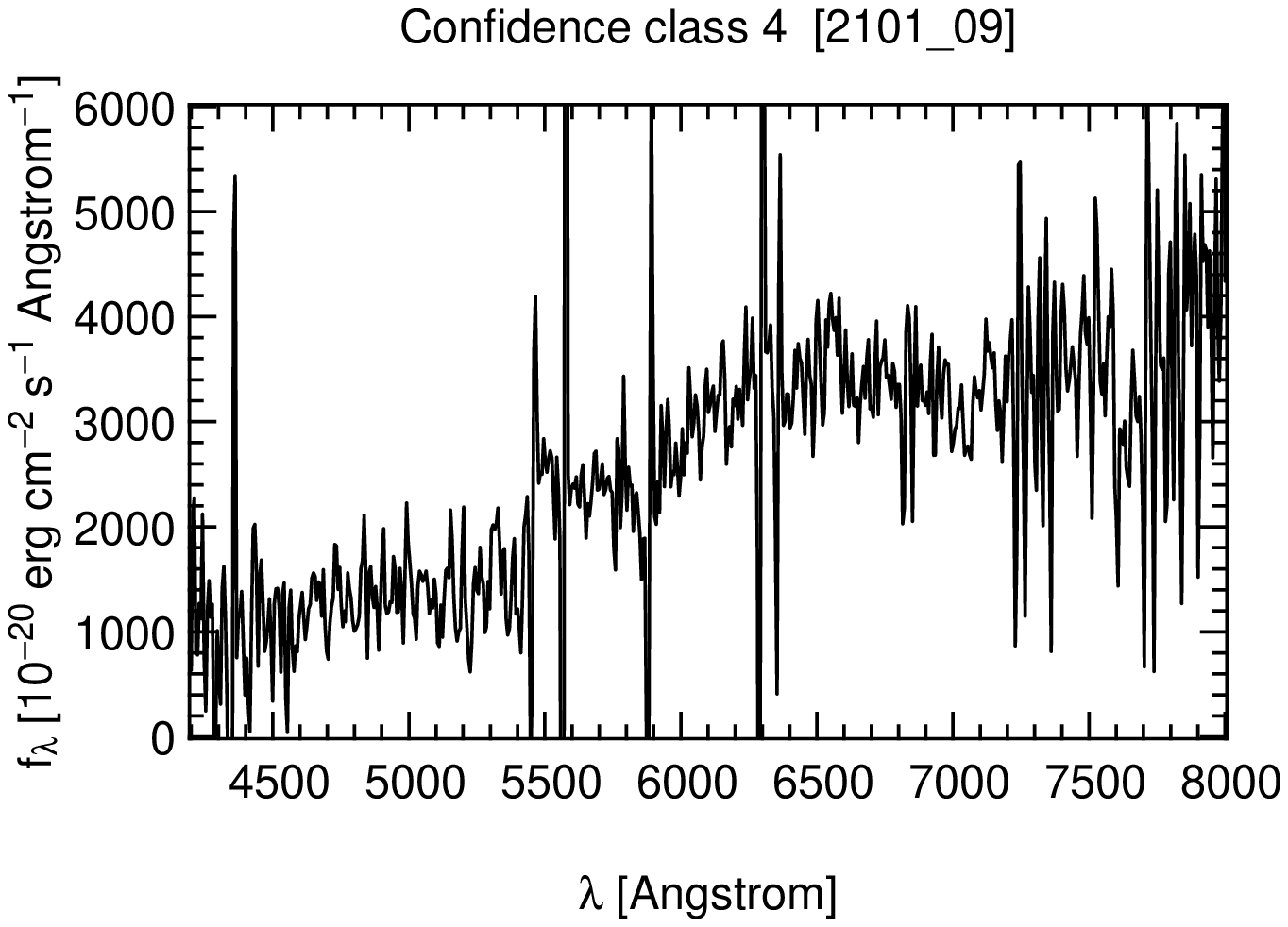,width=5.25cm}
  \hfill
  \epsfig{figure=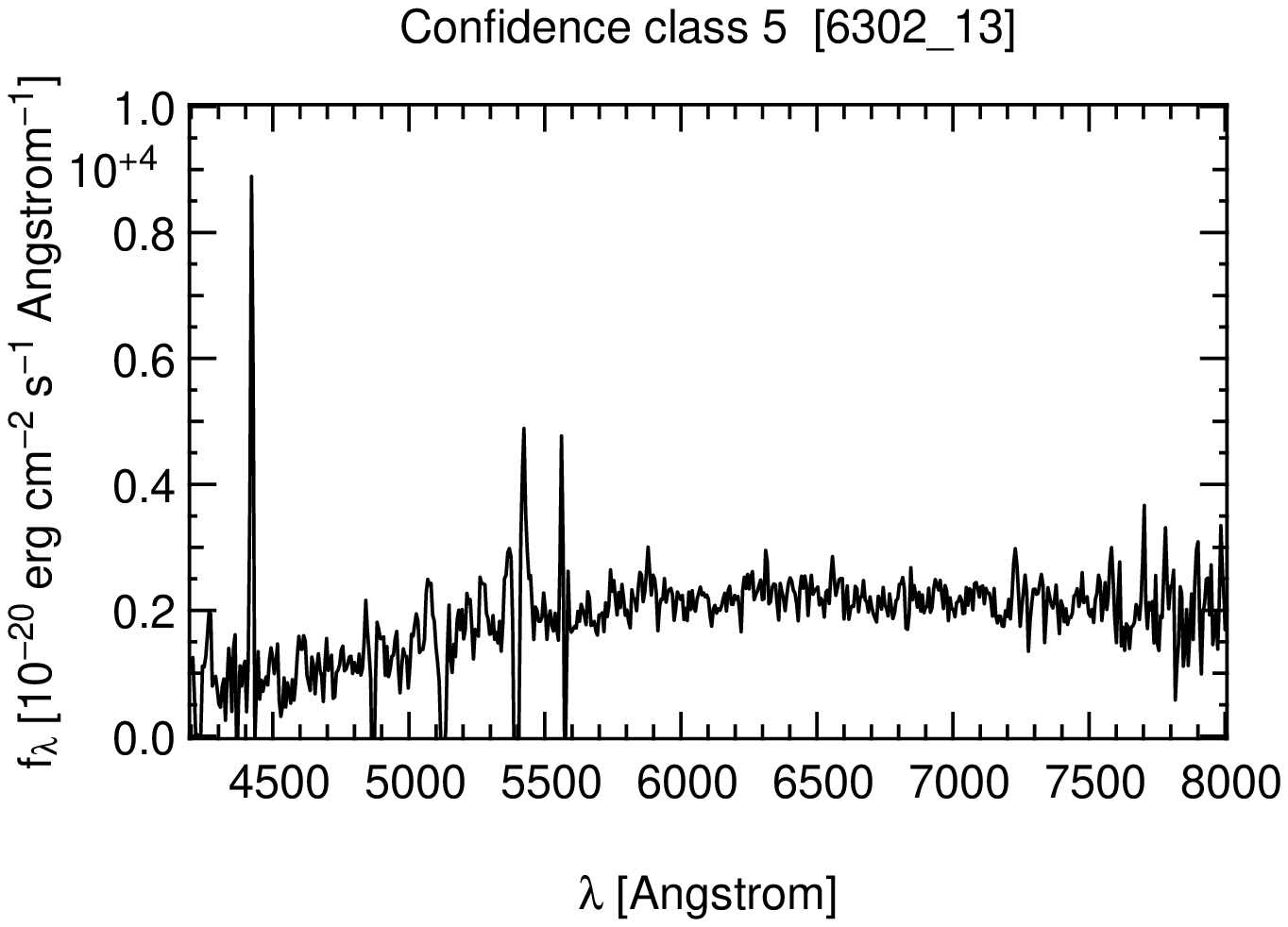,width=5.25cm}
  \hfill
  \epsfig{figure=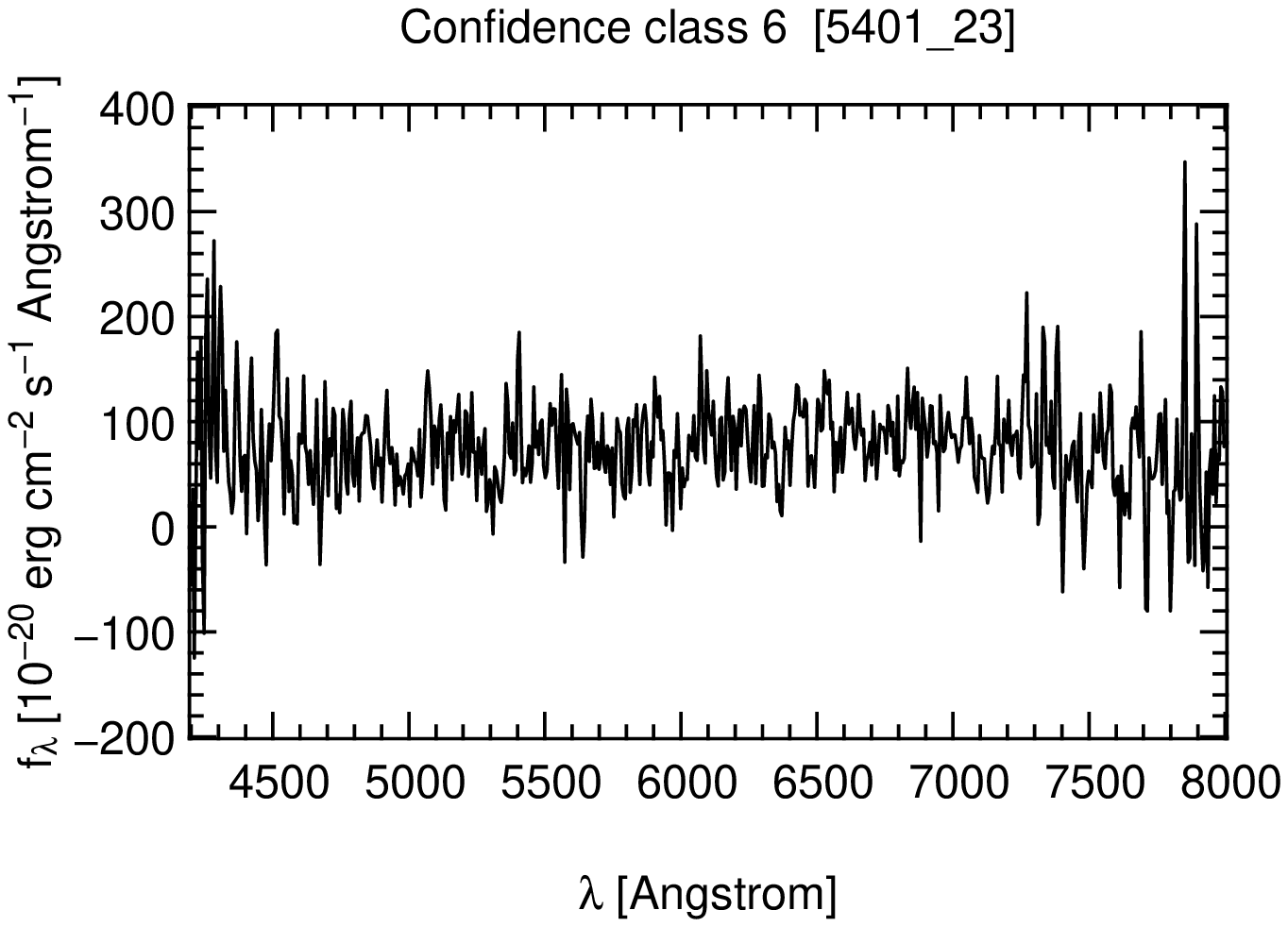,width=5.25cm}

  \caption{Examples for the assignment of confidence classes, ranging
  from ``1'' (upper left) to ``6'' (lower right). The numbers in square
  brackets denote the internal identification number of the object.}

\label{f:conf}

\end{figure*}

The final spectroscopic catalogue contains the celestial coordinates,
the pixel coordinates form the photometric catalogue, the instrumental
identification (mask and slit number), the measured redshift, the
object classification based on the observed spectrum (galaxy, AGN,
star, or unidentified object), a confidence class for the redshift
determination, and a list of spectral features used for measuring the
object's redshift. The confidence class assigned to each object ranges
from ``1'' to ``6'', where ``1'' means highest confidence, and ``6''
means that no redshift could be determined. Descriptions of the
confidence classes can be found in Table~\ref{t:conf}, although these
can give only a crude impression of the meaning of the confidence
classes. For all further analysis objects with confidence classes
``5'' and ``6'' are excluded. To check the assignment of confidence
classes, we have made use of the 44 repeat observations of 41 objects
described in detail in Section~\ref{s:error}. The agreement between
confidence classes attributed to the same object from different
observations is in general very good. Deviations can be explained by
different signal-to-noise ratios due to weather conditions or
technical reasons. Not a single object with confidence class ``1'' to
``4'' has got a deviant redshift from a repeat observation. Typical
examples for spectra and the appropriate confidence classes are shown
in Fig.~\ref{f:conf}.

\begin{table}
\caption{Description of the confidence classes for the redshift
measurement used in the final spectroscopic catalogue.}
\label{t:conf}
\begin{center}
\begin{tabular}{cp{6cm}}
\hline Class & Description \\ \hline 
1 & Many spectral features securely identified, very high
    signal-to-noise ratio \\
2 & Many spectral features securely identified, high signal-to-noise
    ratio \\ 
3 & Many spectral features securely identified, intermediate
    signal-to-noise ratio \\
4 & Several spectral features identified, lower signal-to-noise ratio \\
5 & Tentative identification of only a few lines or objects with
    single emission line, low signal-to-noise ratio \\
6 & no redshift determination possible (either very low
    signal-to-noise ratio or technical problem) \\ \hline
\end{tabular}
\end{center}
\end{table}

%
%

\section{Properties of the spectroscopic sample}
\label{s:prop}

\subsection{Object classes}

The final catalogue contains 830 objects. 725 (or 87.3 per cent) are
galaxies, 3 (0.4 per cent) are active galactic nuclei, 100 (12.0 per
cent) are stars, and 2 objects (0.2 per cent) remain unidentified due
to bad quality of the spectra or a lack of prominent spectral features
in the wavelength range covered by the spectra. Redshifts could be
determined for 500 objects (confidence classes 1 to 4), or 68.7 per
cent of all extragalactic objects.

\subsection{Confidence classes}

The number of objects in the confidence classes described above are
given in Table~\ref{t:nconf}. The first three confidence classes
contain 57.1 per cent of all objects selected for spectroscopy; these
objects have high-quality spectra. Redshifts could be determined for
68.7 per cent of the objects.

\begin{table}
\caption{Distribution of extragalactic objects into the 6 confidence
classes used in the final spectroscopic catalogue.}
\label{t:nconf}
\begin{center}
\begin{tabular}{crr}
\hline
Class & Number of objects & Per cent \\
\hline
1     & 162 &  22.25 \\
2     & 137 &  18.82 \\
3     & 117 &  16.07 \\
4     &  84 &  11.54 \\
5     &   8 &   1.10 \\
6     & 220 &  30.22 \\
\hline
Total & 728 & 100.00 \\
\hline
\end{tabular}
\end{center}
\end{table}

\subsection{Redshift distribution}

\begin{figure}
  \centerline{\epsfig{figure=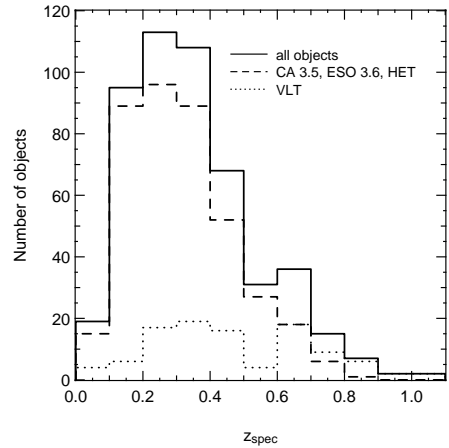,width=5.75cm}} \caption{The
  overall redshift distribution of all galaxies and active galactic
  nuclei in the final spectroscopic catalogue (solid line), divided
  into objects with spectroscopy from the Calar Alto 3.5-m telescope,
  the ESO 3.6-m telescope or the HET (dashed line) and from the VLT
  (dotted line). Note the excess of objects at redshifts above 0.6 due
  to the VLT observations.}
\label{f:zhist}
\end{figure}

The redshift distribution of the spectroscopic sample is shown in
Fig.~\ref{f:zhist}. The histogram has the shape expected for a
magnitude limited spectroscopic sample, except for the slight excess
of objects with redshifts $z \ga 0.6$ which, of course, are due to the
sparse sample of faint objects observed at the VLT. The lack of
objects with redshifts from VLT spectroscopy in the range $0.5 < z <
0.6$ is most likely due to the limits of the instruments' spectral
range. Objects with these redshifts have their 4000\AA\ break around
the lower end of the spectral range and the H$\alpha$ line around the
upper end of the spectral range. Thus redshifts of objects with low
signal-to-noise ratio may be difficult to determine due to the absence
of those prominent spectral features in the observed spectra. This
kind of bias can be corrected for by looking at the fraction of
galaxies with securely determined redshift as a function of two
colours, as will be shown below.

\subsection{Accuracy of redshift determination}
\label{s:error}

For objects with secure redshifts, where we consider all spectral
features to be correctly identified, possible sources of error in the
determination of redshift are firstly a possible displacement between
the object and the slit centre, secondly the error of the wavelength
calibration, and finally the error in the measurement of the line
centres.

\begin{enumerate}

  \item An offset of the object with respect to the slit centre causes
  an error of the redshift measurement. We estimate the RMS error to
  be 0.5\arcsec , corresponding to an inaccuracy of $\sigma \simeq 6
  \times 10^{-4}$ in redshift space. This value is estimated from the
  accuracy of astrometry in the MUNICS catalogue (see MUNICS~I), the
  typical errors of mask alignment and under the reasonable assumption
  that errors in the mechanical production of the masks (or in the
  slit-let positioning for FORS1 and the LRS) can be neglected.

  \item The random errors of the wavelength calibration, the RMS of
  the residuals of the fit to the identified lines, are of the order
  0.3~\AA , or, in redshift, $\sigma \simeq 4 \times 10^{-5}$. Compared
  to the other sources of error, this inaccuracy can be neglected. For
  MOSCA, the maximum shift of the spectrum of the calibration lamp
  between zenith and airmass 2 was measured to be 1 pixel,
  corresponding to a systematic error of $\sigma \la 4 \times 10^{-4}$
  in redshift.

  \item Typical errors of the measurement of the line centres for the
  redshift determination are $\sigma \la 5 \times 10^{-4}$.

\end{enumerate}

The overall limit to the accuracy estimated from these errors is
$\sigma \simeq 8 \times 10^{-4}$ in redshift space.

We can also empirically determine the error of the redshift
measurements from multiple observation of objects. A total of 41
galaxies -- about 8 per cent of the sample -- has been observed
repeatedly, usually in order to fill gaps on the slit masks used for
observation, thus enabling us to estimate typical errors of the
redshift measurement by comparing the redshifts obtained from
different observations. Note that 13 of these objects have been
observed both with MOSCA and FORS, and 5 using MOSCA and the LRS. Thus
it is possible to check for any systematics between redshift
determinations with different spectrographs, and, in particular, for
systematic offsets between different means of wavelength calibration,
since we used the lines of the night sky for the FORS observations,
whereas spectra from calibration lamps were used for the other
spectrographs.

\begin{figure}
  \centerline{\epsfig{figure=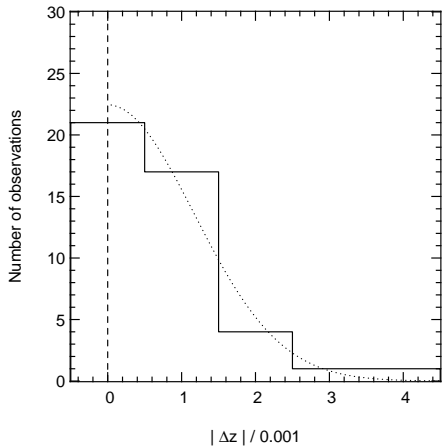,width=5.75cm}}
  \caption{Histogram of the absolute values of pairwise differences of
  the measured redshifts of objects with multiple spectroscopic
  observations in the catalogue (solid line). The dotted line shows a
  Gaussian approximation to the histogram for which $\sigma' = 12
  \times 10^{-4}$, corresponding to a an RMS error of single
  measurements of $\sigma = 8 \times 10^{-4}$.}  \label{f:deltaz}
\end{figure}

The histogram of the absolute values of pairwise differences for the
44 repeated redshift measurements for the 41 objects is shown in
Fig.~\ref{f:deltaz}. Also shown is a Gaussian approximation to the
histogram having a width characterised by $\sigma' = 12 \times
10^{-4}$, corresponding to a an RMS error of single measurements of
$\sigma = \sigma' / \sqrt{2} = 8 \times 10^{-4}$, exactly the same
value we have derived from the formal error analysis. Furthermore, we
do not find any systematical differences between redshift
determinations achieved with data from different instruments or
observing runs.

\subsection{Colours, magnitudes, and redshifts}
\label{s:colours}

\begin{figure*}
  \epsfig{figure=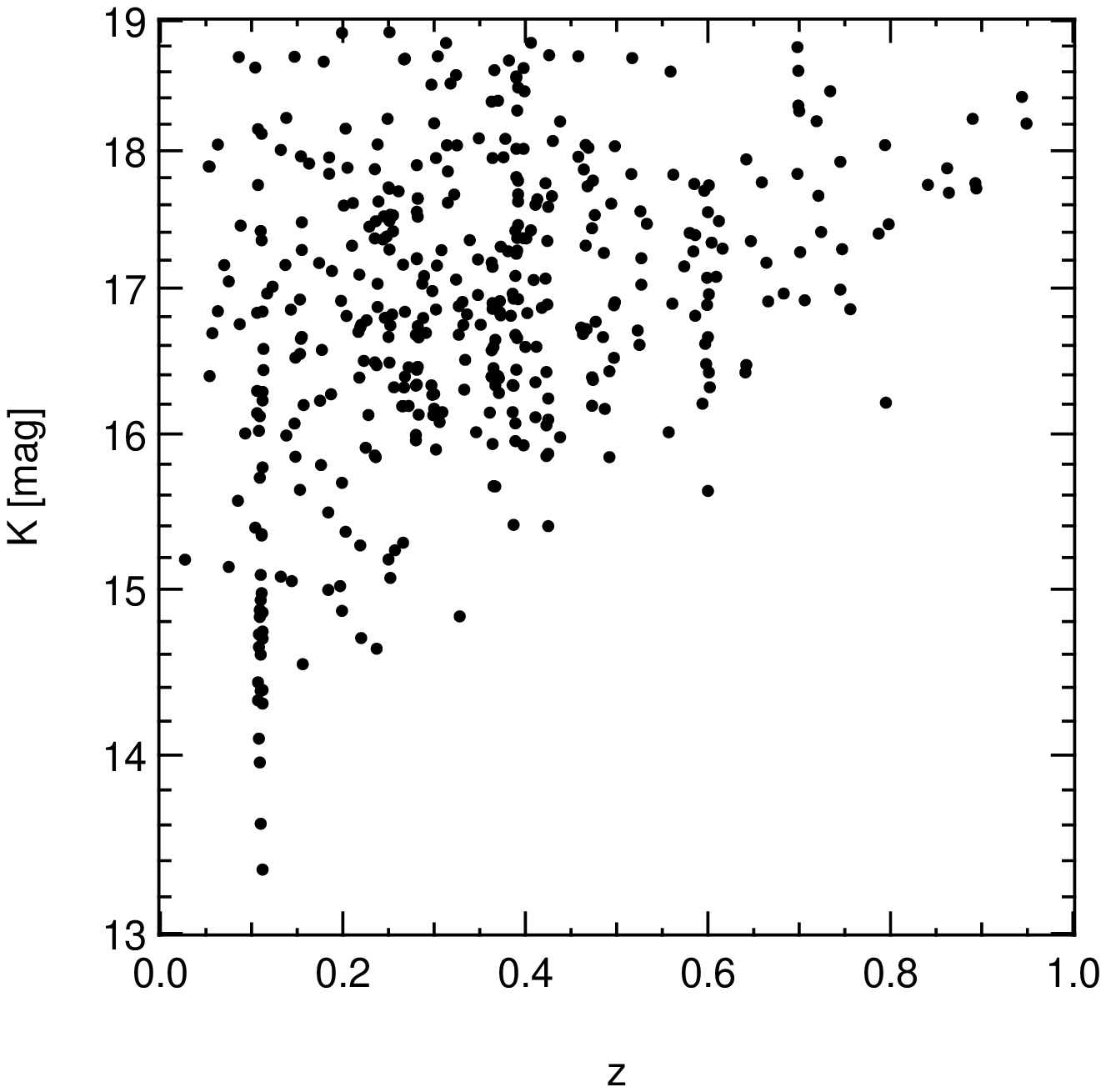,width=5.75cm}
  \hspace*{1cm}
  \epsfig{figure=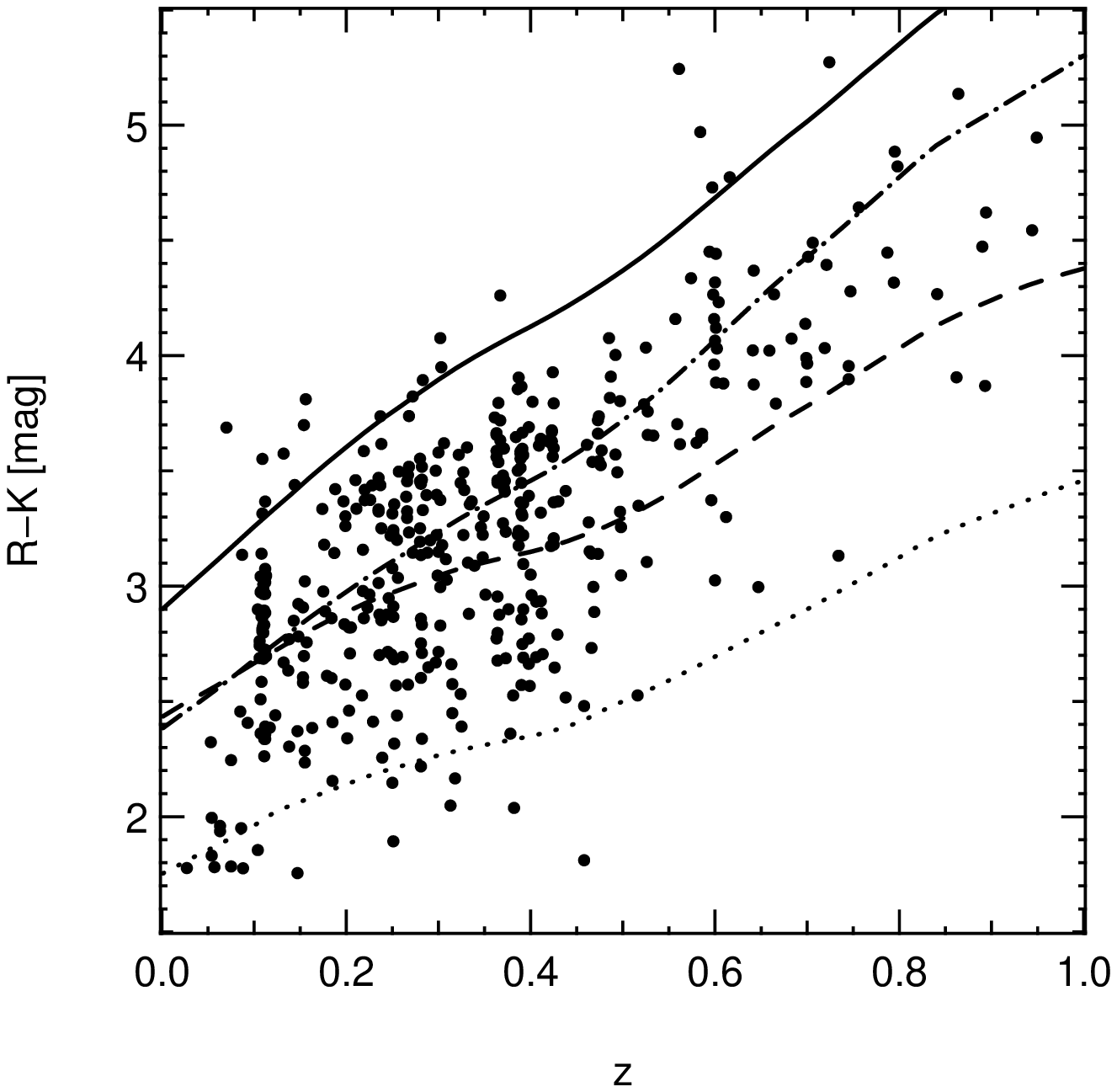,width=5.75cm}

  \vspace*{.5cm}

  \epsfig{figure=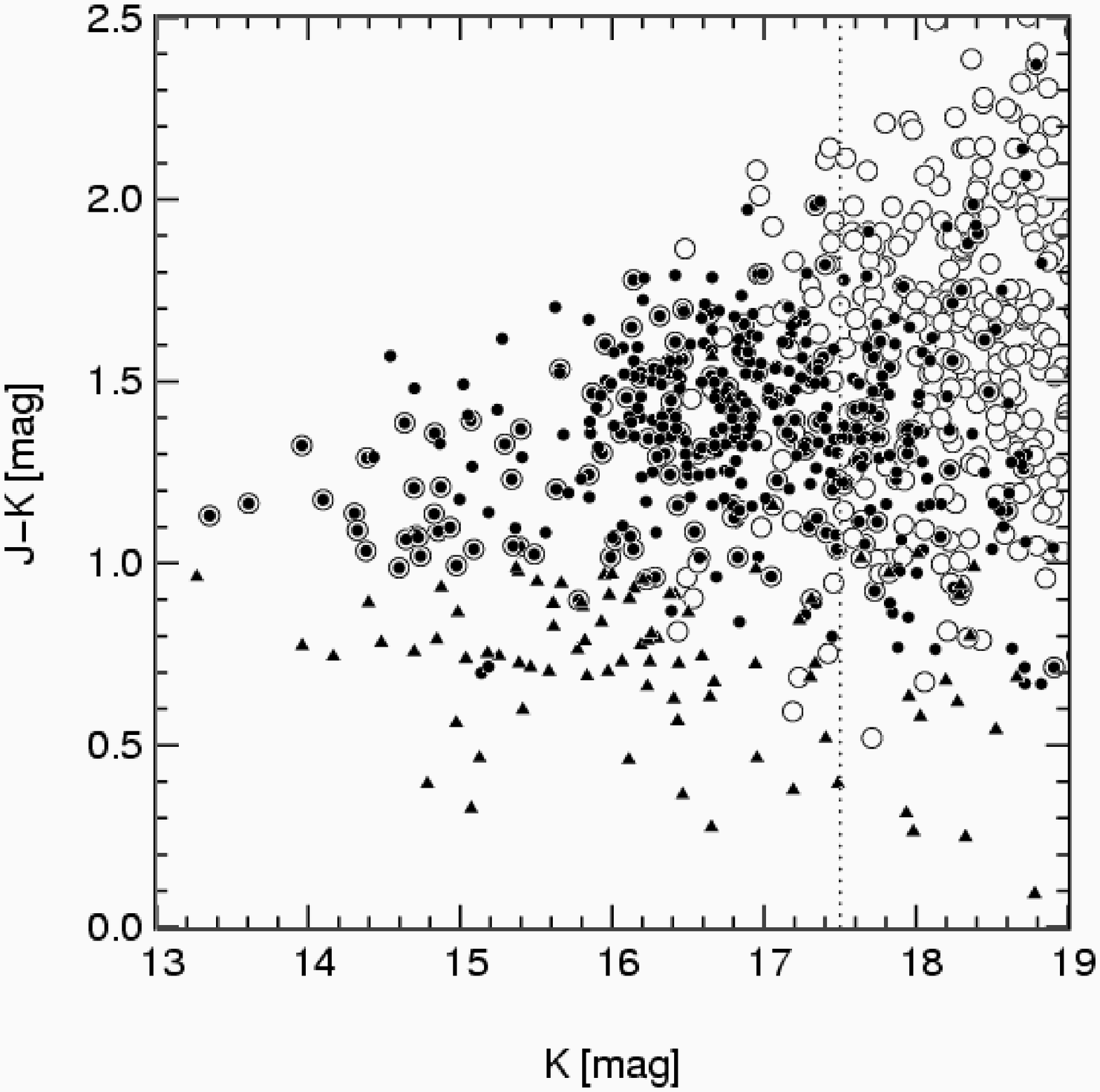,width=5.75cm}
  \hspace*{1cm}
  \epsfig{figure=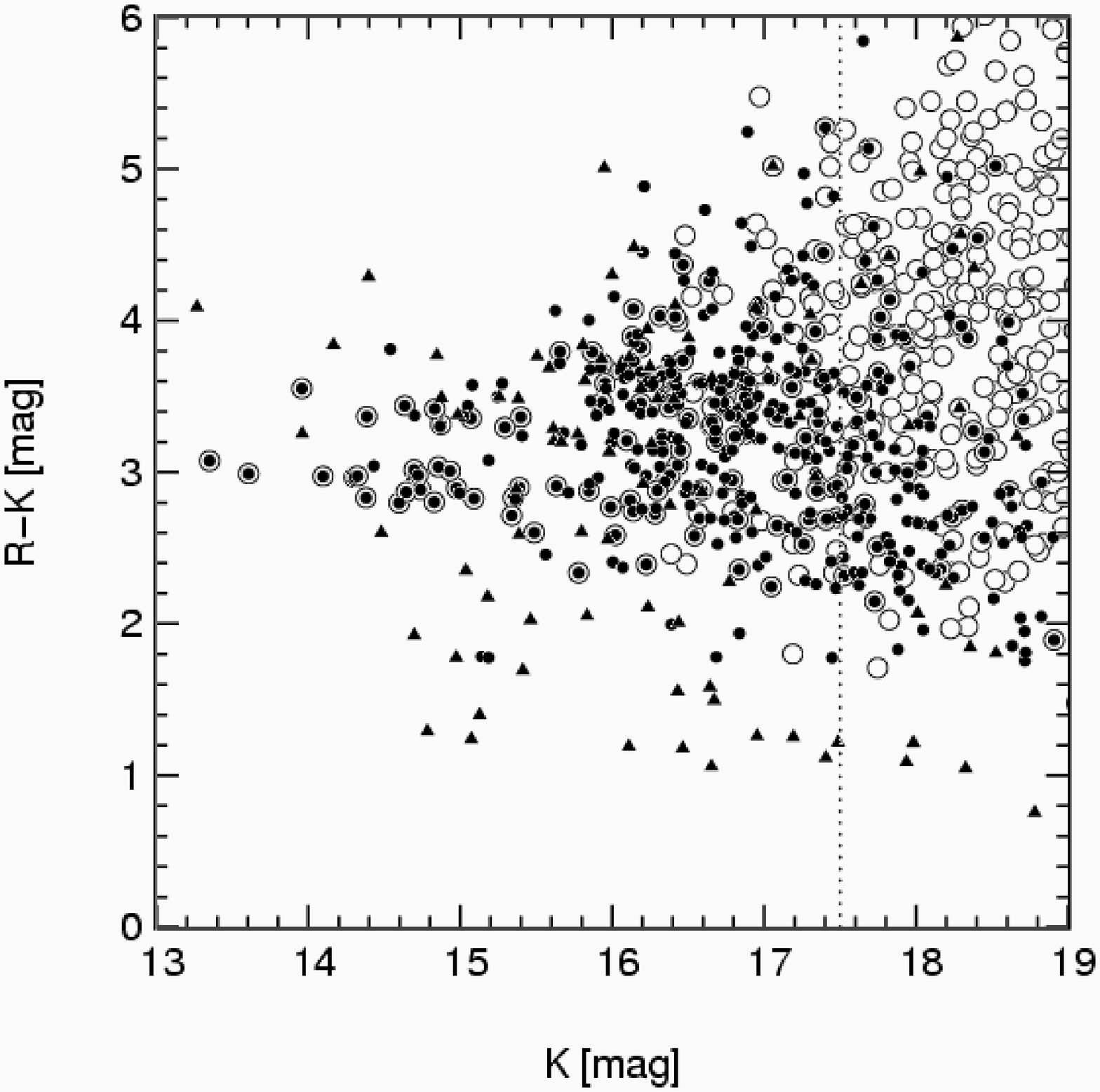,width=5.75cm}

  \caption{\textit{Upper left panel:} The distribution of apparent
  $K$-band magnitudes of galaxies versus redshift in the MUNICS
  spectroscopic sample. \textit{Upper right panel:} The
  $R\!-\!K$-colour versus redshift diagram for galaxies with
  spectroscopic redshifts. Also shown are model spectral energy
  distributions from (see text for details), roughly corresponding to
  Hubble types E (solid line), Sa (dash-dotted line), Sb (dashed
  line), and Sc (dotted line). \textit{Lower left panel:} The
  $J\!-\!K$ versus $K$ colour-magnitude diagram for galaxies (filled
  circles) and stars (filled triangles) with good-quality spectra. The
  open circles show the distribution of objects classified as extended
  in the photometric catalogue of one of the MUNICS fields
  (S2F1). Also indicated is the magnitude limit of the large and
  bright spectroscopic sample at $K = 17.5$ (dotted line).
  \textit{Lower right panel:} Same as lower left panel, but for the
  distribution of $R\!-\!K$ colours versus apparent $K$-band
  magnitude.}

  \label{f:colours}
\end{figure*}

To give a feeling for the distribution of objects in magnitudes,
colours, and redshifts, we present diagrams for various combinations
of these properties in Fig.~\ref{f:colours}.

In the upper panels of the figure, we show distributions of the
apparent $K$-band magnitudes and the $R\!-\!K$ colours versus
redshift. The comparison of model spectral energy distributions from a
combination of empirical spectra and stellar population synthesis
models by \citet{Maraston98} for various Hubble types to the $R\!-\!K$
colours of the objects in the MUNICS spectroscopic catalogue (upper
right panel of Fig.~\ref{f:colours}) shows good agreement, although
we seem to miss blue objects at higher redshifts, a fact which must be
accounted for in any analysis of this dataset.

The two lower panels of Fig.~\ref{f:colours} shows the distribution
of extended objects in the photometric catalogue together with the
colour distribution of galaxies with spectroscopic redshifts. Comparison
shows that the spectroscopic catalogue is a fair representation of the
distribution of objects in the photometric sample. Clearly, there is
some incompleteness for faint red objects, but this can be corrected
for, as will be shown in the following Section.

\subsection{Redshift sampling rate and sky coverage}
\label{s:sampling}

\begin{figure*}
  \epsfig{figure=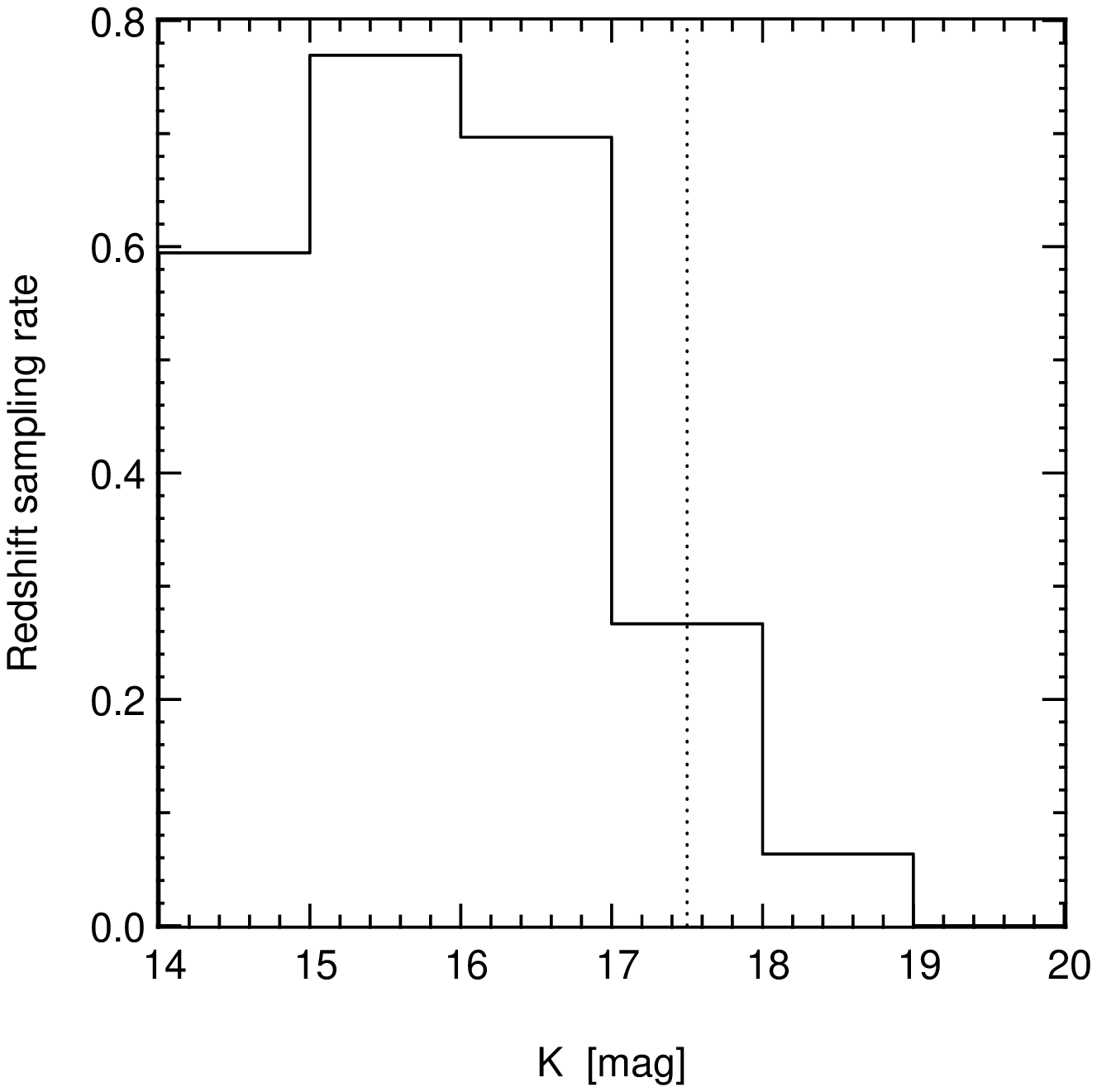,height=5.25cm}
  \hfill
  \epsfig{figure=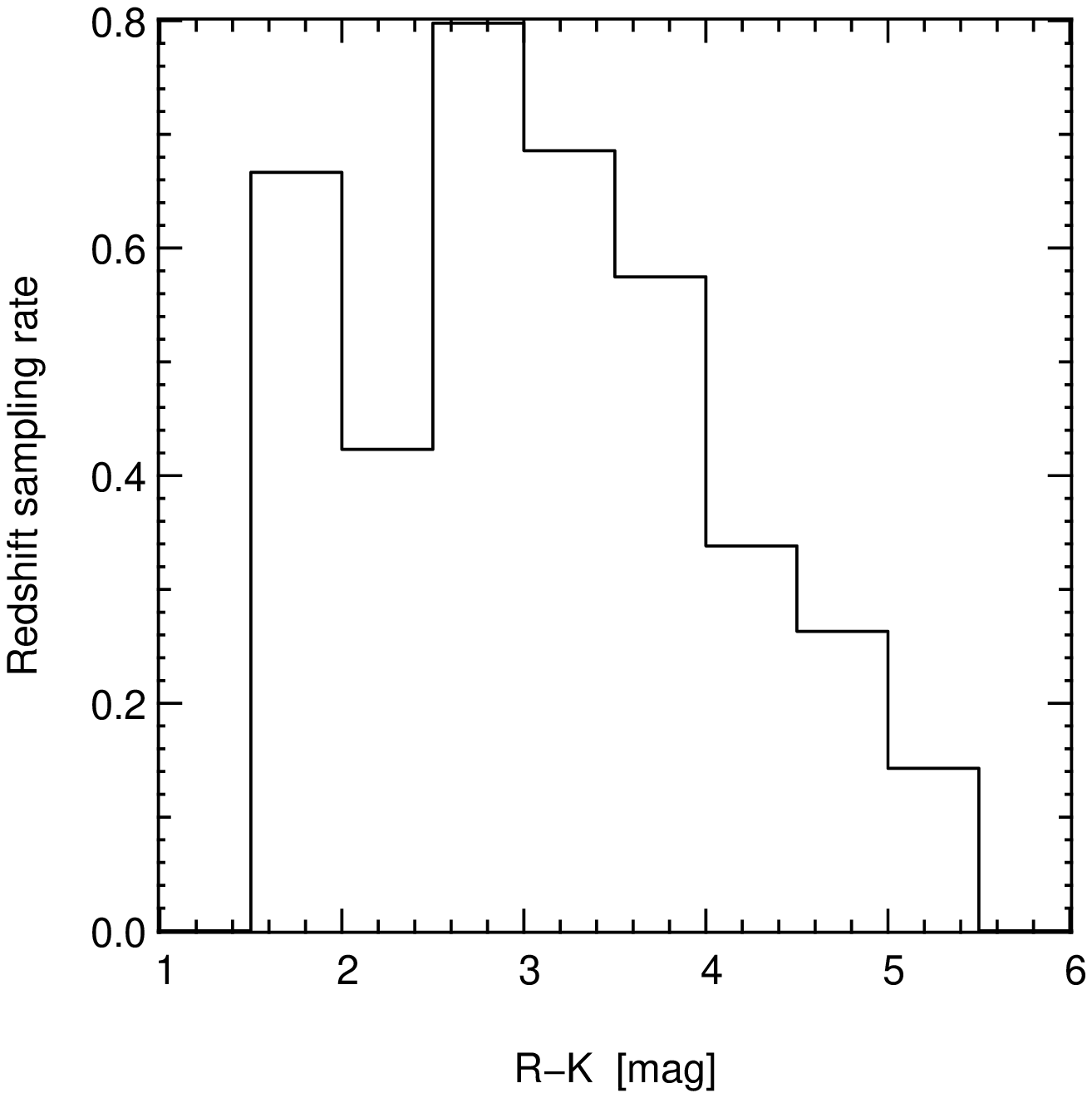,height=5.25cm}
  \hfill
  \epsfig{figure=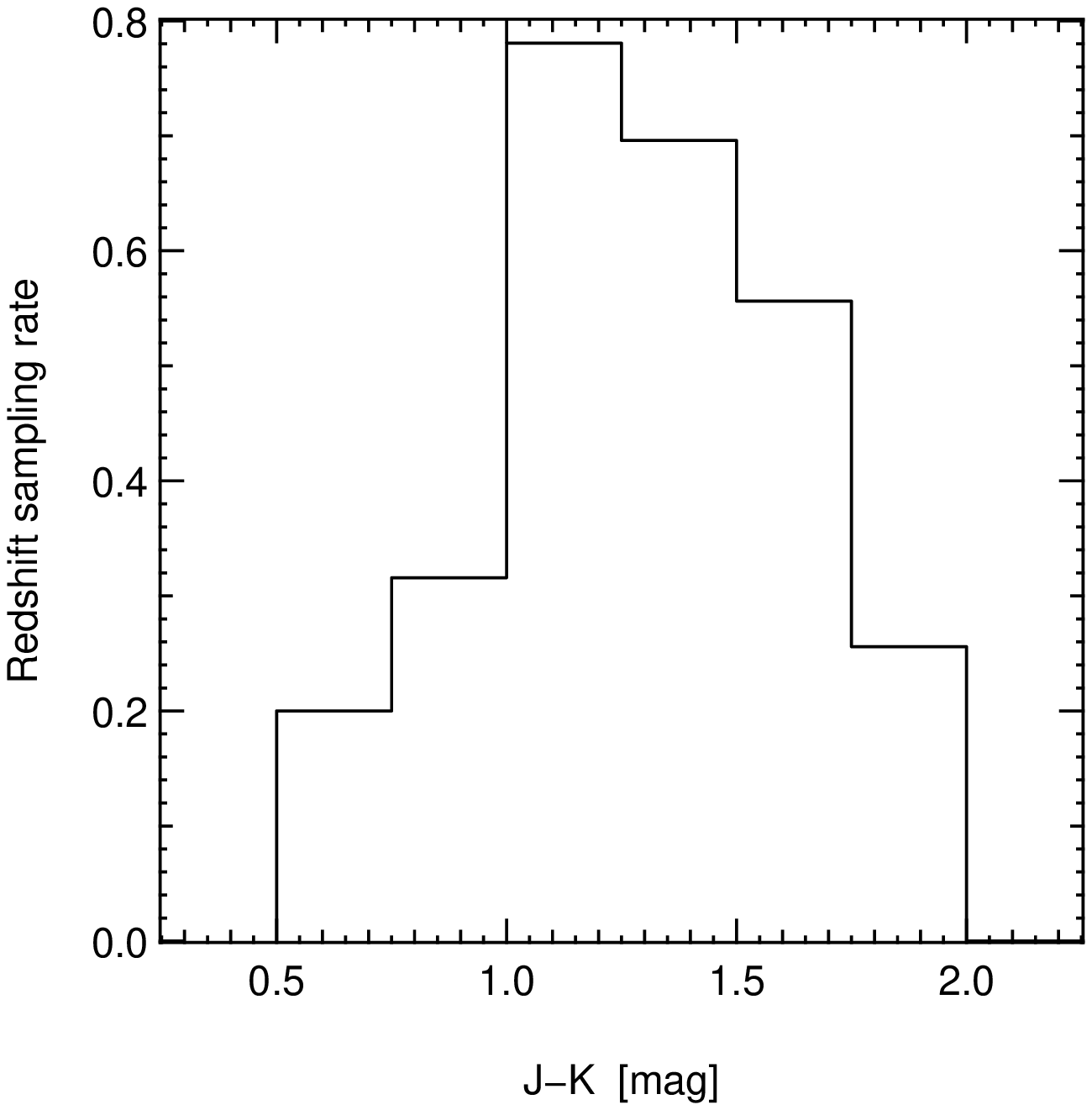,height=5.25cm} 

  \caption{The \textit{redshift sampling rate}, i.e.\ the fraction of
  successful redshift determinations among galaxies in the photometric
  catalogue, as a function of $m_K$ (left panel), $R\!-\!K$ (middle
  panel), and $J\!-\!K$ (right panel) for all galaxies in the survey
  patches S2F1, S5F1, S6F5, and S7F5 (the field S2F5 has been excluded
  from further analysis because of the small number of available
  spectra). The dotted line in the left panel indicates the formal
  limit of $K \le 17.5$ of the main part of the spectroscopic
  survey. The colour distributions in the middle and the right-hand
  panel are those of objects brighter than this limit.}

  \label{f:sampling}
\end{figure*}

\begin{figure*}
  \epsfig{figure=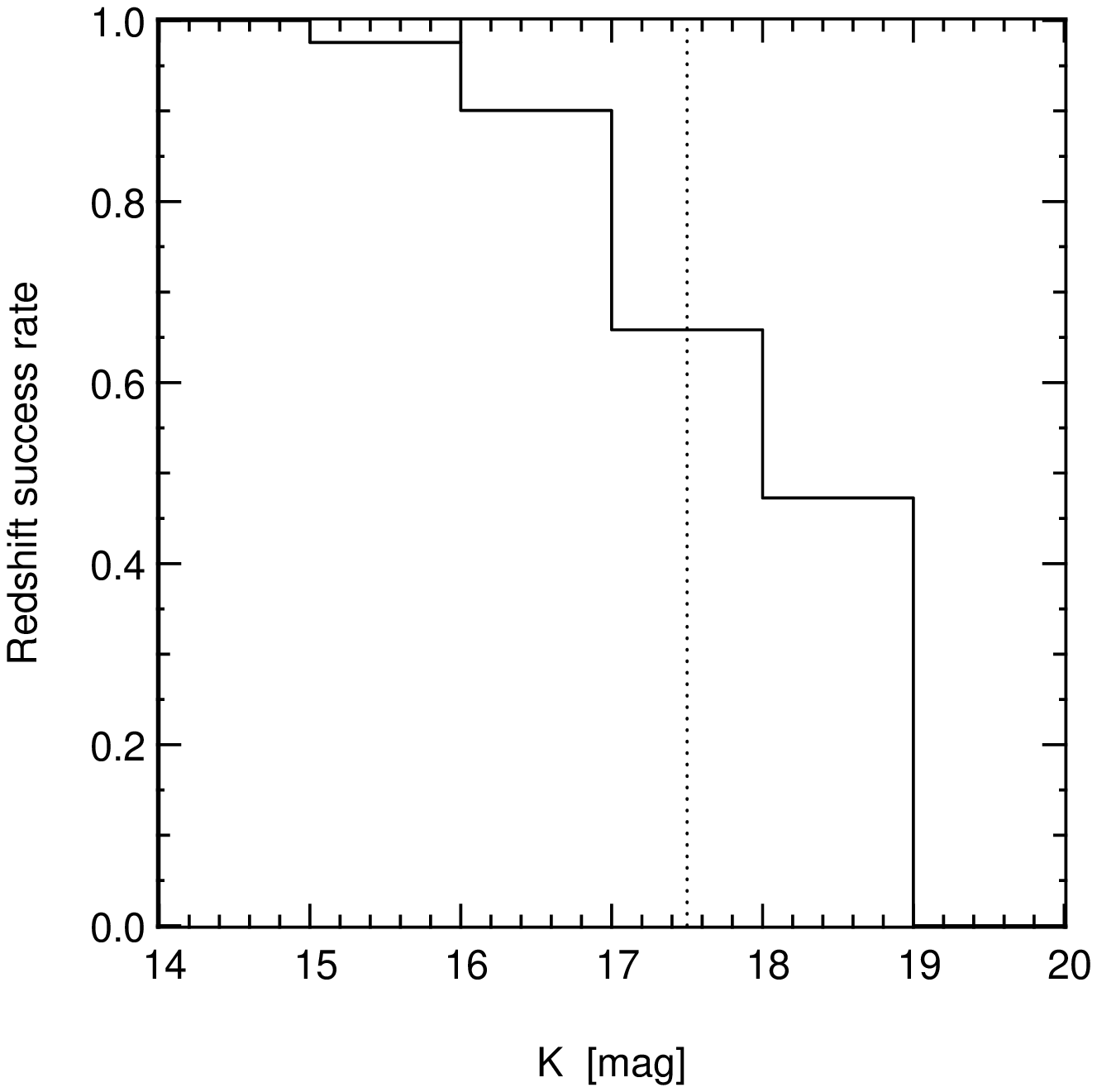,height=5.25cm}
  \hfill
  \epsfig{figure=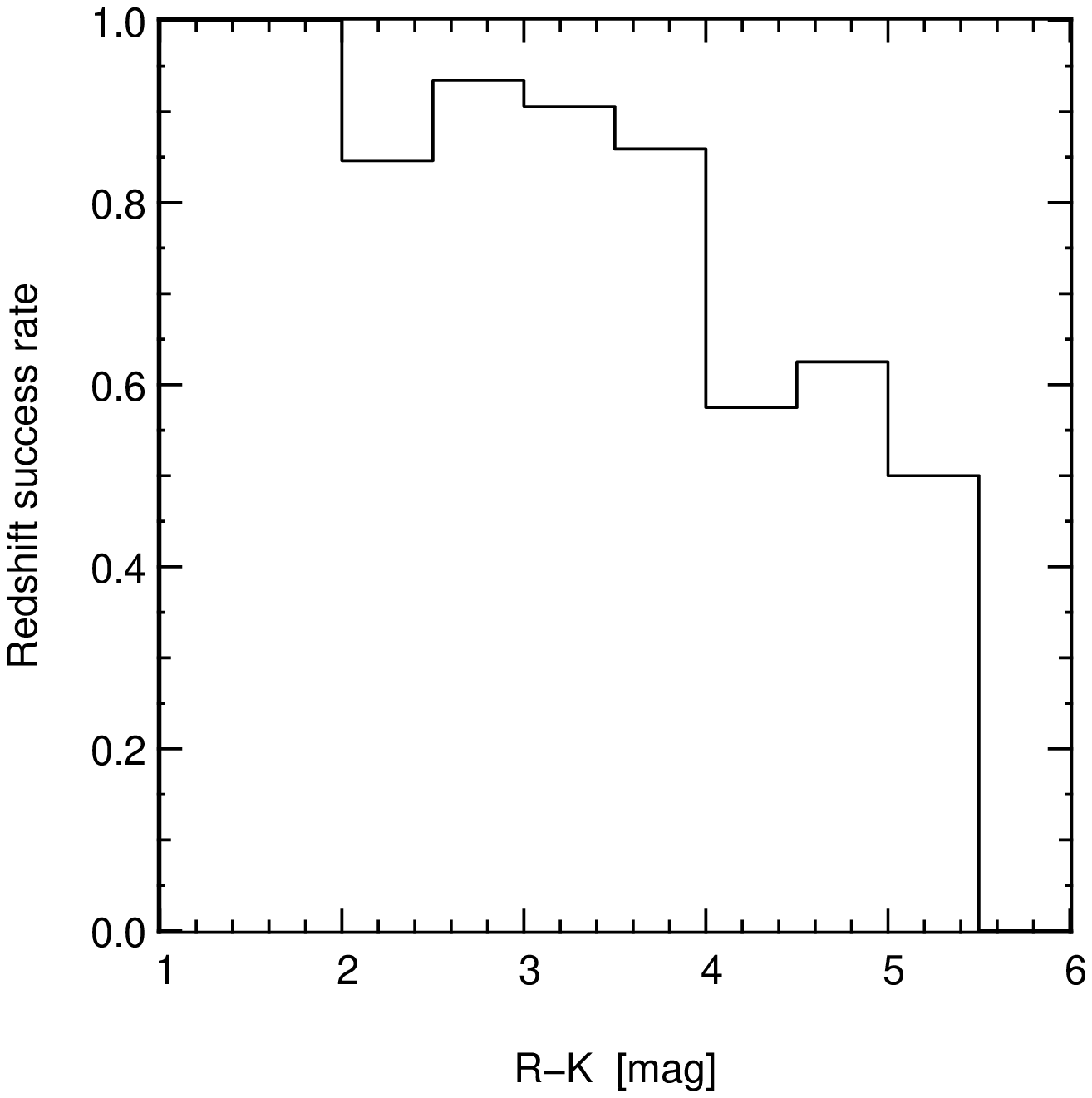,height=5.25cm}
  \hfill
  \epsfig{figure=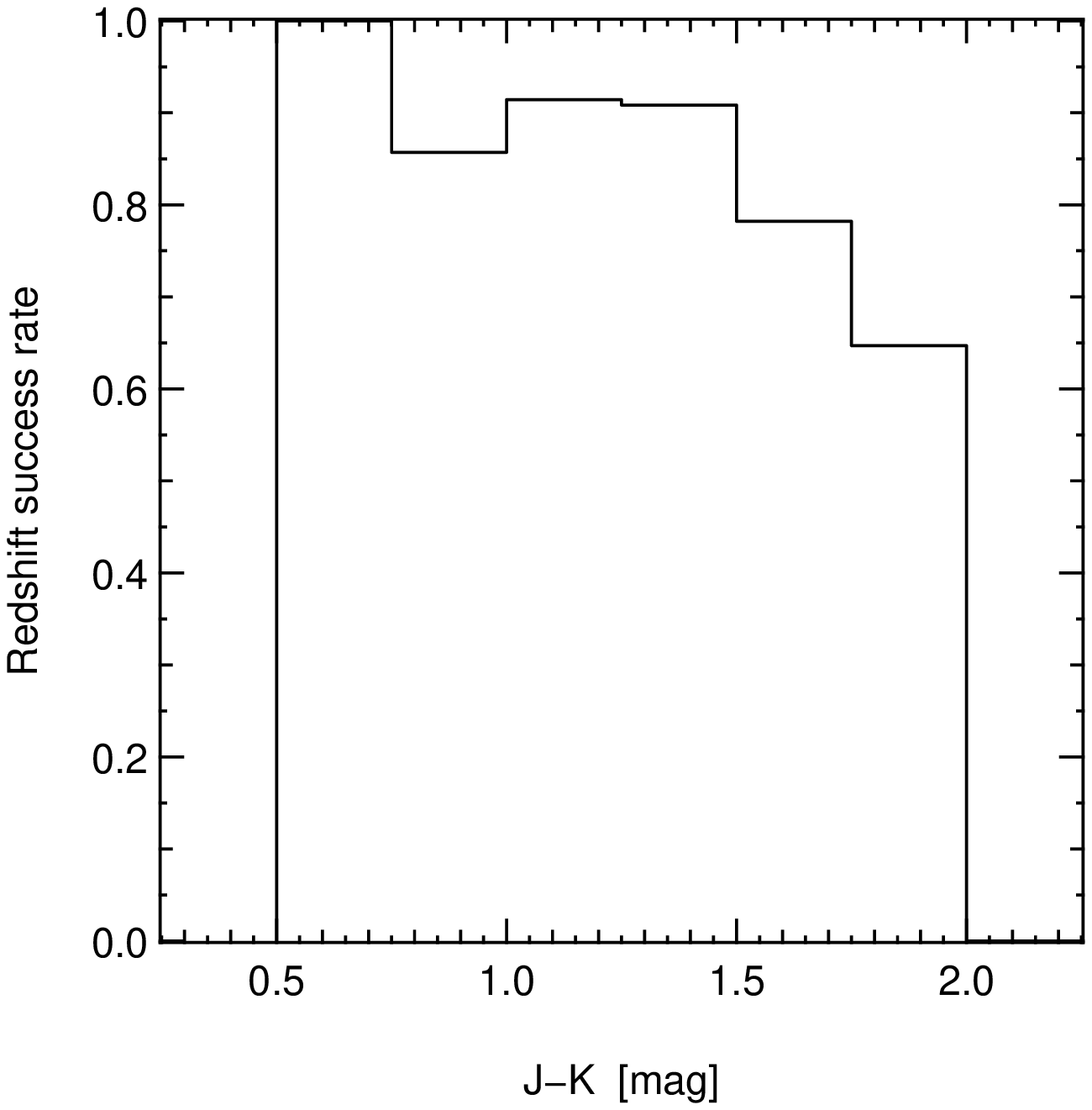,height=5.25cm} 

  \caption{The \textit{redshift success rate}, i.e.\ fraction of
  successful redshift determinations among all galaxies which were
  observed spectroscopically, as a function of $m_K$ (without any
  colour constraint; left panel), $R\!-\!K$ (middle panel), and
  $J\!-\!K$ (right panel) for all galaxies in the survey patches S2F1,
  S5F1, S6F5, and S7F5). The dotted line in the left panel indicates
  the formal limit of $K \le 17.5$ of the main part of the
  spectroscopic survey. The colour distributions in the middle and the
  right-hand panel are those of objects brighter than this limit.}

  \label{f:success}
\end{figure*}

Observations for the spectroscopy of objects from the photometric
MUNICS catalogue are almost complete, and most of the fields have been
observed with reasonable completeness. The fraction of galaxies with
redshift among all galaxies in the photometric catalogue is usually
called the \textit{redshift sampling rate} (e.g.\
\citealt{CNOC299}). As an example, Fig.~\ref{f:sampling} shows the
redshift sampling rate of objects in all survey fields with
spectroscopic data. The redshift sampling rate is the fraction of
objects with successful redshift determination among all galaxies in
the photometric catalogue, and, of course, is different from the
\textit{redshift success rate}, which is the fraction of all galaxies
with redshift among all galaxies in the spectroscopic catalogue. The
redshift success rate depending on apparent magnitude and colours is
shown in Fig.~\ref{f:success} and discussed in
Section~\ref{s:success}. These colour distribution should be compared
to the distributions of objects from the photometric and spectroscopic
catalogue shown in Fig.~\ref{f:colours}.

\begin{figure*}
  \epsfig{figure=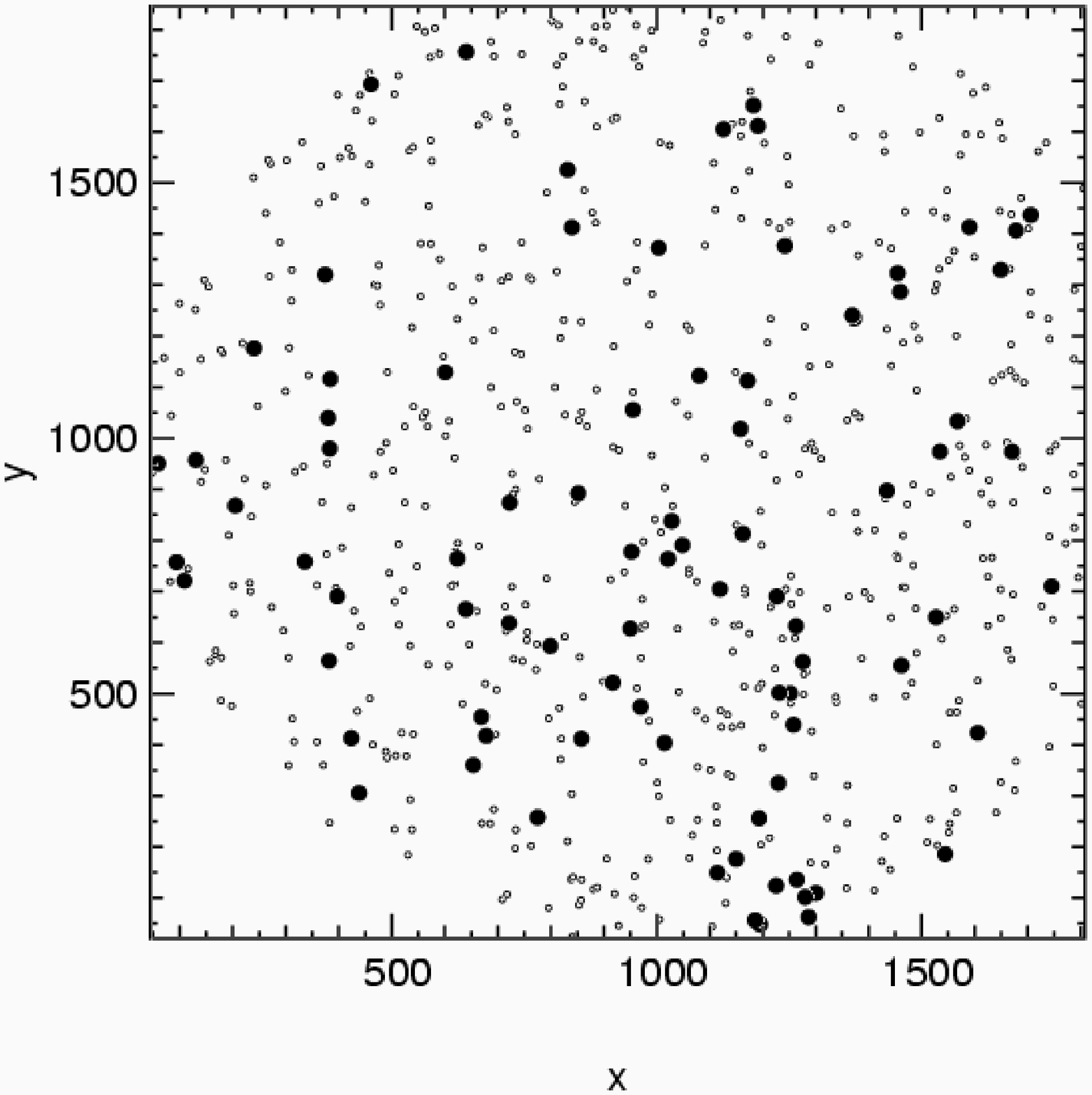,width=5.75cm}
  \hspace*{1cm}
  \epsfig{figure=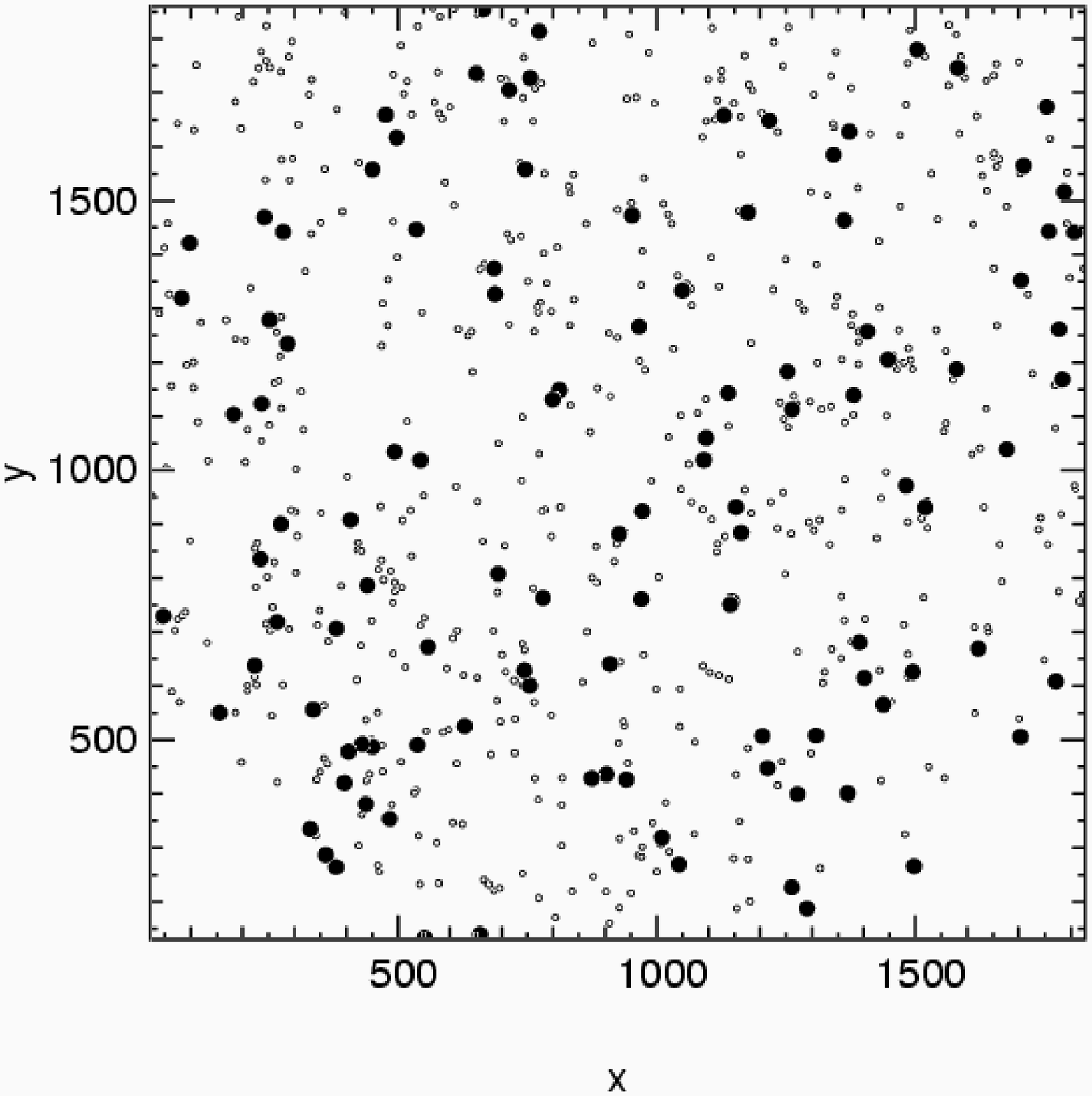,width=5.75cm}

  \vspace*{.5cm}

  \epsfig{figure=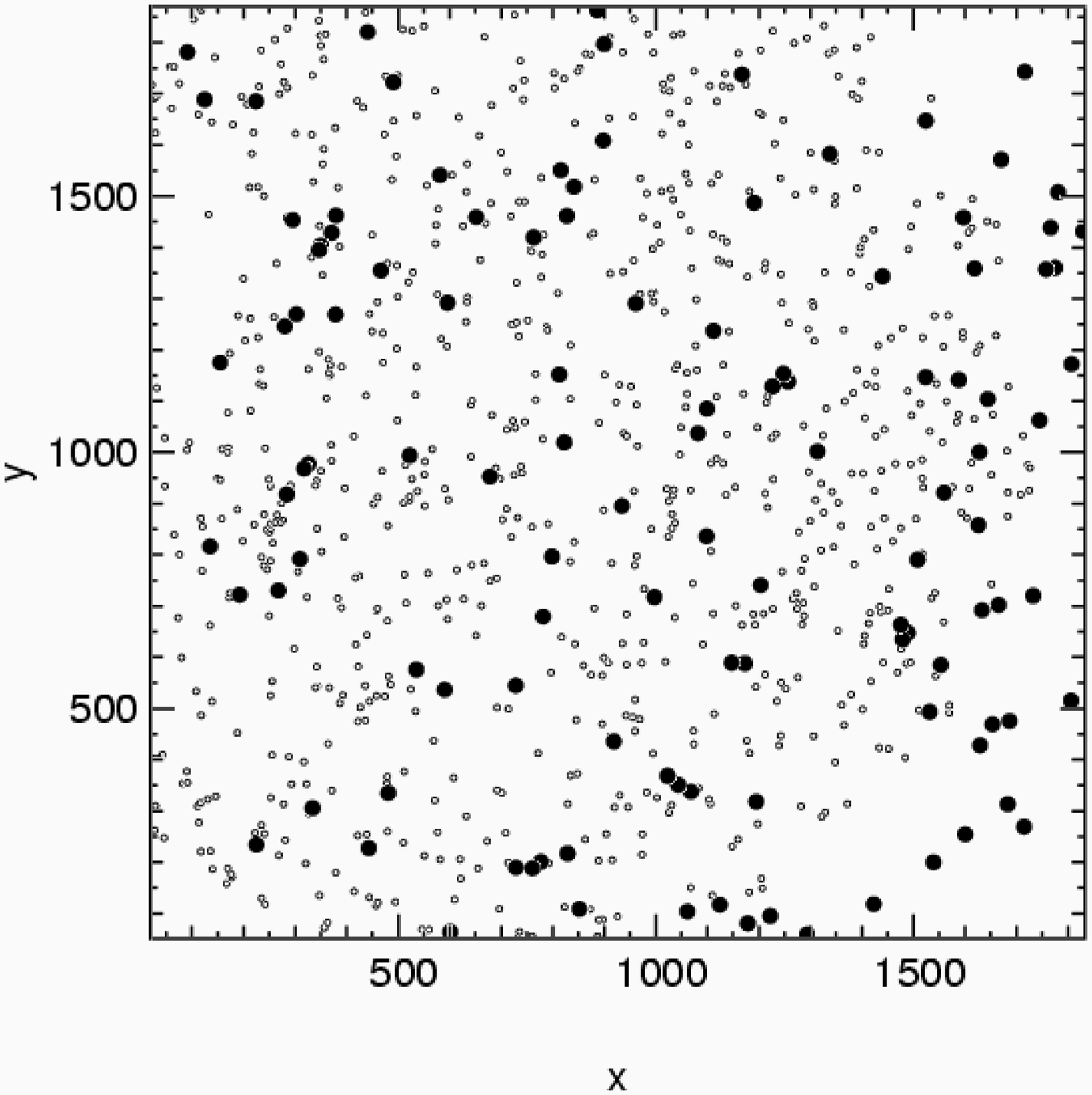,width=5.75cm}
  \hspace*{1cm}
  \epsfig{figure=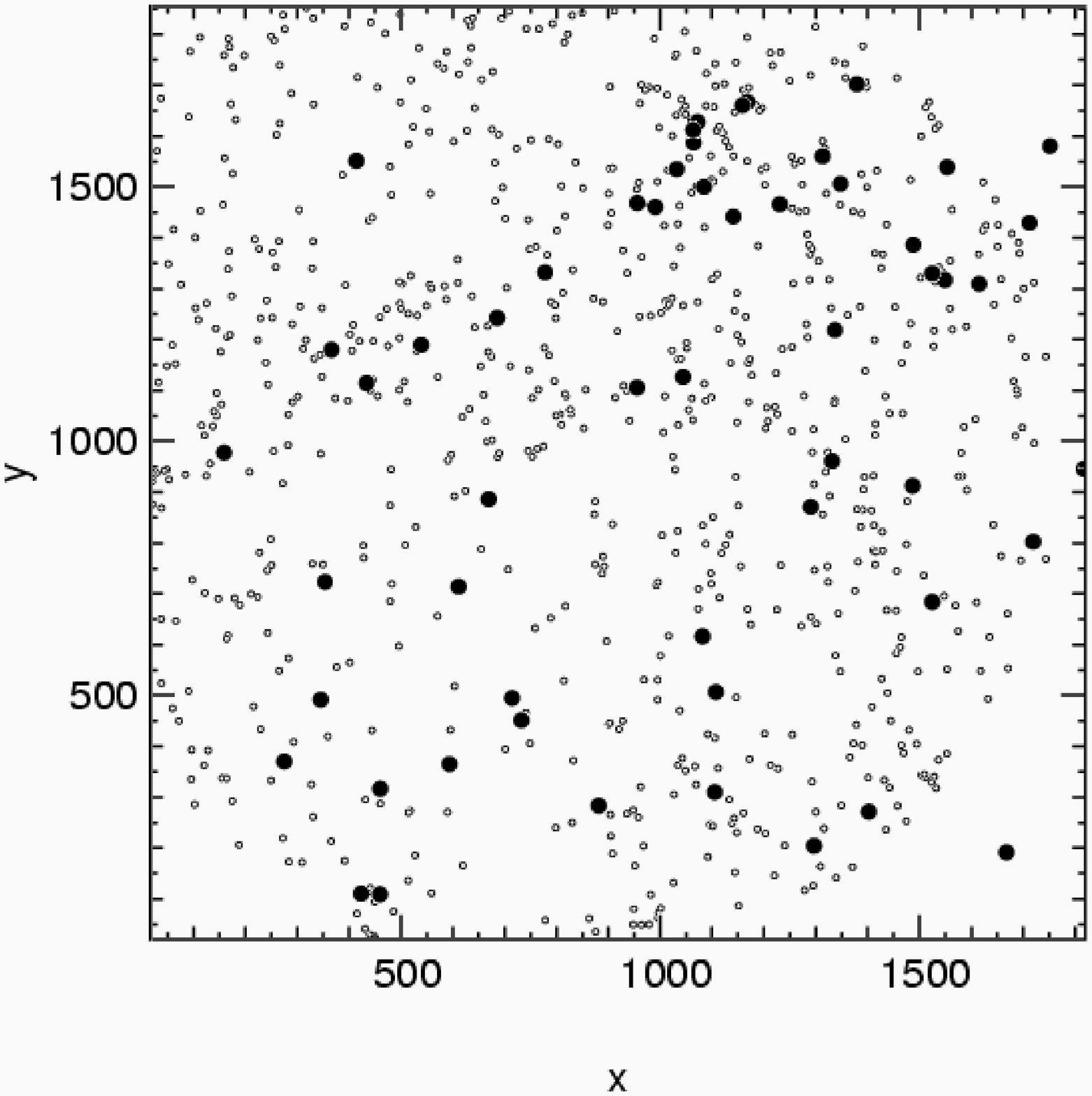,width=5.75cm}

  \caption{Sky coverage of spectroscopy in the MUNICS fields S2F1
  (upper left panel), S5F1 (upper right panel), S6F5 (lower left
  panel), and S7F5 (lower right panel). The small open circles are
  objects in the photometric catalogue, the distribution of which is
  governed by the MUNICS field geometry (see MUNICS~I). All galaxies
  with $K' \le 17.5$ and successful redshift determination are marked
  by filled circles; they seem to be uniformly distributed on the
  sky. Note that spectroscopy of the galaxies in the field S7F5 is not
  complete yet. Coordinates are pixel coordinates in the $K'$-band
  MUNICS frame.}

  \label{f:sky}
\end{figure*}

Although a few of the masks for the fields still have to be observed,
this does not necessarily inhibit the use of the sample for analysis
of, e.g., the luminosity function, as long as the sample of objects
with successful redshift determination is a fair representation of the
total sample, and as long as any systematic incompleteness effects can
be corrected for. We will show that this is the case for our catalogue
because of the procedure we followed during mask preparation and
observation. Firstly, we have tried to select \textit{all} objects
with $K' \le 17.5$ in five survey fields for spectroscopy at Calar
Alto. Since the field of view of the spectrograph (MOSCA) is almost as
large as the size of our survey field, and since we usually have
several masks per magnitude bin (at least for the fainter objects), we
are not limited by geometrical constraints from the arrangement of
slits on each mask. For the brighter magnitude bins we sometimes had
to drop a few objects during the preparation of a mask, but we have
tried to include these objects in fainter mask setups. Hence we expect
that the small fraction of objects on to which no slit could be put is
statistically similar to the distribution of objects to the
appropriate magnitude limit. Secondly, all masks which could be
observed so far were selected randomly from the masks prepared for
observations, so that we do not expect any selection bias here
either. Fig.~\ref{f:sky} shows the distribution of objects with
successful spectroscopy as compared to objects in the photometric
catalogue for the survey patches called S2F1, S5F1, S6F5, and
S7F5. These particular fields have highest completeness. The field
S2F1 contains the sparse sample of fainter objects observed with the
VLT. The figure nicely illustrates the statistically homogeneous
distribution of the objects in the spectroscopic catalogue. The excess
of objects with successful redshift determination visible in the lower
half of the field S2F1 is due to foreground structure which can
readily be seen in the image itself, showing again the influence of
cosmic variance and the importance of having \textit{several} fields
with a relatively \textit{large solid angle} for the analysis of
statistical quantities, one of the huge advantages of the MUNICS
project compared to smaller surveys.

\subsection{Redshift success rate}
\label{s:success}

The efficiency of redshift determination of spectroscopic observations
is described by the \textit{redshift success rate} rather than the
redshift \textit{sampling} rate. The former is the fraction of objects
with secure redshift among all galaxies which were observed
spectroscopically and is shown in Fig.~\ref{f:success}. As expected,
the redshift success rate drops off for very faint and very red
objects, but is in general very high due to the good quality of the
spectra.

\subsection{Extremely red objects}
\label{s:eros}

During the last years there has been a lot of research on `extremely
red objects (EROs)', usually defined in terms of their very red
optical--near-infrared colour ($R\!-\!K > (5\dots 6)$ mag; see, for
instance, \citet{Martini2001} and references therein). While the
MUNICS catalogue obviously contains such objects, they are not aimed
at with the spectroscopic observations described in this paper, mainly
because it is very difficult to obtain optical spectra of faint
EROs. This is evident from Figures~\ref{f:sampling} and
\ref{f:success} showing the redshift sampling rate and the redshift
success rate for the MUNICS spectroscopic observations. Nevertheless,
there are 15 such objects in the spectroscopic catalogue, all having
colours of $5.0 \; \mathrm{mag} \; \le R\!-\!K \le 5.5 \;
\mathrm{mag}$. Among these we find 7 objects for which no redshift
could be determined, 5 galaxies with redshifts $0.46 \le z \le 1.01$,
all having a spectral energy distribution characteristic of early-type
galaxies, as well as 3 stars of spectral type M. First results of
near-infrared spectroscopy of EROs selected from the MUNICS catalogue
are described in \citet{eros1}.

\section{The near-infrared luminosity function of galaxies}
\label{s:lf}

\subsection{Calculation of the luminosity function}

The luminosity function is computed using the non-parametric
$V_\mathrm{max}$ formalism \citep{Schmi68}. This method has been shown
to yield an unbiased estimate of the luminosity function if the sample
is not affected by strong clustering \citep{Takeuchi}. Because of
our field selection and the relatively large area of the survey,
divided into several individual fields, we believe that this
assumption is valid for our sample.

The $V_\mathrm{max}$ formalism accounts for the fact that some fainter
galaxies are not visible in the whole survey volume. Each galaxy $i$
in a given redshift bin $[z_\mathrm{lower},z_\mathrm{upper}]$
contributes to the number density an amount inversely proportional to
the volume $V_i$ in which the galaxy is detectable in the survey:

\begin{equation}
  V_i \; = \;
  \int\limits_{z_\mathrm{lower}}^{\mathrm{min}(z_\mathrm{upper},
  z_\mathrm{max})} \: \frac{dV}{dz} \: dz ,
\end{equation}

where $dV = d\Omega \, r^2 \, dr$ is the comoving volume element,
$d\Omega$ is the solid angle covered by the survey, and
$z_\mathrm{max}$ is the maximum redshift at which galaxy $i$ having
absolute magnitude $M_i$ is still detectable given the limiting
apparent magnitude of the survey. We have made sure that the effect of
the volume correction is of importance only in the faintest bin in
absolute magnitude, and that even in this case the correction is
small.

Additionally, the contribution of each galaxy $i$ is weighted by the
inverse of the detection probability $P(m_{K,i})$ of the $K'$-band
selected photometric catalogue, where we assume that the detection
probability is independent of the galaxy type and can be approximated
by that of point-like sources. Completeness simulations for realistic
galaxy profiles at various redshifts have shown that this
approximation is indeed sufficient for galaxies at redshifts $z < 1$
(\citealt{munics4}, hereafter MUNICS~IV). However, since the objects
under consideration here are comparatively bright, the influence of
this correction is negligible anyway.

In addition to the correction for the incompleteness of the
\textit{photometric} MUNICS catalogue described above, we have to
correct for the incompleteness of the \textit{spectroscopic} catalogue
with respect to the photometric sample, described by the redshift
sampling rate (see Section~\ref{s:sampling}). In principle, this
correction depends on the apparent magnitude of the objects (faint
objects may produce a lower signal-to-noise ratio or might not have
been considered for spectroscopy)\footnote{Note that we can neglect
any influence of the \textit{size} of the objects (large objects might
have larger losses of light at the slit of the spectrograph), since --
especially at faint magnitudes -- objects appear to be almost
point-like.}, on their intrinsic type (it is easier to determine a
redshift for objects showing prominent emission lines, for example)
and on the redshift of the source (influencing the position of
prominent spectral features with respect to the spectral range or
bright night-sky emission lines, for example). However, it is
difficult to determine a completeness ratio depending on spectral type
and redshift, because this information is lacking for all objects
without secure redshift measurement.

Hence the redshift sampling rate is often quantified in terms of its
dependence on apparent magnitude and two colours instead of spectral
type and redshift (see \citet{CNOC299}, for example). In this work we
compute the redshift sampling rate $C_i$,

\begin{equation}
  C_i \; = \; \frac{N_{z,i}}{N_i} ,
\end{equation}

depending on apparent $K'$-band magnitude $m_{K}$, and the colours
$R\!-\!K$ and $J\!-\!K$. Specifically, for each galaxy $i$ with a
redshift, we determine the number $N_i$ of galaxies in the photometric
sample and the number $N_{z,i}$ of galaxies with successful redshift
determination in joint bins of apparent magnitude and colours, where
we count all galaxies $j$ obeying

\begin{eqnarray}
  \nonumber
  \left| m_{K,i} - m_{K,j} \right| & \le & 0.70 \; \mathrm{mag} , \\
  \label{e:scorr}
  \left| (R\!-\!K)_i - (R\!-\!K)_j \right| & \le & 0.70 \; \mathrm{mag} , \\
  \nonumber
  \left| (J\!-\!K)_i - (J\!-\!K)_j \right| & \le & 0.35 \; \mathrm{mag} .
\end{eqnarray}

On the one hand, the size of the intervals should be large enough to
contain a reasonable number of objects to avoid large fluctuations due
to small-number statistics. On the other hand, the intervals should
not be too large in order to be able to follow the change of the
sampling rate. The numbers given in equation (\ref{e:scorr}) have been
chosen after careful tests.  Note that we have used larger intervals
for $m_{K}$ and $R\!-\!K$ since these distributions are broader than
the one for $J\!-\!K$.  To illustrate the general behaviour of the
redshift sampling rate, we show projections of the function in
Fig.~\ref{f:sampling}. Note that we have to apply a correction
factor of 0.93 to the sampling rate because of stellar contamination
of the class of extended objects, a bias introduced by our image-based
object classification algorithm, described in detail in
Section~\ref{s:selection}.

The near-infrared luminosity function $\Phi (M)$ is then computed
according to the formula

\begin{equation}
  \Phi (M) \: dM \; = \;\sum\limits_i \: \frac{1}{V_i} \:
  \frac{1}{P (m_{K,i})} \: \frac{1}{C_i} \:
  dM ,
\end{equation}

where the sum runs over all objects $i$ in the redshift range for which
we want to calculate the luminosity function.

\subsection{Converting to absolute magnitudes}

For the conversion of apparent magnitudes $m$ into absolute magnitudes
$M$ we need $k(z)$ corrections for the galaxies, defined by

\begin{equation}
  M \; = \; m \: - 5 \, \log \left( \frac{d_L (z) }{10 \, \mathrm{pc}}
  \right) \: - 2.5 \, \log \left( 1 + z \right) - k (z) ,
\end{equation}

where $d_L (z)$ is the cosmological luminosity distance of the galaxy,
and $z$ is the redshift.

\begin{figure*}
  \epsfig{figure=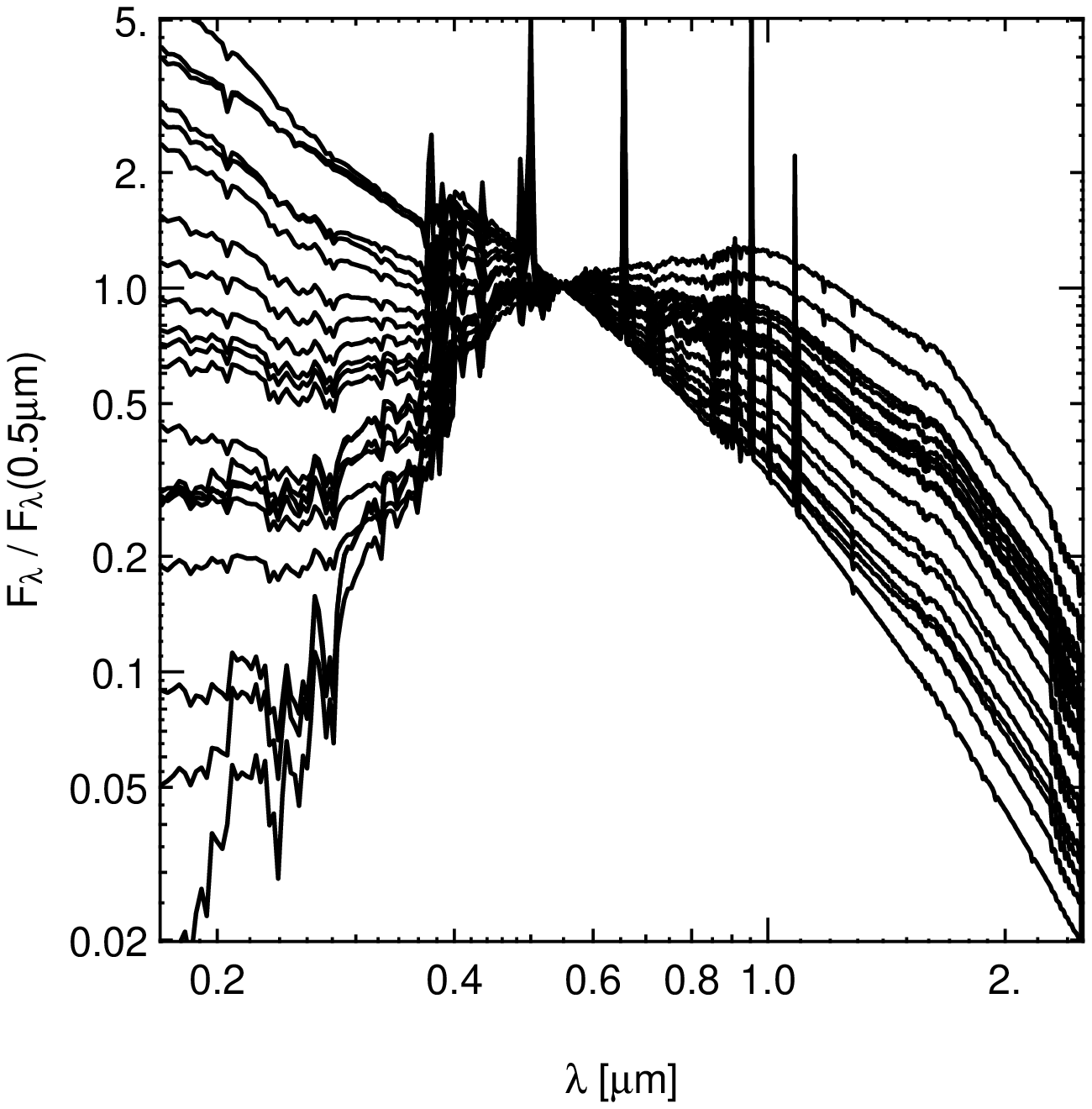,height=5.cm}
  \hfill
  \epsfig{figure=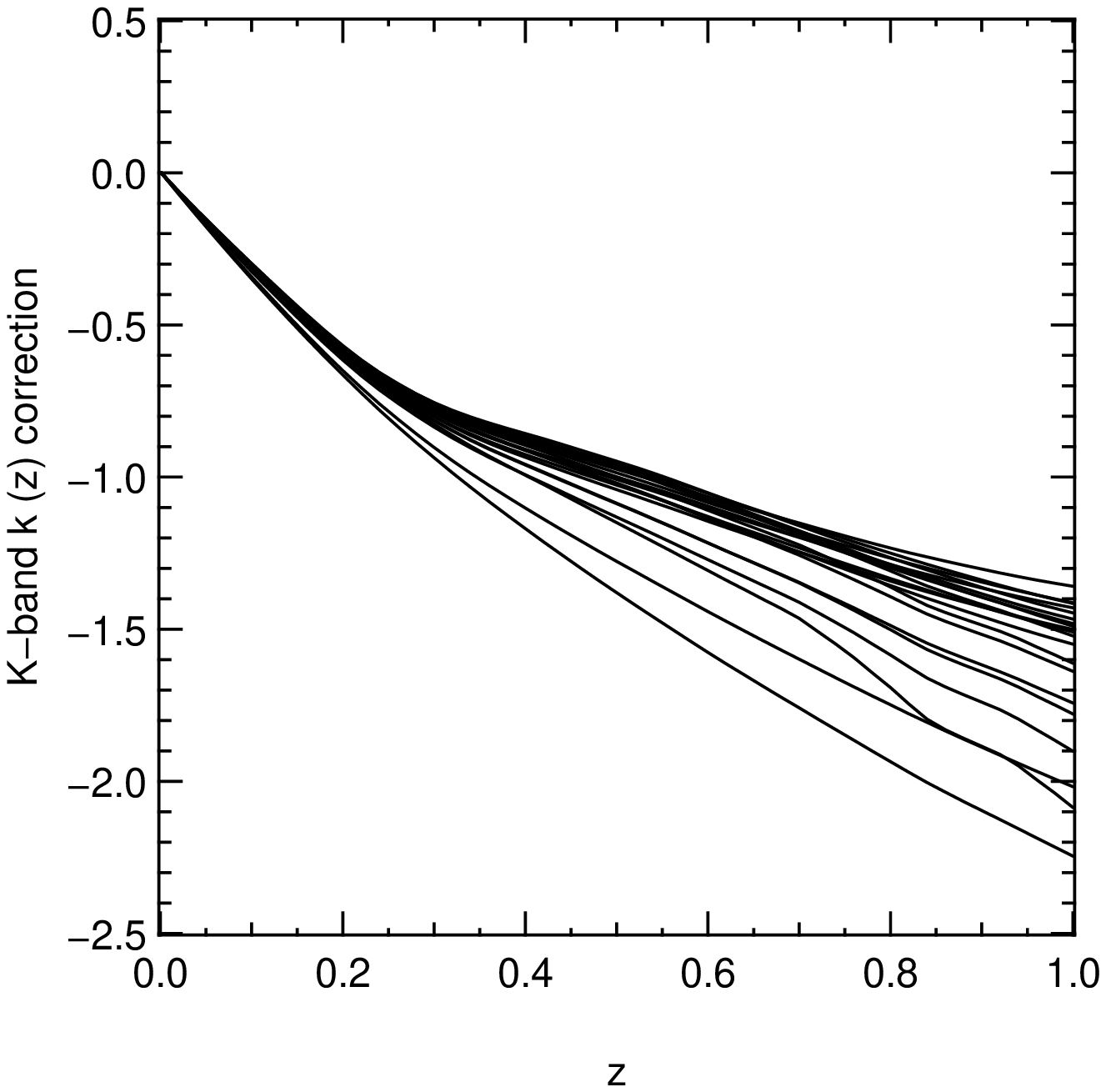,height=5.cm}
  \hfill
  \epsfig{figure=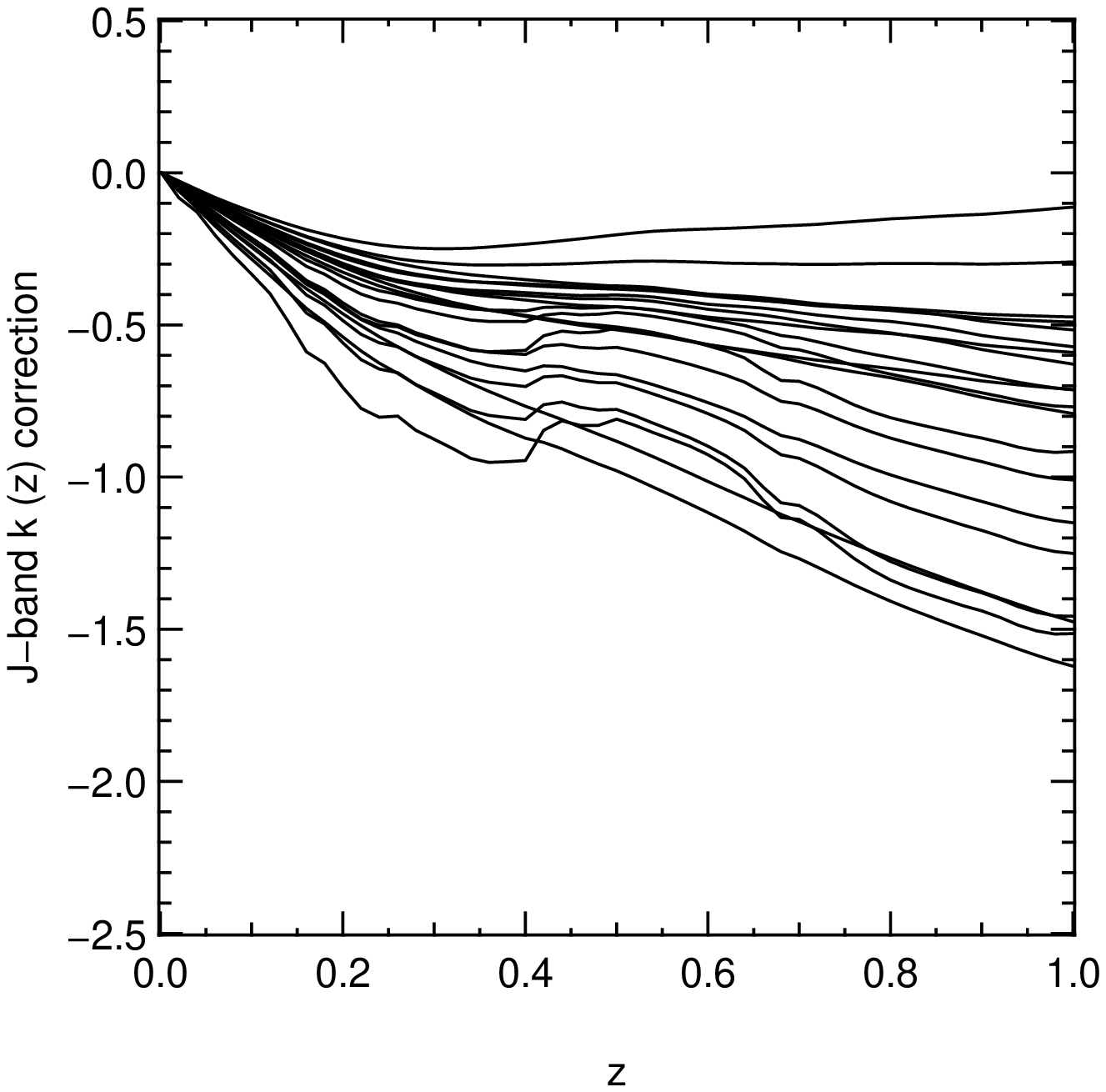,height=5.cm}

  \caption{\textit{Left panel:} Model spectral energy distributions
  used to compute $k(z)$ corrections (see text for
  details). \textit{Middle panel:} $K$-band $k(z)$ corrections as a
  function of redshift for the spectral energy distributions shown in
  the left panel. \textit{Right panel:} Same as middle panel, but for
  the $J$ band.}

  \label{f:kcorr}
\end{figure*}

The $k(z)$ corrections for the objects in the spectroscopic catalogue
are obtained by fitting the model spectral energy distributions shown
in the left panel of Fig.~\ref{f:kcorr} to the broad-band photometry
of the objects. The template spectral energy distributions are
constructed by combining empirical spectra and stellar population
synthesis models by \citet{Maraston98}. The detailed description of
these models will be found in MUNICS~II. The $k(z)$ corrections for
the $K$ band and the $J$ band are presented in the middle and right
panel of Fig.~\ref{f:kcorr}, respectively, while Fig.~\ref{f:flux}
shows a comparison of flux-calibrated spectrum, broad-band photometry,
and fitted model spectral energy distribution for two objects in the
spectroscopic catalogue.

We have also tested the influence of using only one intermediate-type
model spectral energy distributions for deriving the $k(z)$
corrections. As is obvious from Fig.~\ref{f:kcorr}, this should work
well for the $K$ band, where the spread between $k(z)$ corrections for
different models is small, but a bit less so well for the $J$ band
with its larger variations. Indeed, we do see hardly any difference for
the $K$-band luminosity function, and only a small difference in the
$J$ band. However, it is important to note that in the following we do
\textit{not} use $k(z)$ corrections from one model, but those from the
fitting of template spectral energy distributions to the six-filter
broad-band photometry of the MUNICS catalogue.

\subsection{The $K$-band luminosity function}

The rest-frame $K$-band luminosity function of galaxies drawn from the
spectroscopic sample was constructed in the redshift intervals $0.1
\le z \le 0.3$ (median redshift $z = 0.2$), $0.3 \le z \le 0.6$
(median redshift $z = 0.4$, and $0.6 \le z \le 0.9$ (median redshift
$z = 0.7$). The two lower redshift bins comprise the majority of
objects in the sample, as can be seen in Fig.~\ref{f:zhist}.

The centres of the bins in absolute magnitude were chosen in a way
which ensures a fair representation of the bright end of the
luminosity function, i.e.\ the bin centres at the bright end
correspond roughly to the absolute magnitudes of the brightest objects
in that bin.

\begin{figure*}
  \epsfig{figure=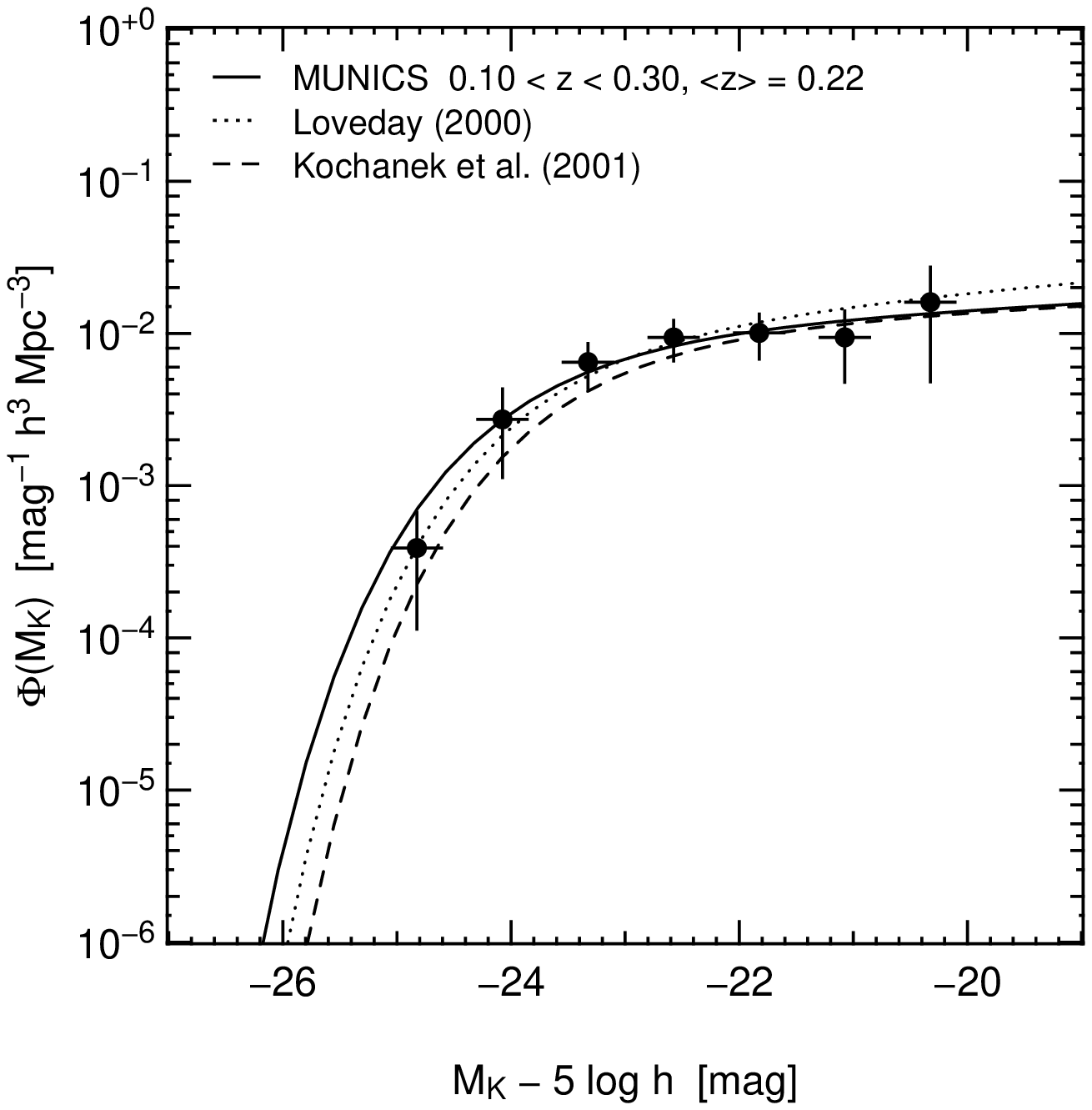,height=5.75cm}
  \hspace*{1cm}
  \epsfig{figure=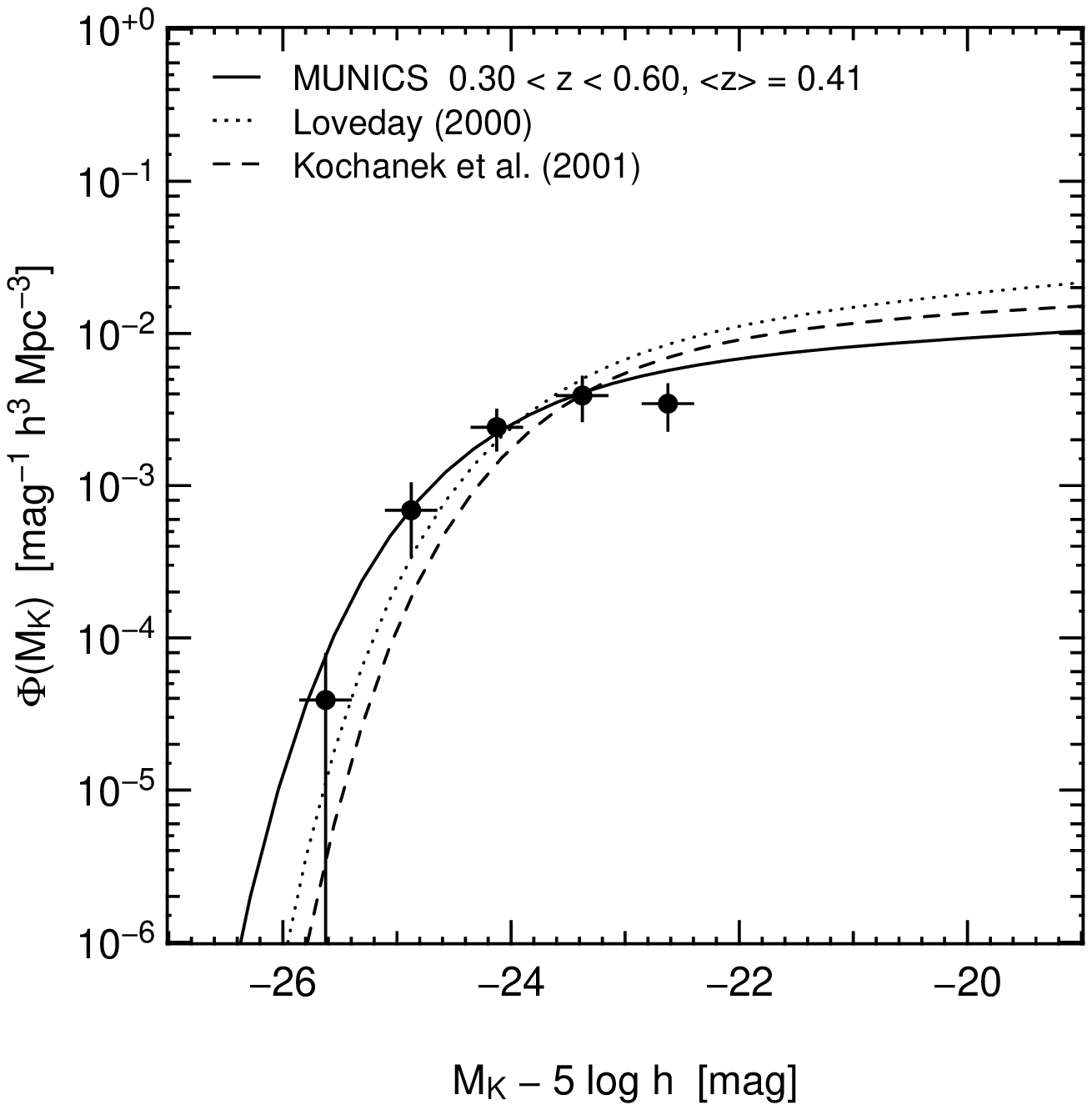,height=5.75cm}

  \vspace*{.5cm}

  \epsfig{figure=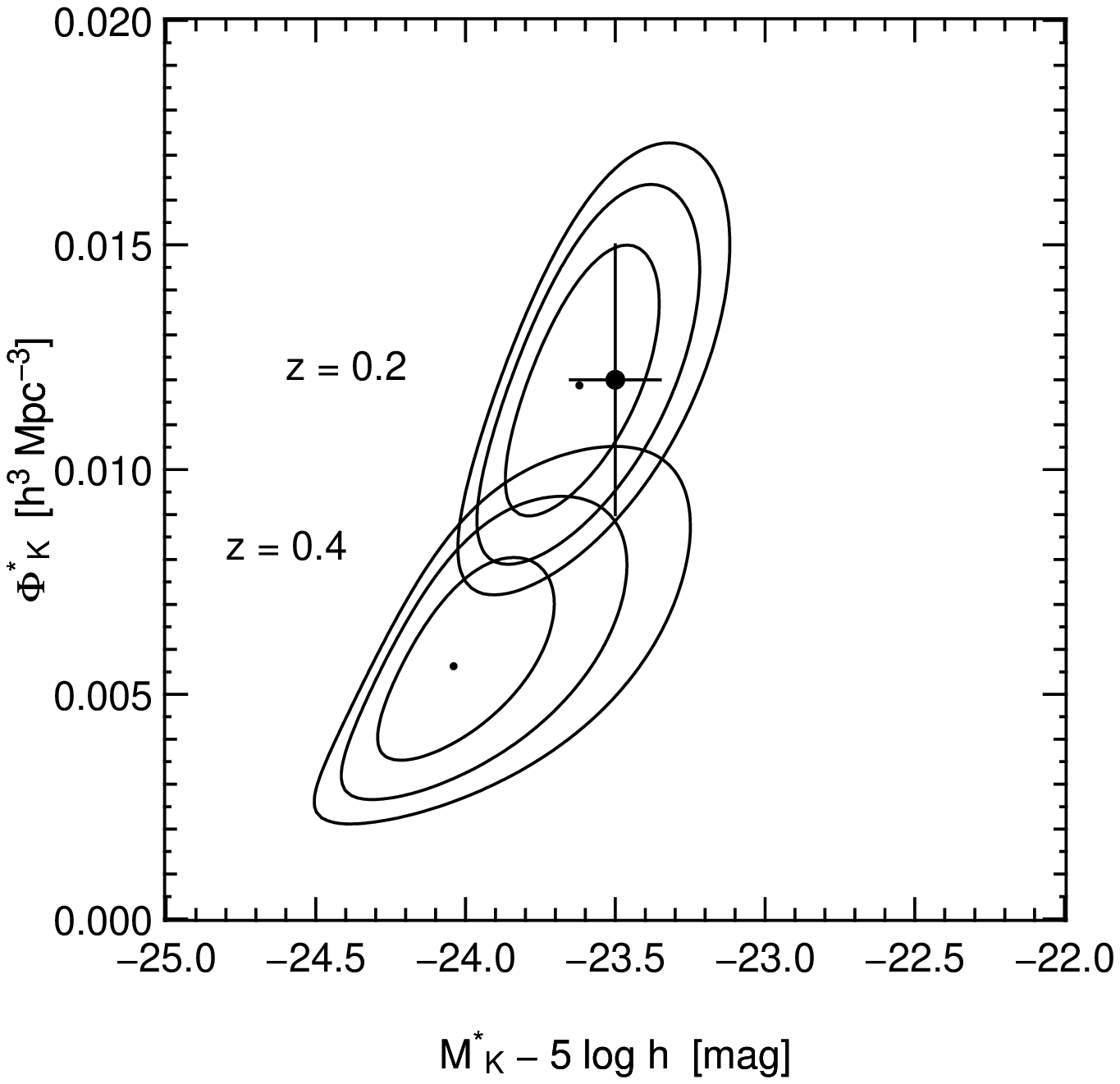,height=5.7cm}
  \hspace*{1cm}
  \epsfig{figure=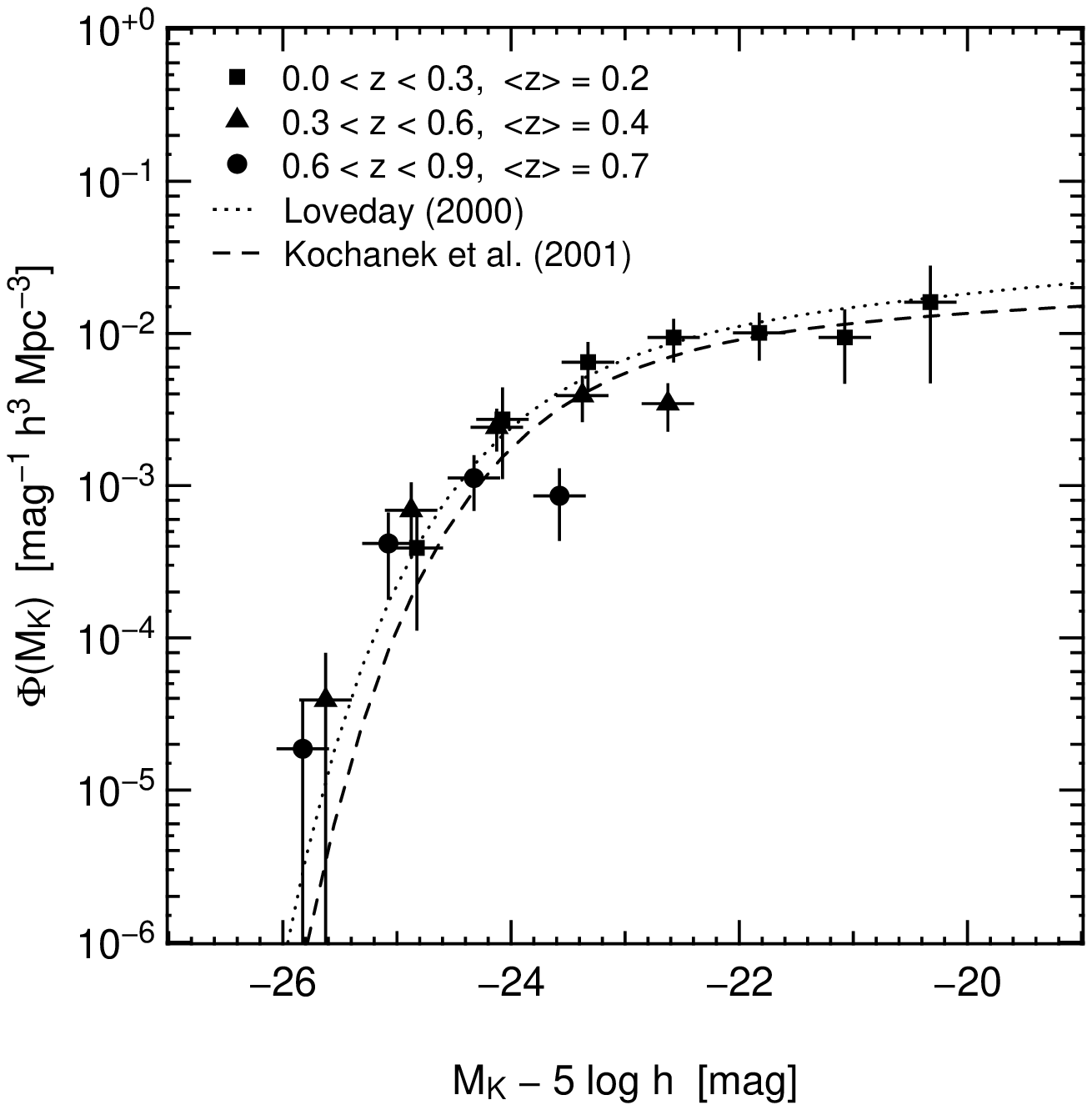,height=5.75cm}

  \caption{The $K$-band luminosity function of galaxies from
  spectroscopic observations of MUNICS galaxies. \textit{Upper left
  panel:} The luminosity function in the redshift range $0.1 \le z \le
  0.3$ (median redshift $z=0.2$; filled symbols). The vertical error
  bars give the Poissonian error, while the binning error was
  estimated as $b/\sqrt{12}$ with the size $b$ of the bin. Also shown
  are the measurements of the local $K$-band luminosity function by
  \citeauthor{Loveday00} (\citeyear{Loveday00}; dotted line) and
  \citeauthor{Kochaneketal01} (\citeyear{Kochaneketal01}; dashed
  line), as well as a Schechter approximation to the MUNICS data
  (solid line). The parameters of the Schechter fit are \mklt , \pklt
  , and $\alpha_K = -1.10$ (fixed). \textit{Upper right
  panel:} The same as before, but in the redshift range $0.3 \le z \le
  0.6$ (median redshift $z=0.4$), with Schechter parameters \mkmt ,
  \pkmt , and $\alpha_K = -1.10$ (fixed). \textit{Lower left panel:}
  Error contours ($1\sigma$, $2\sigma$, and $3\sigma$) for the
  Schechter parameters $M_K^*$ and $\Phi_K^*$ from the $\chi^2$
  distribution. The filled circle indicates the average value of local
  measurements from Table~\ref{t:litk}; the ``error bars'' give an
  idea of the variation between different authors'
  measurements. \textit{Lower right panel:} The luminosity function in
  the interval $0.6 \le z \le 0.9$ (median redshift $z=0.70$; filled
  circles), compared to the results from the two lower redshift ($z =
  0.2$, squares; $z = 0.4$, triangles) as well as the local
  measurements.}\label{f:lfk}
\end{figure*}

Fig.~\ref{f:lfk} shows the results for the $K$-band luminosity
function of galaxies from spectroscopic observations in different
redshift bins compared to the luminosity functions determined from the
local samples of \citet{Loveday00} and \citet{Kochaneketal01}. Clearly
there is a good general agreement between the luminosity function
measured at $z = 0.2$ and the ones from local galaxy samples, although
our data seem to suggest a somewhat smaller value for $M_K^*$. Fitting
a Schechter function (\citealt{Schechter76}; see also equations
(\ref{e:schechter}) and (\ref{e:schechter2}) for the parametrisation
of the function) to the luminosity function yields the parameters
\mklt\ and \pklt . We used a fixed value of $\alpha_K = -1.10$, close
to the values derived for local galaxies by \citet{Loveday00} and
\citet{Kochaneketal01}. Contrary to the $J$-band luminosity function
discussed below, our normalisation also nicely agrees with the
\citet{Coleetal2001_2} measurement, although they derive a
considerably shallower faint-end slope of $\alpha = -0.93 \pm
0.04$. The errors were derived by running Monte-Carlo simulations with
100~000 iterations, taking into account the errors due to the binning
in absolute magnitudes. Excluding the faintest magnitude bin does not
change the result of the fit significantly.

Note that the agreement between our measurement and the local
Schechter functions is also very good at the faint end, although with
poor statistics and large correction factors due to the
incompleteness. From this we draw the conclusion that our method to
correct for incompleteness yields reliable estimates.

At redshifts $0.3 \le z \le 0.6$, we find a mild evolution of the
luminosity function, with a higher characteristic luminosity and a
lower normalisation, yielding Schechter parameters of \mkmt\ and \pkmt
. The error contours for the Schechter parameters for the luminosity
function in the two lower redshift bins are shown in the same
figure. These contours were computed from the $\chi^2$
distribution. Finally, we show a comparison of the measurements of the
$K$-band luminosity function in all three redshift bins. Although the
statistics becomes rather poor in the highest redshift bin with median
redshift $z = 0.7$, it confirms the trend for the evolution of the
luminosity function. However, it is evident that the total evolution
out to redshifts around 0.7 is not dramatically large.

Comparison of our result with the measurements by \citet{CSHC96} who
derived the $K$-band luminosity function at various redshifts $0 < z <
1$ from a spectroscopic sample in two small fields shows good
agreement with our results. Their values for the Schechter parameters
$M_K^*$, $\Phi_K^*$, and $\alpha_K$ are similar to measurements for
local samples. Again, our data favour a slightly larger characteristic
luminosity, but the trend of a falling normalisation with redshift is
found both in the MUNICS and in the Cowie et al.\ measurements of the
luminosity function.

\subsection{The $J$-band luminosity function}

\begin{figure*}
  \epsfig{figure=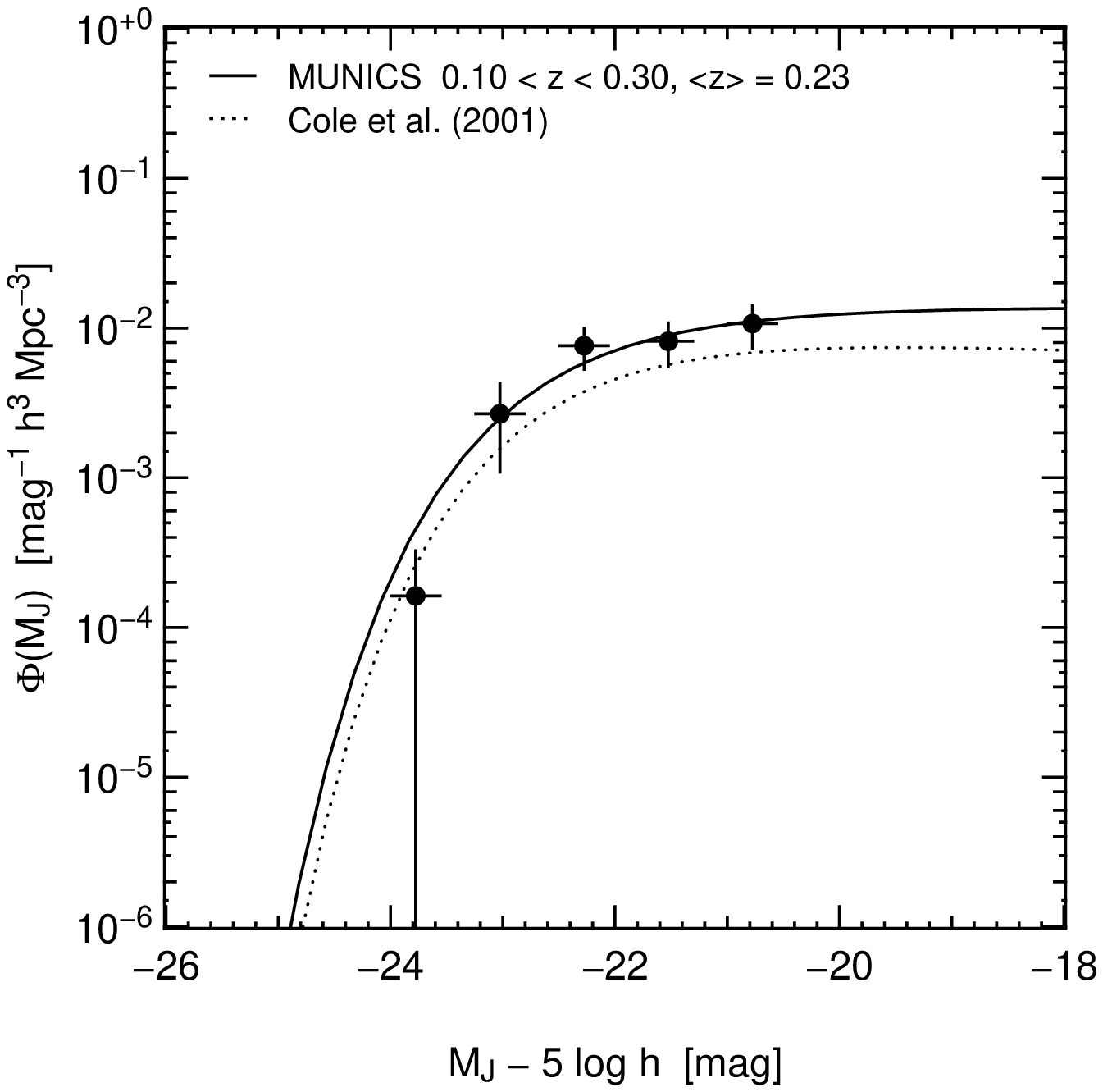,height=5.75cm}
  \hspace*{1cm}
  \epsfig{figure=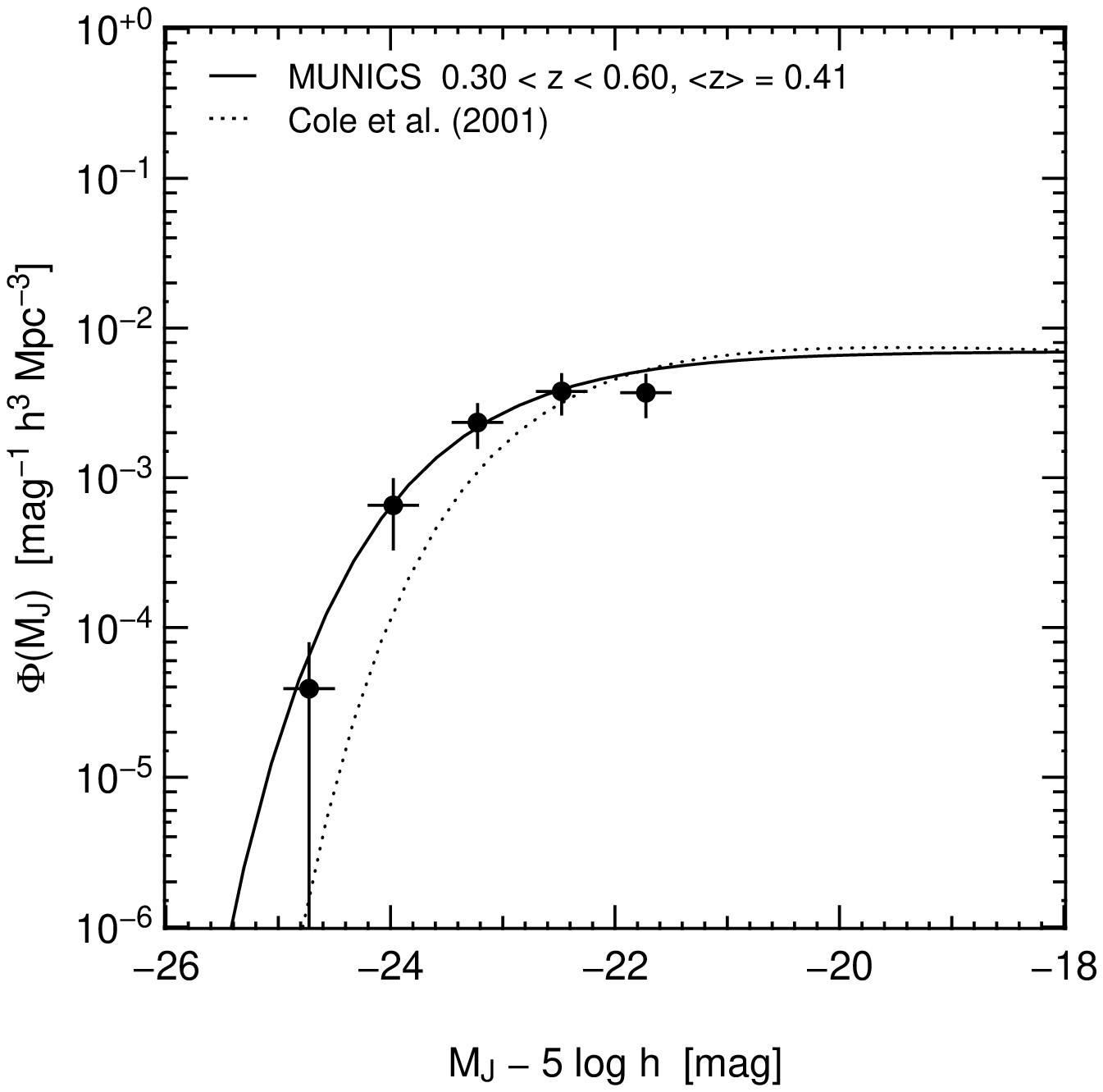,height=5.75cm}

  \vspace*{.5cm}

  \epsfig{figure=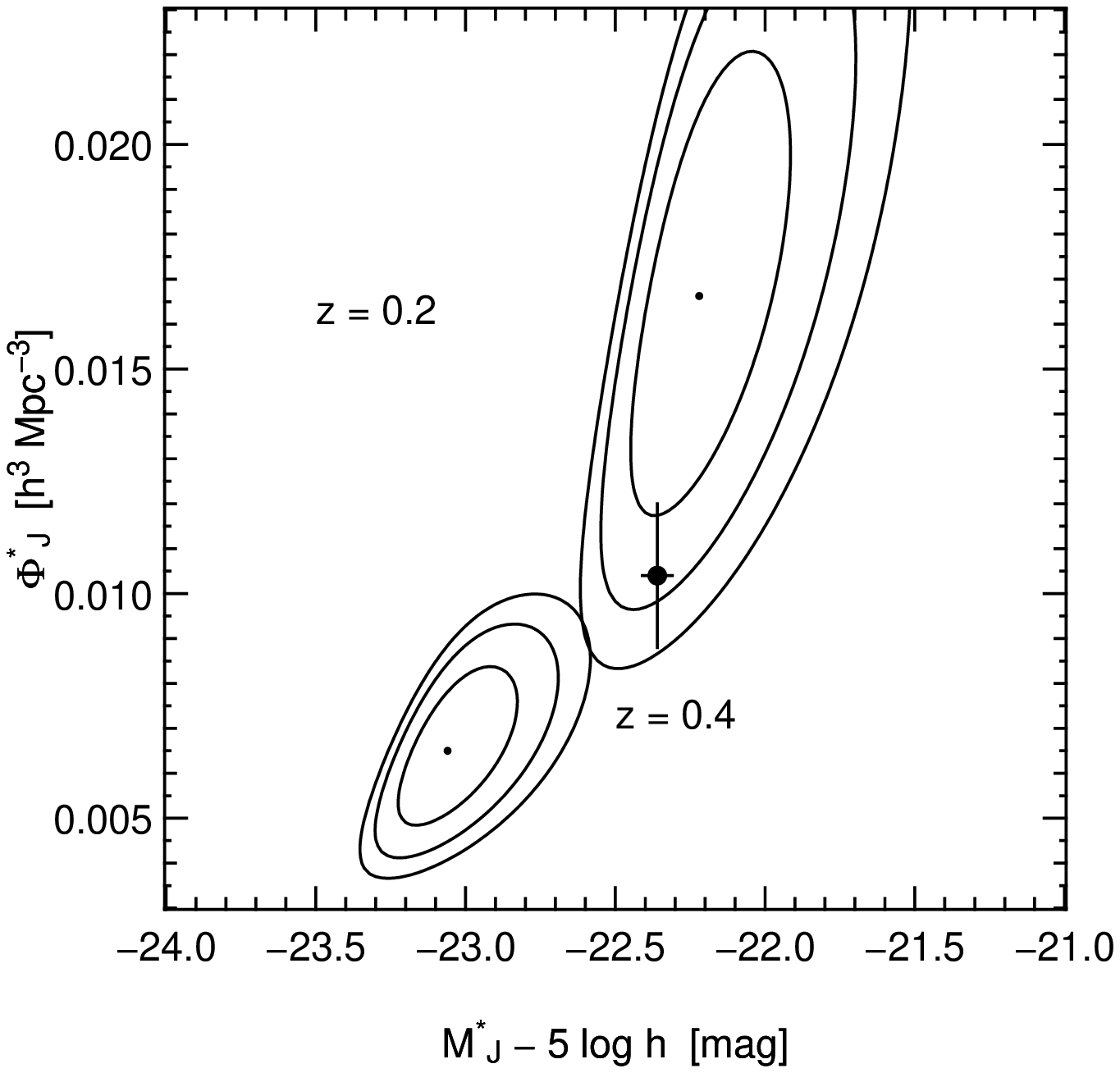,height=5.7cm}
  \hspace*{1cm}
  \epsfig{figure=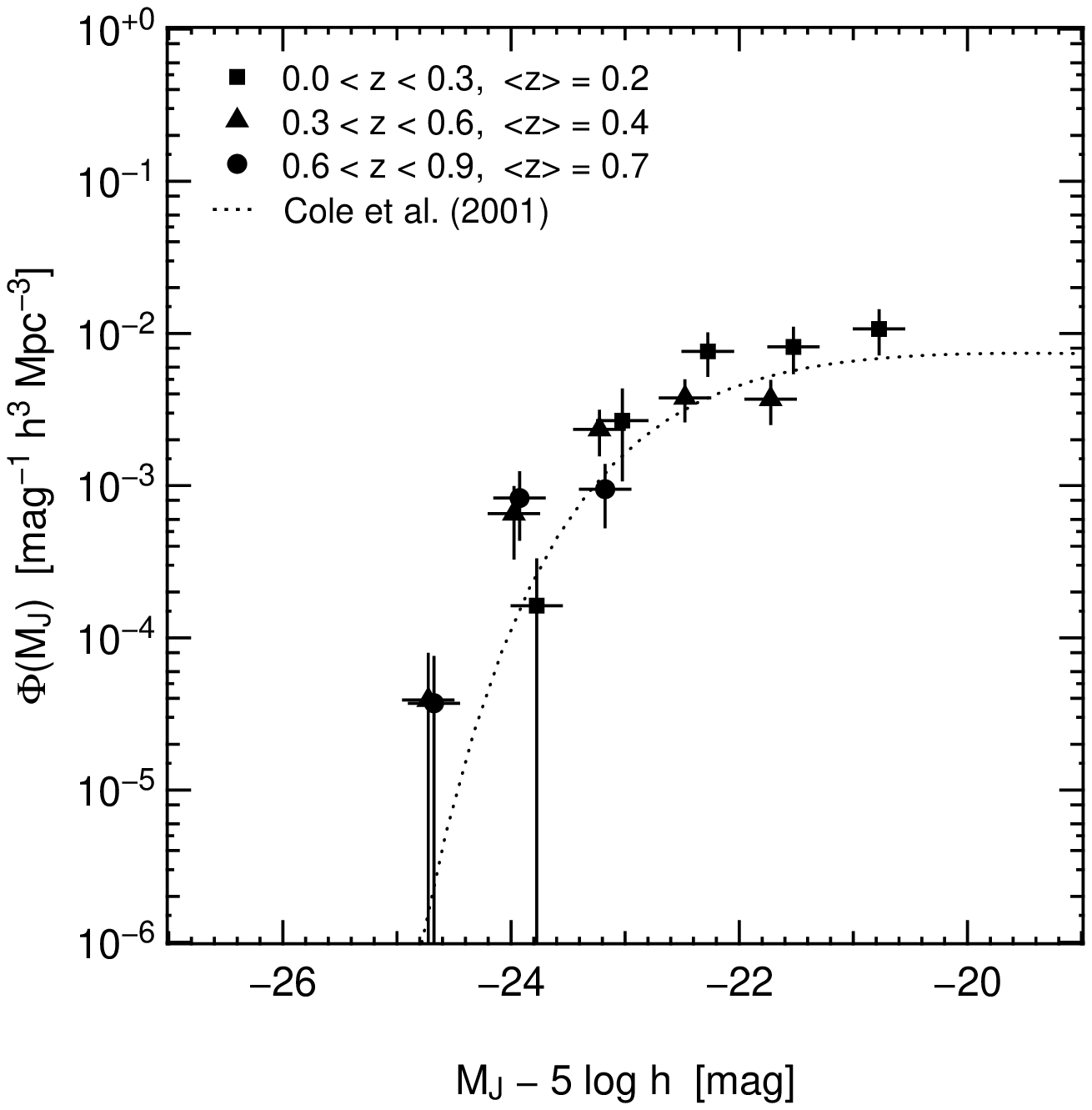,height=5.75cm}

  \caption{The $J$-band luminosity function of galaxies from
  spectroscopic observations of MUNICS galaxies. \textit{Upper left
  panel:} The luminosity function in the redshift range $0.1 \le z \le
  0.3$ (median redshift $z=0.2$; filled symbols). The vertical error
  bars give the Poissonian error, while the binning error was
  estimated from $b/\sqrt{12}$ with the size $b$ of the bin. Also
  shown are the measurement of the local $J$-band luminosity function
  by \citet{Coleetal2001_2} as well as a Schechter approximation to
  the MUNICS data (solid line). The parameters of the Schechter fit
  are \mjlt , \pjlt , and $\alpha_J = -1.00$ (fixed). \textit{Upper
  right panel:} The same as before, but in the redshift range $0.3 \le
  z \le 0.6$ (median redshift $z=0.4$), with Schechter parameters
  \mjmt , \pjmt , and $\alpha_J = -1.00$ (fixed). \textit{Lower left
  panel:} Error contours ($1\sigma$, $2\sigma$, and $3\sigma$) for the
  Schechter parameters $M_J^*$ and $\Phi_J^*$ in the two lower
  redshift bins from the $\chi^2$ distribution. The filled circle
  indicates the value of the local measurement by
  \citet{Coleetal2001_2} with the appropriate error bar. \textit{Lower
  right panel:} The luminosity function in the interval $0.6 \le z \le
  0.9$ (median redshift $z=0.70$; filled circles) compared to the
  luminosity function in the two lower redshift bins ($z = 0.2$,
  squares; $z = 0.4$, triangles) and the measurements of the local
  $J$-band luminosity function by \citet{Coleetal2001_2}.}
  \label{f:lfj}
\end{figure*}

The results for the rest-frame $J$-band luminosity function of
galaxies from our spectroscopic catalogue are shown in
Fig.~\ref{f:lfj}. Note that this is the first determination of the
$J$-band luminosity function at higher redshifts following the local
measurements by \citet{Baloghetal2001} and \citet{Coleetal2001_2}.

The Schechter parameters (see equations~(\ref{e:schechter}) and
(\ref{e:schechter2}) for the definition of the Schechter function)
derived in the lowest redshift bin ($0.1 \le z \le 0.3$, median
redshift $z = 0.2$) are \mjlt , \pjlt\ with the faint-end slope
$\alpha_J$ set to $\alpha_J = -1.00$. The errors were derived by
running Monte-Carlo simulations with 100~000 iterations, taking into
account the errors due to the binning in absolute magnitudes.
Compared to the \citet{Coleetal2001_2} sample, our measurement seems
to favour a similar characteristic magnitude $M_J^*$, but a somewhat
larger value of the normalisation $\Phi_K^*$. Unfortunately,
\citet{Baloghetal2001}, the only other work on the $J$-band luminosity
function available so far, do not derive the normalisation, thus the
reason for the discrepancy remains unclear. In the $K$-band, the value
of the normalisation derived by \citet{Coleetal2001_2} is in good
agreement with other measurements, although maybe a bit on the low
side.

For the intermediate redshift bin $0.3 \le z \le 0.6$ (median redshift
$z = 0.4$), the Schechter parameters derived from our data are \mjmt ,
\pjmt , where $\alpha_J$ was again set to a value of $-1.00$. Thus,
similar to the $K$-band, we see a mild evolution of the luminosity
function with a higher characteristic luminosity and a smaller
normalisation. The error contours for the Schechter parameters of the
luminosity function in the two lower redshift bins are shown in the
same Figure. These contours were derived from the $\chi^2$
distribution. At the highest redshifts probed by our sample, the
evolution of the $J$-band luminosity function seems to be confirmed,
although the statistics becomes rather poor in this case.

\begin{table}

\caption{The uncorrected values $\Phi_\mathrm{u} (M_K)$ of the
$K$-band luminosity function, the completeness-corrected values $\Phi
(M_K)$, and their errors $d\Phi (M_K)$ for the individual redshift
intervals and magnitude bins. Here, `completeness correction' refers to
the correction of incompleteness of the photometric catalogue, the
incompleteness of the spectroscopic sample with respect to the
photometric one, and to the $V/V_\mathrm{max}$ correction.}\label{t:lfk}

\begin{center}
\begin{tabular}{cccc}
\hline
$M_K - 5 \log h$ & $\Phi_\mathrm{u} (M_K)$ & $\Phi (M_K)$ & $d\Phi (M_K)$\\
$\mbox{}$[mag] & & [$\mathrm{mag}^{-1} \, h^3 \, \mathrm{Mpc}^{-3}$] & \\
\hline
\multicolumn{4}{c}{Median redshift $\langle z \rangle = 0.2$:}\\
$-24.825$ & $3.25 \times 10^{-4}$ & $3.89 \times 10^{-4}$ & $2.75 \times 10^{-4}$ \\
$-24.075$ & $1.63 \times 10^{-3}$ & $2.56 \times 10^{-3}$ & $1.55 \times 10^{-3}$ \\
$-23.325$ & $5.69 \times 10^{-3}$ & $6.78 \times 10^{-3}$ & $2.22 \times 10^{-3}$ \\
$-22.575$ & $7.64 \times 10^{-3}$ & $9.68 \times 10^{-3}$ & $2.88 \times 10^{-3}$ \\
$-21.825$ & $5.85 \times 10^{-3}$ & $1.05 \times 10^{-2}$ & $3.49 \times 10^{-3}$ \\
$-21.075$ & $3.25 \times 10^{-3}$ & $1.01 \times 10^{-2}$ & $4.97 \times 10^{-3}$ \\
$-20.325$ & $1.14 \times 10^{-3}$ & $1.75 \times 10^{-2}$ & $1.22 \times 10^{-2}$ \\
\hline
\multicolumn{4}{c}{Median redshift $\langle z \rangle = 0.4$:}\\
$-25.625$ & $2.93 \times 10^{-5}$ & $3.90 \times 10^{-5}$ & $3.90 \times 10^{-5}$ \\
$-24.875$ & $4.69 \times 10^{-4}$ & $6.91 \times 10^{-4}$ & $3.42 \times 10^{-4}$ \\
$-24.125$ & $1.41 \times 10^{-3}$ & $2.44 \times 10^{-3}$ & $7.14 \times 10^{-4}$ \\
$-23.375$ & $1.50 \times 10^{-3}$ & $4.12 \times 10^{-3}$ & $1.32 \times 10^{-3}$ \\
$-22.625$ & $8.51 \times 10^{-4}$ & $3.86 \times 10^{-3}$ & $1.27 \times 10^{-3}$ \\
\hline
\multicolumn{4}{c}{Median redshift $\langle z \rangle = 0.7$:}\\
$-25.825$ & $1.50 \times 10^{-5}$ & $1.87 \times 10^{-5}$ & $1.87 \times 10^{-5}$ \\
$-25.075$ & $1.65 \times 10^{-4}$ & $4.20 \times 10^{-4}$ & $2.37 \times 10^{-4}$ \\
$-24.325$ & $2.84 \times 10^{-4}$ & $1.14 \times 10^{-3}$ & $4.42 \times 10^{-4}$ \\
$-23.575$ & $1.35 \times 10^{-4}$ & $9.63 \times 10^{-4}$ & $4.59 \times 10^{-4}$ \\
\hline
\end{tabular}
\end{center}

\end{table}

\begin{table}

\caption{The uncorrected values $\Phi_\mathrm{u} (M_J)$ of the
$J$-band luminosity function, the completeness-corrected values $\Phi
(M_J)$, and their errors $d\Phi (M_J)$ for the individual redshift
intervals and magnitude bins. Here, `completeness correction' refers to
the correction of incompleteness of the photometric catalogue, the
incompleteness of the spectroscopic sample with respect to the
photometric one, and to the $V/V_\mathrm{max}$ correction.}\label{t:lfj}

\begin{center}
\begin{tabular}{cccc}
\hline
$M_J - 5 \log h$ & $\Phi_\mathrm{u} (M_J)$ & $\Phi (M_J)$ & $d\Phi (M_J)$\\
$\mbox{}$[mag] & & [$\mathrm{mag}^{-1} \, h^3 \, \mathrm{Mpc}^{-3}$] & \\
\hline
\multicolumn{4}{c}{Median redshift $\langle z \rangle = 0.2$:}\\
$-23.775$ & $1.63 \times 10^{-4}$ & $1.63 \times 10^{-4}$ & $1.63 \times 10^{-4}$ \\
$-23.025$ & $1.95 \times 10^{-3}$ & $2.88 \times 10^{-3}$ & $1.70 \times 10^{-3}$ \\
$-22.275$ & $6.34 \times 10^{-3}$ & $7.55 \times 10^{-3}$ & $2.30 \times 10^{-3}$ \\
$-21.525$ & $6.50 \times 10^{-3}$ & $8.26 \times 10^{-3}$ & $2.68 \times 10^{-3}$ \\
$-20.775$ & $6.50 \times 10^{-3}$ & $1.13 \times 10^{-2}$ & $3.55 \times 10^{-3}$ \\
\hline
\multicolumn{4}{c}{Median redshift $\langle z \rangle = 0.4$:}\\
$-24.725$ & $2.93 \times 10^{-5}$ & $3.90 \times 10^{-5}$ & $3.90 \times 10^{-5}$ \\
$-23.975$ & $4.40 \times 10^{-4}$ & $6.58 \times 10^{-4}$ & $3.21 \times 10^{-4}$ \\
$-23.225$ & $1.20 \times 10^{-3}$ & $2.37 \times 10^{-3}$ & $7.57 \times 10^{-4}$ \\
$-22.475$ & $1.61 \times 10^{-3}$ & $3.94 \times 10^{-3}$ & $1.16 \times 10^{-3}$ \\
$-21.725$ & $9.68 \times 10^{-4}$ & $4.02 \times 10^{-3}$ & $1.24 \times 10^{-3}$ \\
\hline
\multicolumn{4}{c}{Median redshift $\langle z \rangle = 0.7$:}\\
$-24.675$ & $2.99 \times 10^{-5}$ & $3.67 \times 10^{-5}$ & $3.67 \times 10^{-5}$ \\
$-23.925$ & $2.84 \times 10^{-4}$ & $8.41 \times 10^{-4}$ & $3.89 \times 10^{-4}$ \\
$-23.175$ & $2.24 \times 10^{-4}$ & $9.72 \times 10^{-4}$ & $4.21 \times 10^{-4}$ \\
\hline
\end{tabular}
\end{center}

\end{table}

\section{Discussion}
\label{s:discussion}

To quantify the redshift evolution of the near-infrared luminosity
functions, we performed both Kolmogorov-Smirnov tests for the
cumulative distributions of absolute magnitudes and a $\chi^2$
analysis for the redshift evolution of the Schechter parameters.

\subsection{Kolmogorov-Smirnov tests}
\label{s:ks}

One method to compare the measurements of the near-infrared luminosity
functions at various redshifts is to apply the Kolmogorov-Smirnov test
to the cumulative distribution of objects in absolute magnitudes. Of
course, completeness and $1/V_\mathrm{max}$ corrections have to be
applied to each individual object entering the distribution. The
advantage of this method is that it uses the absolute magnitudes as
measured in the sample \textit{without binning}. Furthermore, it is
independent of the \textit{relative} normalisation of the samples one
wants to compare.

\begin{figure*}

  \epsfig{figure=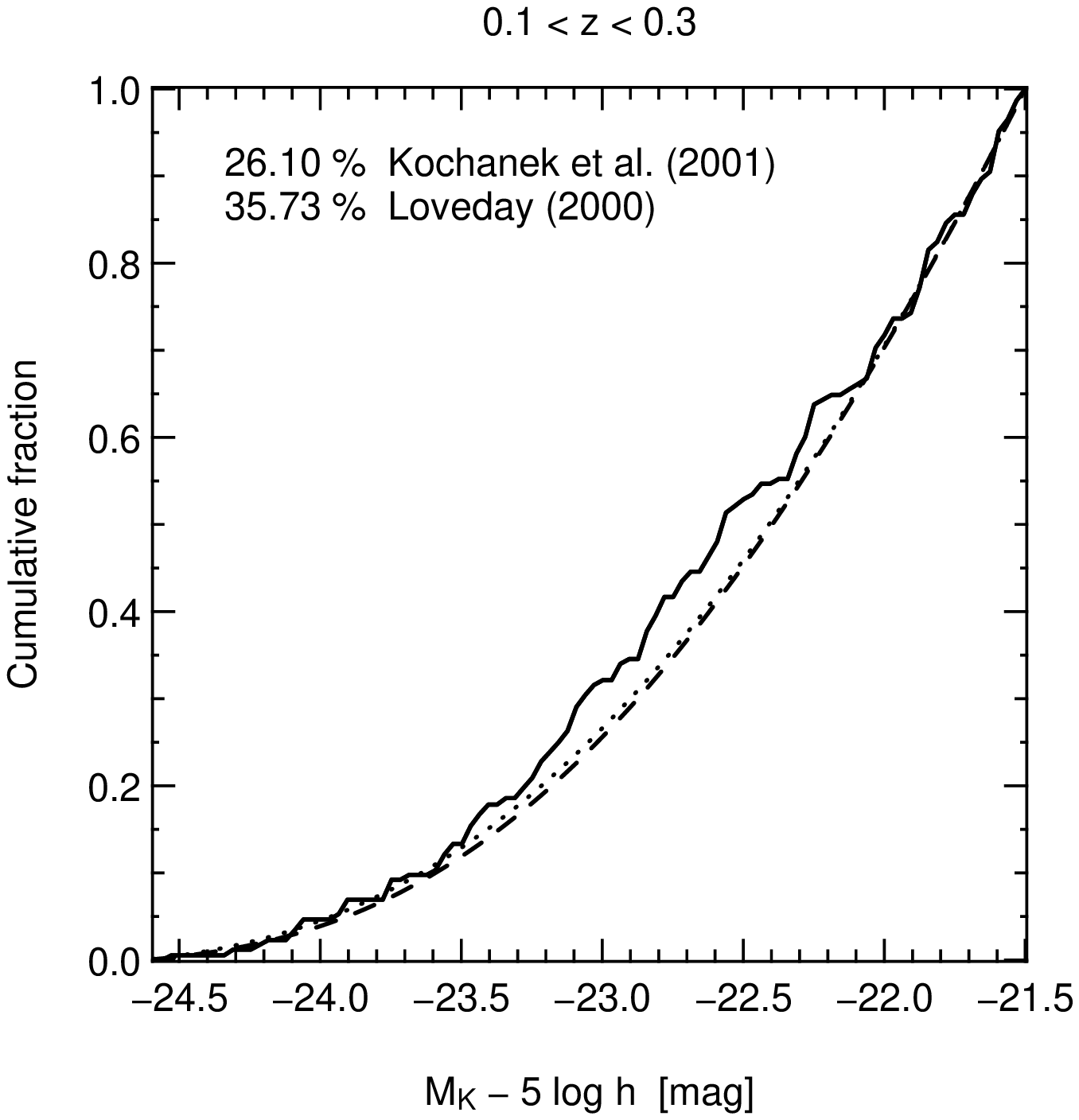,width=5.25cm}
  \hfill
  \epsfig{figure=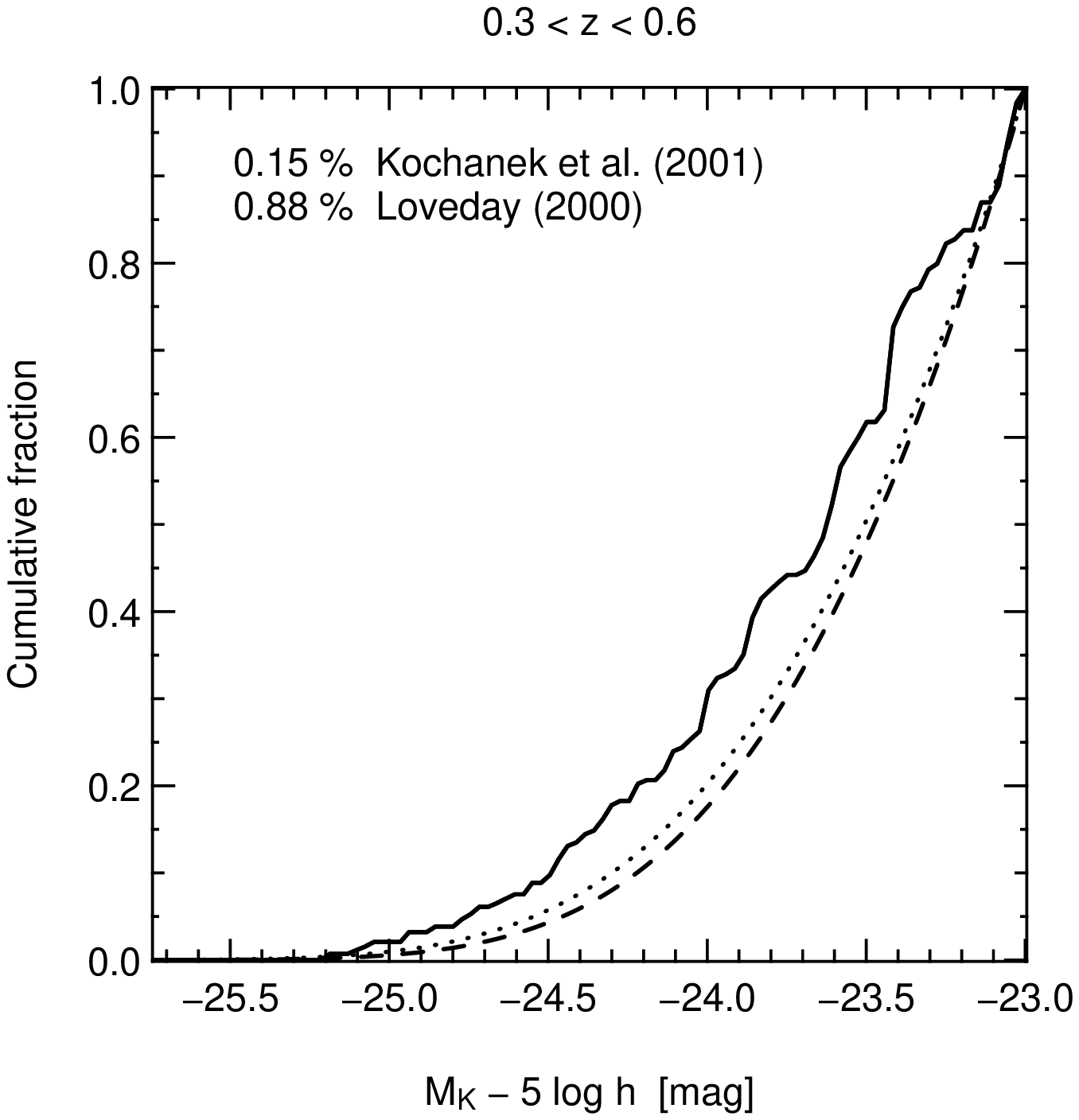,width=5.25cm}
  \hfill
  \epsfig{figure=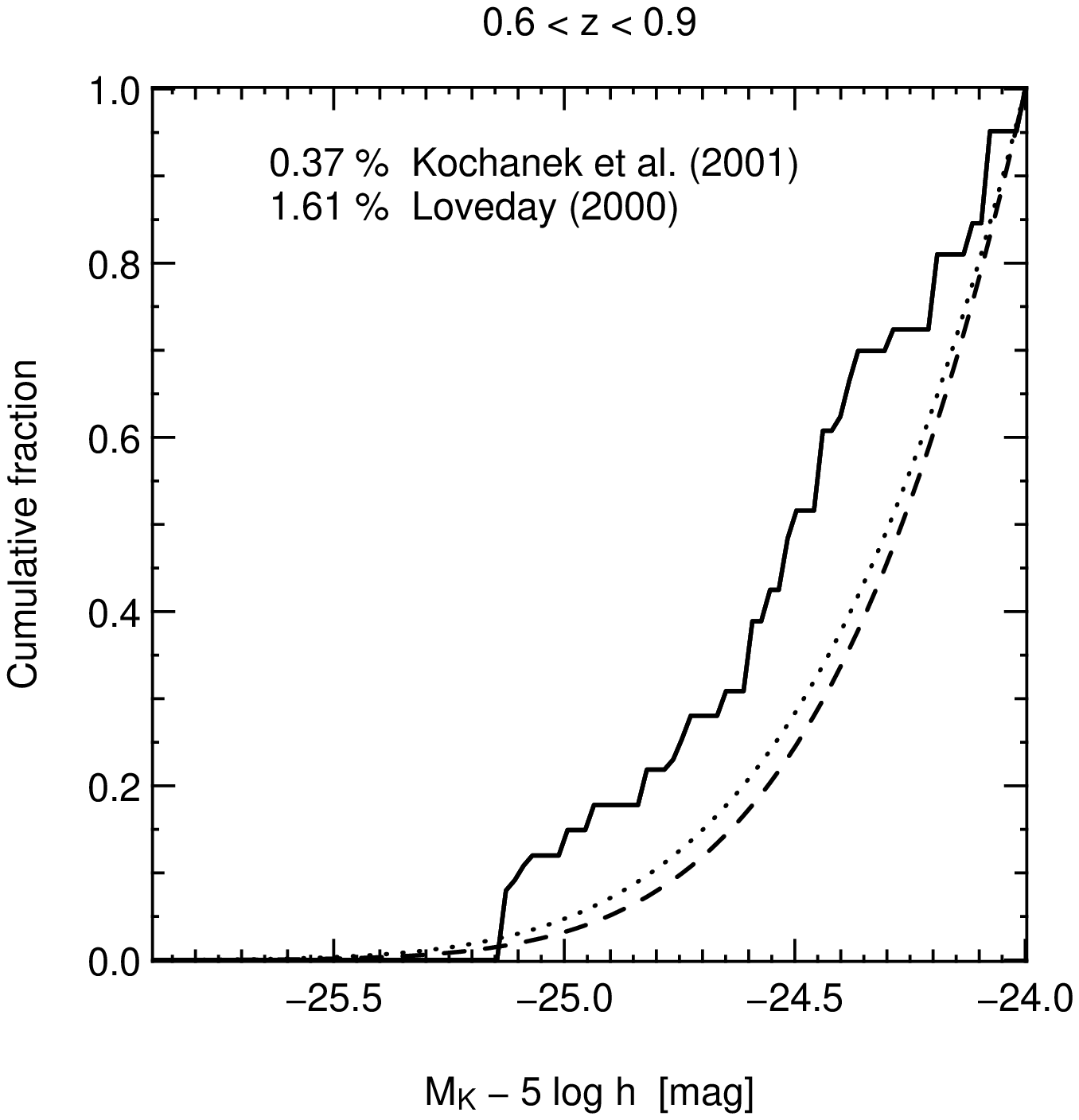,width=5.25cm}

\caption{Resulting cumulative distribution functions for the absolute
$K$-band magnitudes from the MUNICS spectroscopic catalogue (solid
line) as compared to the local measurements by \citeauthor{Loveday00}
(\citeyear{Loveday00}; dotted line) and \citeauthor{Kochaneketal01}
(\citeyear{Kochaneketal01}; dashed line). The diagram on the left-hand
side compares the $0.1 < z < 0.3$ redshift bin to the local one, the
middle diagram shows the same for the $0.3 < z < 0.6$ redshift
interval, and the diagram on the right-hand side compares the
distributions in the $0.6 < z < 0.9$ interval to the local ones. The
values quoted in the diagrams give the probabilities that both
cumulative distributions are drawn from the same
population.}\label{f:ks1k}

\end{figure*}

\begin{figure*}

  \epsfig{figure=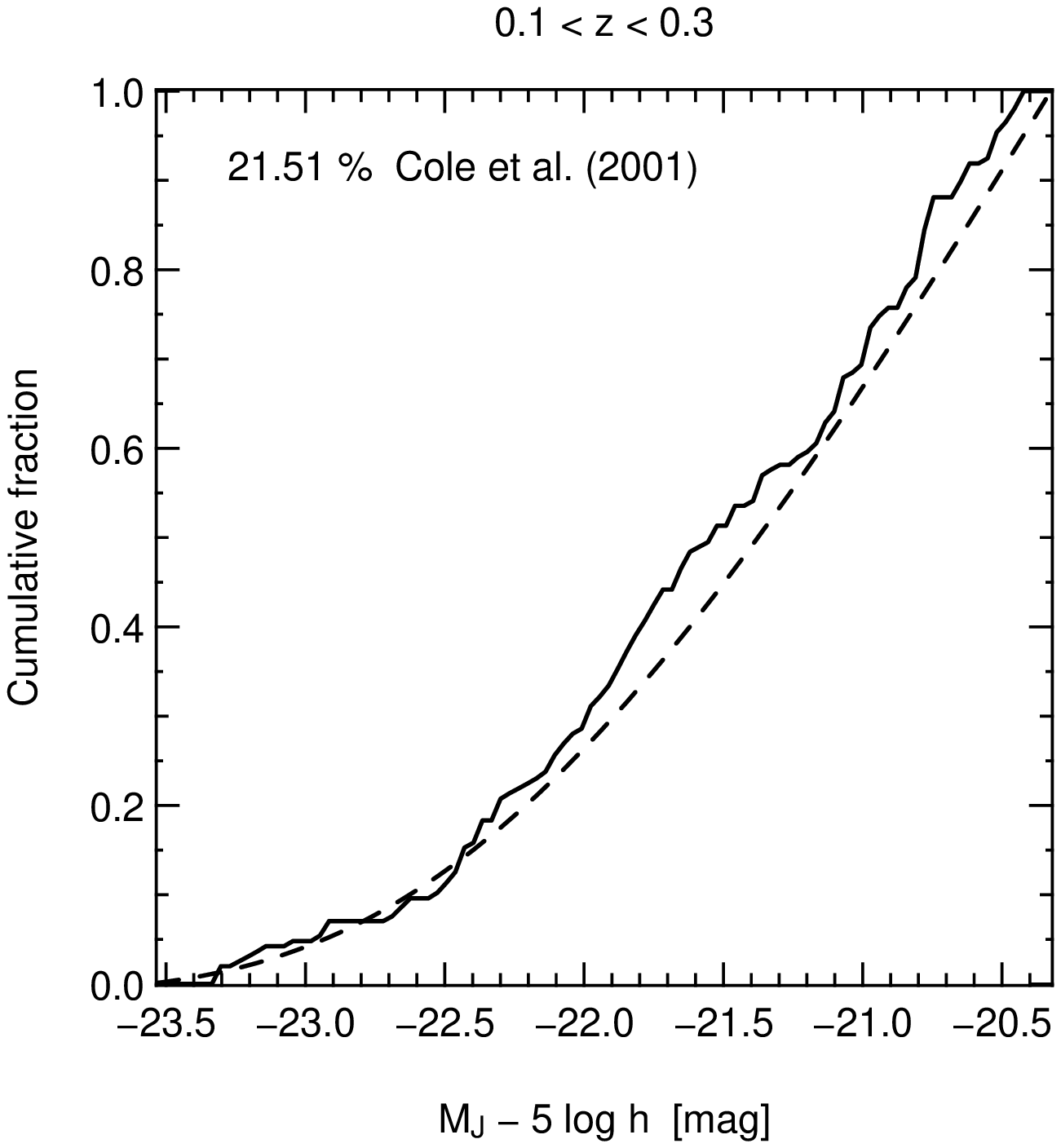,width=5.25cm}
  \hfill
  \epsfig{figure=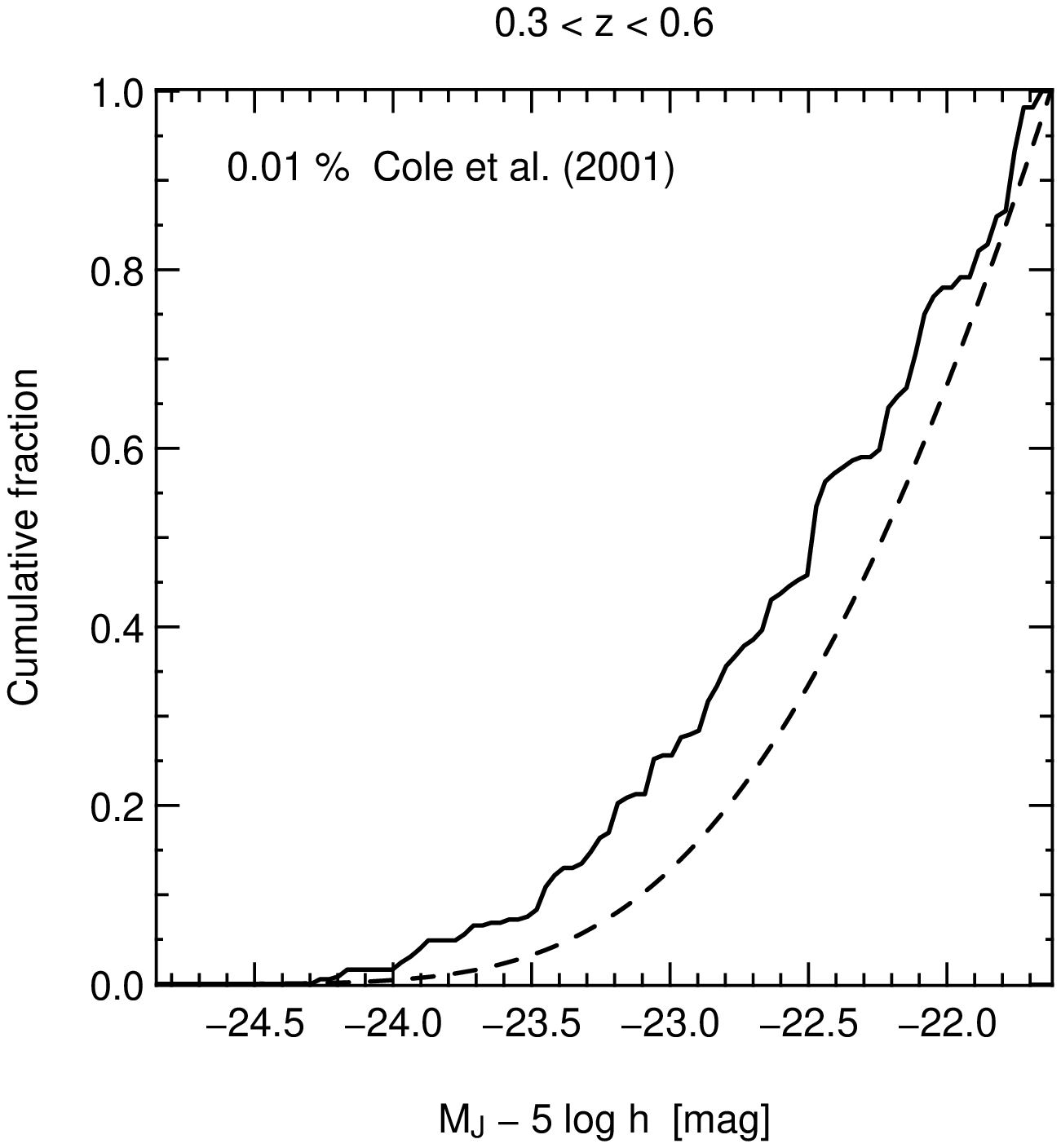,width=5.25cm}
  \hfill
  \epsfig{figure=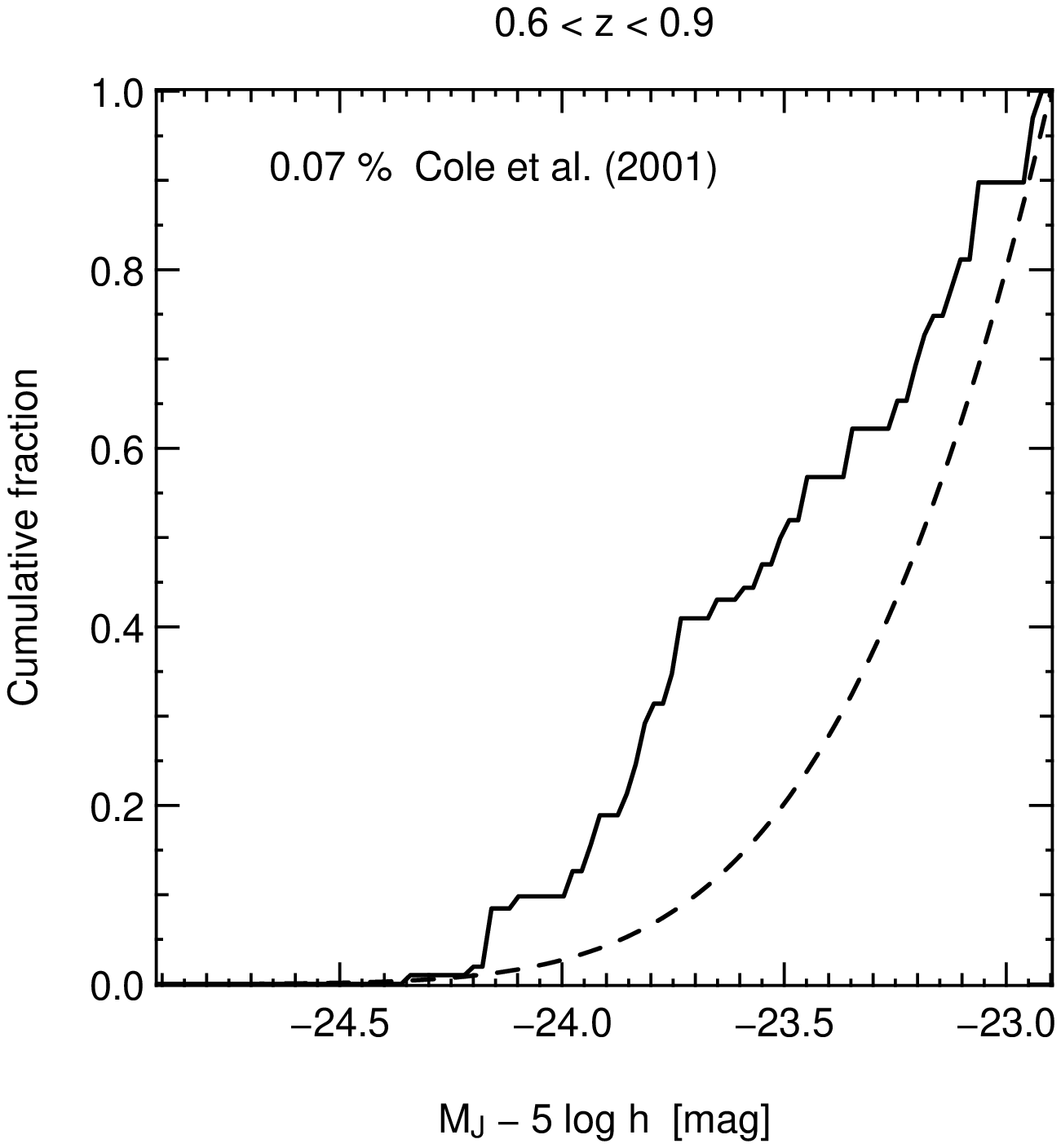,width=5.25cm}

\caption{Resulting cumulative distribution functions for the absolute
$J$-band magnitudes from the MUNICS spectroscopic catalogue (solid
line) as compared to the local measurements by
\citeauthor{Coleetal2001_2} (\citeyear{Coleetal2001_2}; dashed
line). The diagram on the left-hand side compares the $0.1 < z < 0.3$
redshift bin to the local one, the middle diagram shows the same for
the $0.3 < z < 0.6$ redshift interval, and the diagram on the
right-hand side compares the distribution in the $0.6 < z < 0.9$
interval to the local one. The values quoted in the plots give the
probabilities that both cumulative distributions are drawn from the
same population.}\label{f:ks1j}

\end{figure*}

We firstly have run this test by comparing the cumulative luminosity
distributions from our data in the three redshift bins to the local
measurements of the luminosity function. The results are shown in
Table~\ref{t:ks1k} for the $K$-band, and Table~\ref{t:ks1j} for the
$J$-band, respectively. Furthermore, we show the resulting cumulative
distributions in Figures~\ref{f:ks1k} and \ref{f:ks1j} for the $K$
band and the $J$ band, respectively.

While the cumulative distributions are in fair agreement for the lower
redshift bin ($0.1 < z < 0.3$), they are significantly different for
the two higher redshift bins ($0.3 < z < 0.6$ and $0.6 < z < 0.9$,
respectively), confirming the result found from the fitting of
Schechter parameters $M^*$ and $\Phi^*$.

\begin{table}

\caption{Probabilities for compatible cumulative distributions of
absolute magnitudes derived from a Kolmogorov-Smirnov test. The
comparison distributions are derived from the local luminosity
function measurements by \citet{Loveday00} and
\citet{Kochaneketal01}. The magnitudes in the second column are upper
limits to the range of absolute magnitudes for which the distribution
has been computed.}\label{t:ks1k}

\begin{center}
\begin{tabular}{ccrr}
\hline
Redshift & $M_K$ [mag] & \citet{Loveday00} & \citet{Kochaneketal01} \\
\hline
$\langle z \rangle = 0.2$ & $-21.5$ & 35.73 \% & 26.10 \% \\
$\langle z \rangle = 0.4$ & $-23.0$ &  0.88 \% &  0.15 \% \\
$\langle z \rangle = 0.7$ & $-24.0$ &  1.61 \% &  0.37 \% \\
\hline
\end{tabular}
\end{center}
\end{table}

\begin{table}

\caption{Probabilities for compatible cumulative distributions of
absolute magnitudes derived from a Kolmogorov-Smirnov test. The
comparison distribution is derived from the local luminosity function
measurement by \citet{Coleetal2001_2}. The magnitudes in the second
column are upper limits to the range of absolute magnitudes for which
the distribution has been computed.}\label{t:ks1j}

\begin{center}
\begin{tabular}{ccr}
\hline
Redshift & $M_J$ [mag] & \citet{Coleetal2001_2} \\
\hline
$\langle z \rangle = 0.2$ & $-20.3$ & 21.51 \% \\
$\langle z \rangle = 0.4$ & $-21.6$ &  0.01 \% \\
$\langle z \rangle = 0.7$ & $-22.9$ &  0.07 \% \\
\hline
\end{tabular}
\end{center}
\end{table}

\begin{figure*}

  \epsfig{figure=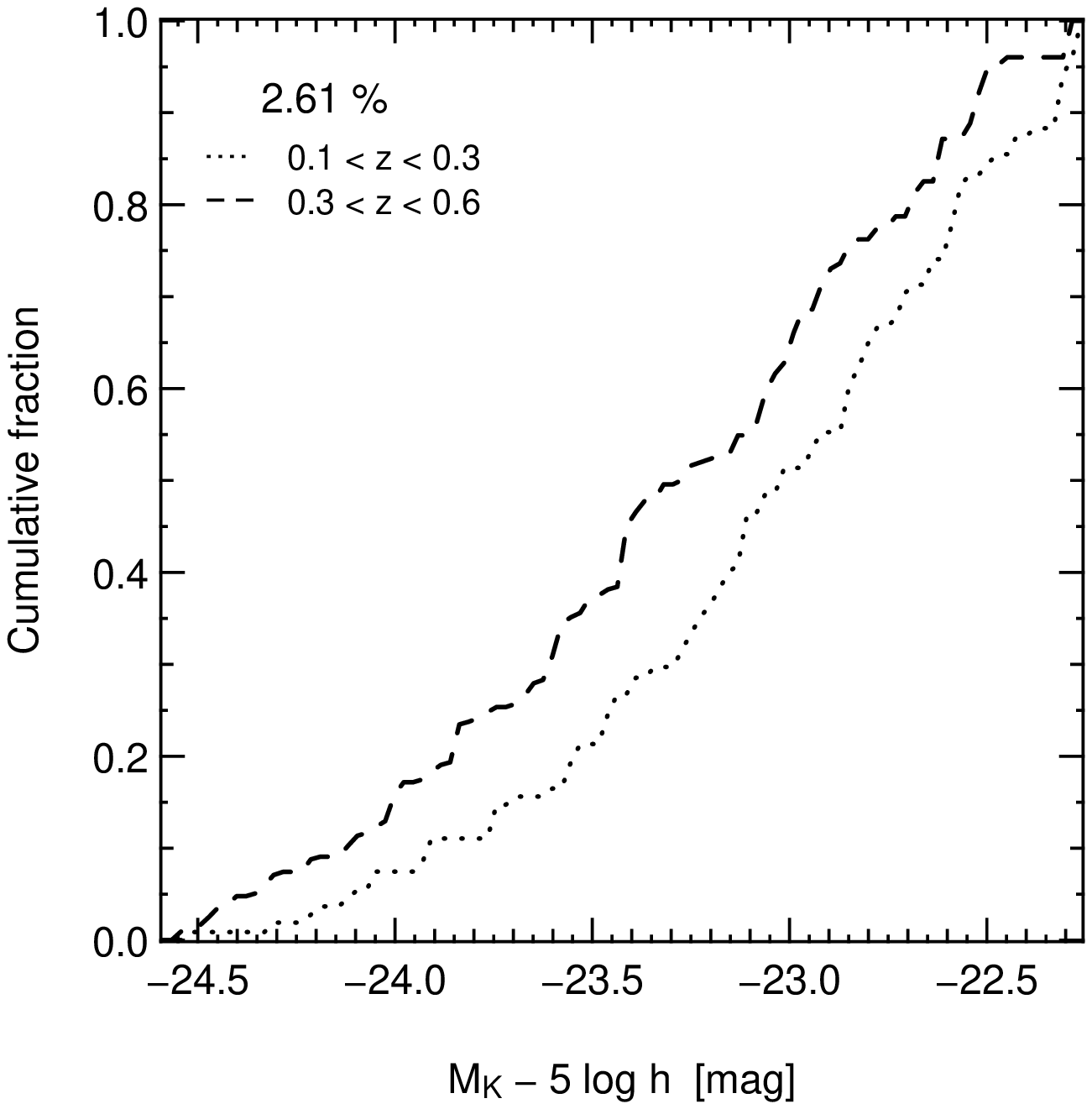,height=5.25cm}
  \hspace*{2em}
  \epsfig{figure=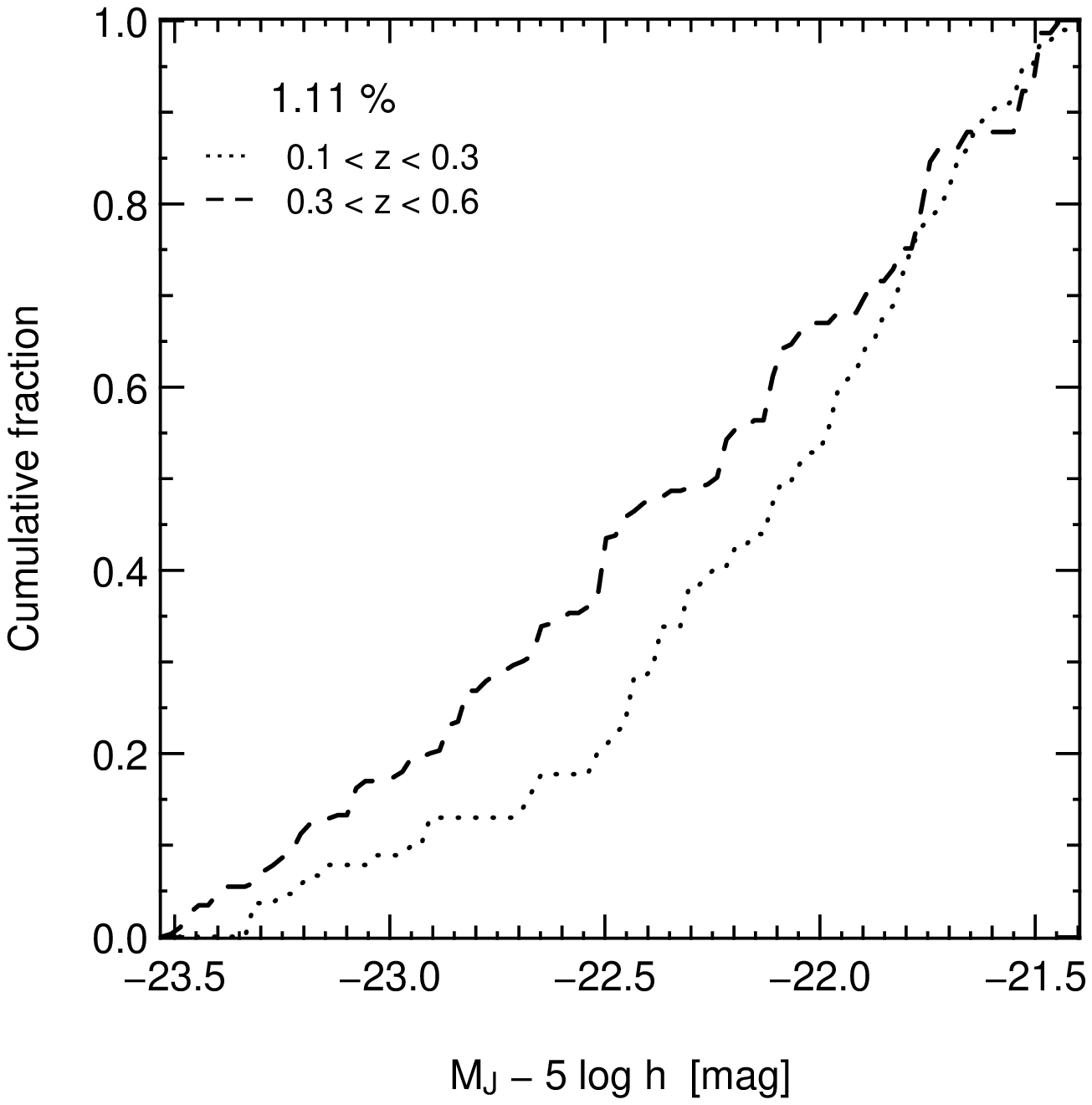,height=5.25cm}

\caption{Cumulative distributions in absolute magnitude for the two
lower redshift bins and results from a Kolmogorov-Smirnov test applied
to these data, both for the $K$ band (left panel) and the $J$ band
(right panel).}\label{f:ks2}

\end{figure*}

As an additional test we have applied the Kolmogorov-Smirnov test to
the cumulative distributions from the two lower redshift
bins. The results of this test are shown in Fig.~\ref{f:ks2}, where
one can see that the probabilities that both distributions are drawn
from the same parent population are very small.

\subsection{Likelihood analysis of luminosity function evolution}

In this Section we will describe a $\chi^2$ analysis of the redshift
evolution of the near-infrared luminosity functions. This test uses
the Schechter parametrisation

\begin{equation}
  \Psi \, (L) \; = \; \frac{\Phi^*}{L^*} \, \left( \frac{L}{L^*}
  \right)^\alpha \, \exp \left( - \frac{L}{L^*} \right)
  \label{e:schechter}
\end{equation}

of the luminosity function, where $L^*$ is the characteristic
luminosity, $\alpha$ the faint-end slope, and $\Phi^*$ the
normalisation of the luminosity function \citep{Schechter76}. The
corresponding equation in absolute magnitudes reads

\begin{eqnarray}
  \nonumber
  \Psi \, (M) & = & \frac{2}{5} \, \Phi^* \, \ln 10 \: 
  10^{0.4(M^*-M)(1+\alpha)} \\
  & \mbox{} & \exp \left( - 10^{0.4 (M^*-M)} \right) .
  \label{e:schechter2}
\end{eqnarray}

To estimate the rate of evolution of the parameters with redshift, we
define evolution parameters $\mu$ and $\nu$ as follows:

\begin{eqnarray}
  \nonumber
  \Phi^* \, (z) & = & \Phi^* \, (0) \: \left( 1 \, + \, \mu z \right)
  , \\
  M^*    \, (z) & = & M^*    \, (0) \, + \, \nu z , \; \mathrm{and} 
\label{e:evol}\\
  \nonumber
  \alpha \, (z) & = & \alpha \, (0) \;\; \equiv \;\; \alpha .
\end{eqnarray}

Note that the faint end of the luminosity function cannot be
determined very well from our data, thus we leave the faint-end slope
$\alpha$ of the Schechter luminosity function fixed, as we have
also done during the fitting of a Schechter function to our data.

To quantify the redshift evolution of $\Phi^*$ and $M^*$ we now
compare our luminosity function data in \textit{all} redshift bins
with the local Schechter function evolved according to equation
(\ref{e:evol}) to the appropriate redshift. We do this for a grid of
values for $\mu$ and $\nu$, and calculate the value of $\chi^2$ for
each grid point according to

\begin{equation}
  \chi^2 \, (\mu, \nu) \; = \; \frac{1}{n} \, \sum_{i=1}^N \, \frac{\left[
  \Phi (M_i, z_i) - \Psi (M_i, \mu, \nu, z_i ) \right]^2}{\sigma_i^2}
  ,
  \label{e:chi}
\end{equation}

where $\Phi (M, z)$ is the measurement of the luminosity function at
median redshift $z$ in the magnitude bin centred on $M$, $\Psi (M,
\mu, \nu, z)$ is the local Schechter function evolved according to the
evolution model defined in equation (\ref{e:evol}) to the redshift
$z$, $\sigma_i$ is the RMS error of the luminosity function value, and
$n$ is the number of free parameters of the approximation, i.e.\ the
number of data points used minus the number of parameters derived from
the fitting.

We want to compare our measurement of the $K$-band luminosity function
with the Schechter approximations to the local determinations. We use
the measurements by \citet{Loveday00} and \citet{Kochaneketal01} for
the $K$-band (since the luminosity function parameters derived from
local samples are very similar anyway), and the local $J$-band
luminosity function is the one by \citet{Coleetal2001_2}. The
Schechter parameters derived by those authors are shown in
Tables~\ref{t:litk} and \ref{t:litj}. Chosing a shallower faint-end
slope in the $K$ band, similar to the one derived by
\citet{Coleetal2001_2}, changes the result slightly, but -- within the
errors -- not significantly.

To avoid that data points with large completeness correction factors
affect the result, we exclude all luminosity function measurements
with a total correction factor (photometric incompleteness,
spectroscopic incompleteness, and $V/V_\mathrm{max}$ correction)
larger than three.

\begin{figure*}

  \epsfig{figure=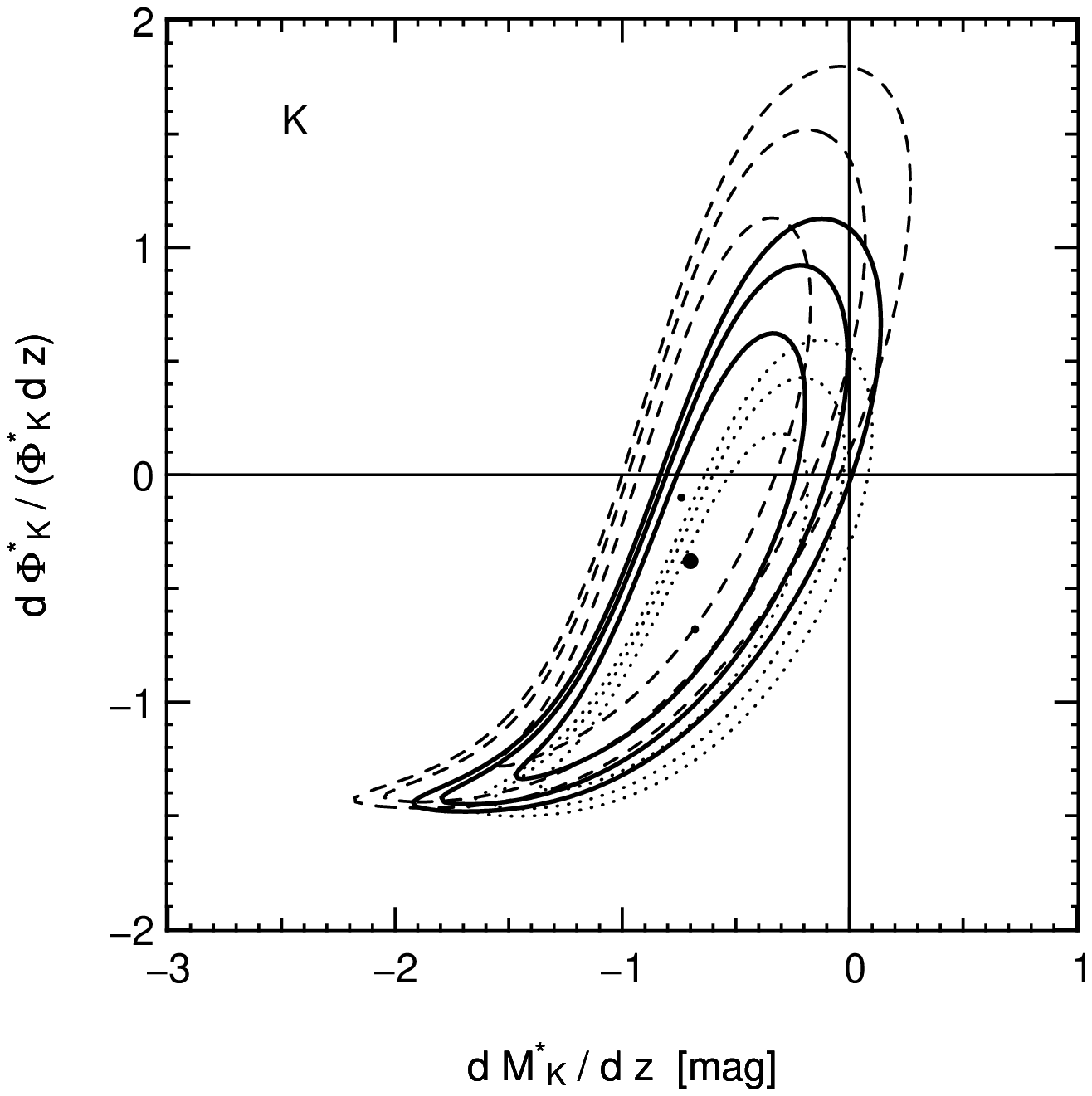,height=5.75cm}
  \hspace*{1.25cm}
  \epsfig{figure=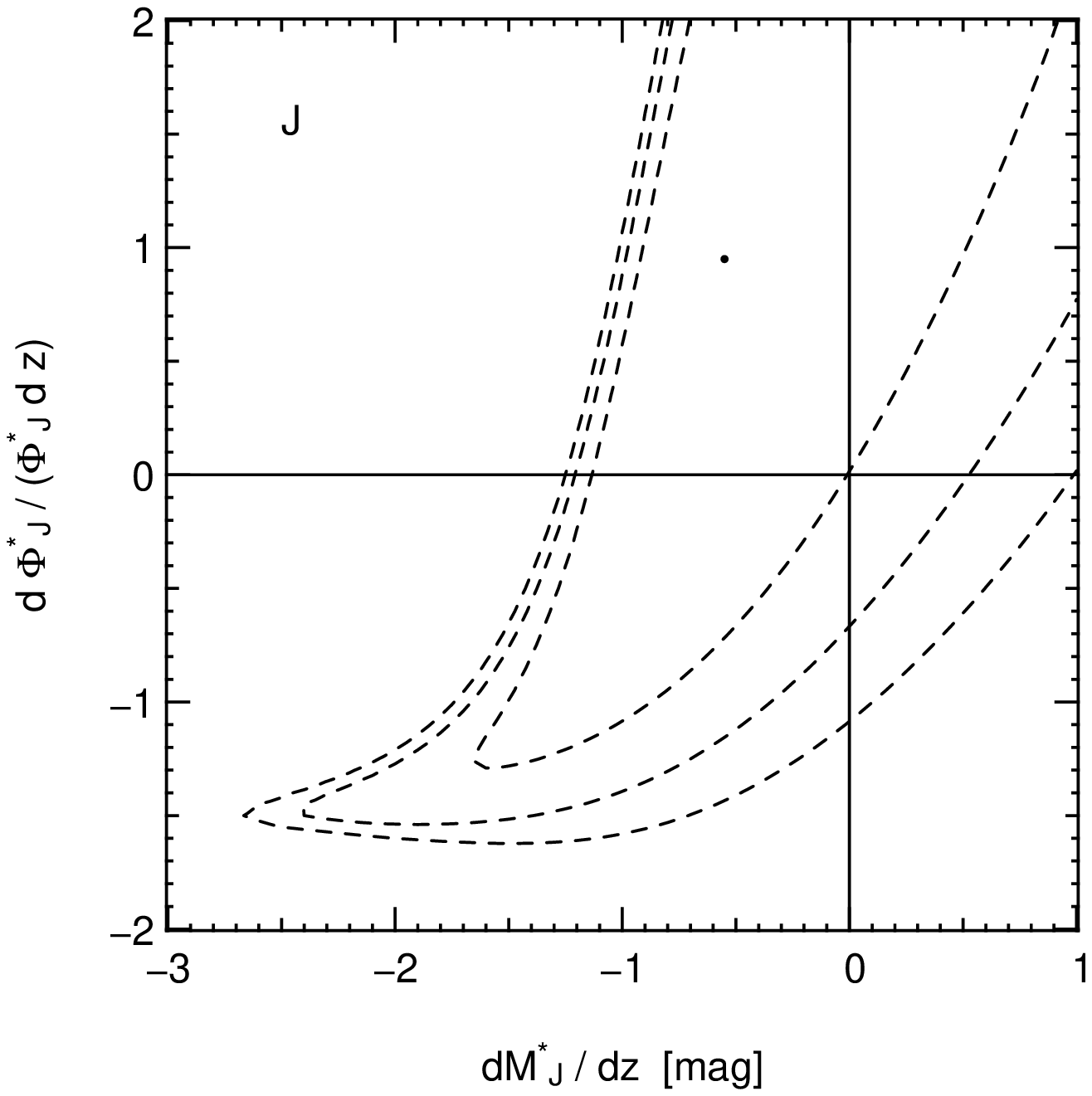,height=5.75cm}

  \caption{\textit{Left panel:} Result of the estimation of the
  redshift evolution of the Schechter parameters $M_K^*$ and
  $\Phi_K^*$ between $z = 0.4$ (MUNICS luminosity function) and the
  local universe from \citeauthor{Loveday00} (\citeyear{Loveday00};
  dotted line), \citeauthor{Kochaneketal01}
  (\citeyear{Kochaneketal01}; dashed line), and for an average of these
  two local measurements (solid line) as derived from a $\chi^2$
  approach (see text for details). The contours correspond to
  $1\sigma$, $2\sigma$, and $3\sigma$ confidence level. The $d \Phi^*
  / d z = 0$ and $d M^* / d z = 0$ lines indicate the non-evolution
  values. \textit{Right panel:} The same for the $J$-band luminosity
  function. In this case, the local measurement is taken from Cole et
  al.\ (\citeyear{Coleetal2001_2}, dotted line), with the appropriate
  Schechter parameters from Table~\ref{t:litj}, and the evolution is
  very badly constrained.}  \label{f:evol}

\end{figure*}

The result of the likelihood analysis is shown in
Fig.~\ref{f:evol}. For the $K$-band, we compare our measurements at
redshifts $0.2$, $0.4$, and $0.7$ to the local measurements by
\citet{Loveday00} and \citet{Kochaneketal01}, and to the average of
their Schechter parameters. We detect a brightening of $\Delta M_K^* /
\Delta z \simeq 0.70$ magnitudes, and a decline of the number density
of objects to redshift one. The decrease of $\Phi_K^*$ with redshift
is obviously quite strongly dependent on the parameters of the local
luminosity function, however, for the average value we derive $\Delta
\Phi_K^* / (\Phi_K^* \Delta z) \simeq - 0.35$. These results also give
quantitative estimates of the evolution which can already be seen in
the Schechter parameters derived from our data, see Table~\ref{t:lfk}
for details. Note that \citet{Huangetal2002} derive a significantly
brighter $M^*$, a slightly larger normalisation, and a steeper
faint-end slope, which they ascribe to redshift selection effects. If
their measurements are valid, the brightening to redshift one would be
smaller, whereas the evolution in number density would be even larger.

Within the errors, the results found in this work agree nicely with
the measurements of the $K$-band luminosity function derived from the
full MUNICS sample based on photometric redshifts. First results are
shown in MUNICS~III and show the same trend for the evolution of the
luminosity function with redshift. A more detailed analysis will be
shown in MUNICS~II, where the evolutionary trend with a brightening of
0.5 to 0.7 mag and a decrease in number density of roughly 25 per cent
to redshift one is confirmed.

In the case of the $J$-band (right panel of Fig.~\ref{f:evol}), the
evolution of the luminosity function is obviously not
constrained. This is also apparent from the error contours of the
Schechter parameters shown in Fig.~\ref{f:lfj} (lower left panel),
where one can see that the local measurement by \citet{Coleetal2001_2}
has a characteristic magnitude similar to the one derived in our
lowest redshift bin, but a normalisation \textit{in between} the ones
derived from our two lower redshift intervals, thus making any
conclusions about evolution with respect to the local sample
difficult. Nevertheless, we note that our $J$-band luminosity function
data seem to confirm the trend seen for the $K$ band, which is also
evident from the Kolmogorov-Smirnov tests presented in
Section~\ref{s:ks}.

\begin{table*}

\caption{Summary of samples used for deriving the $K$-band
field-galaxy luminosity function, giving the source for the data, the
approximate limiting magnitude (either in the optical or in the
near-infrared), the approximate area, the number of objects, the
redshift range of the survey (either as interval or as median
redshift) and the parameters of the Schechter function derived from
the data. Parameters without error estimate were kept fixed during the
fitting procedure. Where necessary, the values quoted have been
converted from the cosmology originally used to the $\Omega_m = 0.3$,
$\Omega_\Lambda = 0.7$ cosmology following \citet{Coleetal2001_2}.}
\label{t:litk}

\begin{tabular}{lcrrcccc}

\hline
Source & Limit & Area & \# & $z$ & $M_K^*-5\log h$
& $\alpha_K$ & $\Phi_K^*$ \\
& [mag] & [arcmin$^2$] & & &  [mag] & & [$10^{-2} h^3 \mathrm{Mpc}^{-3}$] \\
\hline
\citet{MSE93}\footnotemark[1]         & $B_J \le 17$, $K \le$13.0 & & 181
& 0.0\dots0.1 & $-$23.37 $\pm$ 0.30 & $-$1.00 $\pm$ 0.30 & 1.12 $\pm$ 0.16  \\
\citet{SCHG94}\footnotemark[2]        & $K \le$14.5\dots 20.0 &
5544\dots 5 & \\
\citet{GPMC95}\footnotemark[3]        & $K \le$17.0 &   552 & 124 &
0.0\dots0.8 & $-$23.14 $\pm$ 0.23 & $-$1.04 & 2.22 $\pm$ 0.53 \\
\citet{CSHC96}\footnotemark[4]        & $K \le$19.5 &    26 & 254 &
0.2\dots1.0 & $-$23.49 $\pm$ 0.10 & $-$1.25 $\pm$ 0.15 & 0.80 $\pm$ 0.20\\
\citet{GSFC97}\footnotemark[5]        & $K \le$15.0 & 15840 & 567 &
0.14 & $-$23.30 $\pm$ 0.17 & $-$1.00 $\pm$ 0.24 & 1.44 $\pm$ 0.20 \\
\citet{Szokolyetal98}\footnotemark[6] & $R_F \le 18.5$, $K \le$16.5 &
2160 & 175 & 0.0\dots0.4 & $-$23.80 $\pm$ 0.30 & $-$1.30 $\pm$ 0.20 &
0.86 $\pm$ 0.29 \\
\citet{Loveday00}\footnotemark[7]     & $K \le$12.0 &       & 363 & 0.05 &
$-$23.58 $\pm$ 0.42 & $-$1.16 $\pm$ 0.19 & 1.20 $\pm$ 0.08 \\
\citet{Kochaneketal01}\footnotemark[8] & $K_{20} \le 11.25$ & & 4192 &
0.02 & $-$23.43 $\pm$ 0.05 & $-$1.09 $\pm$ 0.06 & 1.16 $\pm$ 0.10 \\
\citet{Coleetal2001_2}\footnotemark[9] & $b_J \le 19.5$, $K \le 13$ &
2.2 10$^6$ & 17173 & 0.05 & $-$23.36 $\pm$ 0.02 & $-$0.93 $\pm$ 0.04 &
1.16 $\pm$ 0.17 \\
\citet{Baloghetal2001}\footnotemark[10] & $R_c \le 15\dots17$, $K \le
13$ & & & 0.0\dots0.18 & $-$23.48 $\pm$ 0.08 & $-$1.10 $\pm$ 0.14 & \\
\citet{Huangetal2002}\footnotemark[11] & $K \le 15$ & & 1065 & 0.138 &
$-$23.70 $\pm$ 0.08 & $-$1.39 $\pm$ 0.09 & 1.30 $\pm$ 0.20 \\\hline

\textbf{MUNICS (this work)} & $K \le 17.5\dots19.0$ & 649 & 157 &
0.1\dots0.3 & \mkl & $-$1.10 & \pkl \\
& & & 145 & 0.3\dots0.6 & \mkm & $-$1.10 & \pkm \\
\hline

\end{tabular}

\footnotesize
\flushleft
\footnotemark[1] Based on the optically-selected Anglo-Australian
Redshift Survey \citep{AARS}. A correction of +0.22 magnitudes is
sometimes applied to their result because of their method of
calculating $k(z)$ corrections \citep{GPMC95}.

\footnotemark[2] The area of the \citet{SCHG94} sample actually
continuously decreases with $K$ magnitude. The values in the table are
intended to give a rough impression of the parameters of the
catalogue. No luminosity function is derived by the authors, however,
the sample has been used by \citet{CSHC96} in combination with their
deep sample to derive the $K$-band luminosity function in various
redshift intervals.

\footnotemark[3] A correction of $-$0.30 magnitudes is often applied
to their result due to the use of fixed-aperture photometry
\citep{GPMC95}.

\footnotemark[4] \citet{CSHC96} determine the luminosity function in
four redshift bins from a combination of their deep sample with the
shallower samples from \citet{SCHG94}. The values for the Schechter
parameters given in the table are from their fit to the luminosity
function over the whole redshift range, with errors estimated from the
dispersion of the values for different redshift intervals.

\footnotemark[5] The spectroscopic catalogue of \citet{GSFC97} is
sparsely selected on the large photometric sample with the parameters
given in the table due to geometric limitations imposed by the fibre
spectrograph used for the observations.

\footnotemark[6] Based on redshift from the optically-selected
Kitt-Peak Galaxy Redshift Survey \citep{KPGRS}.

\footnotemark[7] Based on $K$-band imaging of galaxies from the
optically-selected Stromlo-APM galaxy survey \citep{APM90,APM96}.

\footnotemark[8] Based on a combination of the 2MASS catalogue
\citep{2MASS} with the CfA2 \citep{CfA2} and UZC \citep{UZC} surveys.

\footnotemark[9] Based on a combination of the 2MASS catalogue
\citep{2MASS} with the 2dFGRS \citep{2dFGRS_2}.

\footnotemark[10] Based on a combination of the 2MASS catalogue
\citep{2MASS} with the Las Campanas Redshift Survey \citep{LCRS96}.

\footnotemark[11] Hawaii--AA0 K-band Galaxy Redshift Survey.

\normalsize

\end{table*}

\begin{table*}

\caption{Summary of samples used for deriving the $J$-band
field-galaxy luminosity function, giving the source for the data, the
approximate limiting magnitude (either in the optical or in the
near-infrared), the approximate area, the number of objects, the
redshift range of the survey (either as interval or as median
redshift), and the parameters of the Schechter function derived from
the data. Parameters without error estimate were kept fixed during the
fitting procedure.}
\label{t:litj}

\begin{tabular}{lcrrcccc}

\hline
Source & Limit & Area & \# & $z$ & $M_J^*-5\log h$
& $\alpha_J$ & $\Phi_J^*$ \\
& [mag] & [arcmin$^2$] & & &  [mag] & & [$10^{-2} h^3 \mathrm{Mpc}^{-3}$] \\
\hline

\citet{Coleetal2001_2}\footnotemark[1]\hspace*{3em} & $b_J \le 19.5$,
$K \le 13$ & 2.2 10$^6$ & 17173 & 0.05 & $-$22.36 $\pm$ 0.02 & $-$0.93
$\pm$ 0.04 & 1.08 $\pm$ 0.16 \\ 
\citet{Baloghetal2001}\footnotemark[2] & $R_c \le 15\dots17$, $K \le
13$ & & & 0.0\dots0.18 & $-$22.23 $\pm$ 0.07
& $-$0.96 $\pm$ 0.12 & \\\hline
\textbf{MUNICS (this work)} &
$K \le 17.5\dots19.0$ & 649 & 132 & 0.1\dots0.3 & \mjl
& $-$1.00 & \pjl \\ 
& & & 145 & 0.3\dots0.6 & \mjm & $-$1.00 & \pjm \\
\hline

\end{tabular}

\footnotesize
\flushleft

\footnotemark[1] Based on a combination of the 2MASS catalogue
\citep{2MASS} with the 2dFGRS \citep{2dFGRS_2}.

\footnotemark[2] Based on a combination of the 2MASS catalogue
\citep{2MASS} with the Las Campanas Redshift Survey \citep{LCRS96}.

\normalsize
\end{table*}

%
%

\section{Conclusions}
\label{s:conclusions}

We have presented spectroscopic follow-up observations of galaxies
from the Munich Near-Infrared Cluster Survey (MUNICS), described the
observations, the data-reduction and the properties of the
spectroscopic sample. Furthermore we have presented the rest-frame
$K$-band luminosity function for galaxies at median redshifts of $z =
0.2$, $z = 0.4$, and $z = 0.7$. The Schechter parameters derived at
redshift $z = 0.2$ are \mklt , \pklt\ for fixed $\alpha_K = -1.10$ as
measured locally, in good agreement to values derived at low
redshifts. At redshift $z = 0.4$, however, the Schechter parameters
are \mkmt\ and \pkmt\ (for the same value of $\alpha_K$). Thus the
value for the characteristic luminosity is somewhat larger and the
normalisation smaller than the ones derived locally. This is confirmed
by Kolmogorov-Smirnov tests for the cumulative distributions in
absolute magnitudes, and by a $\chi^2$ analysis for the evolution of
the luminosity function. From the latter we find mild evolution in
magnitudes ($\Delta M_K^* = -0.70 \pm 0.30$~mag) and number densities
($\Delta \Phi_K^*/\Phi_K^* = - 0.35 \pm 0.30$) to redshift one.
Furthermore, we have presented the first measurement of the $J$-band
luminosity function of galaxies at higher redshifts with Schechter
parameters \mjlt , \pjlt\ for $z = 0.2$, and \mjmt , \pjmt\ for $z =
0.4$, showing the same trend of evolution as the $K$-band luminosity
function. The faint-end slope $\alpha_J$ set to a value of $-1.00$ in
both redshift bins. The evolutionary trend of the near-infrared
luminosity of $K$-selected galaxies described above is consistent with
expectations from pure luminosity evolution, while the decrease in
number density with redshift can be understood in the context of
hierarchical galaxy formation models.

\section*{Acknowledgements}

The authors would like to thank the staff at Calar Alto Observatory
for their extensive support during the many observing runs of this
project and especially for carrying out the service observations in
2002, as well as the staff at Paranal Observatory and McDonald
Observatory for assistance during observing runs for this
project. Furthermore we appreciate the efforts of Dr.\ Josef Fried
(MPIA Heidelberg) who organised the production of a very large number
of MOSCA slit masks. We are grateful to Dr.\ Lutz Wisotzki (Potsdam)
for allowing us to use spectroscopic data obtained in the course of a
joint project, and to Dr.\ Donald Hamilton for taking the observations
at the ESO 3.6-m telescope. GF wants to thank Claus G\"ossl for kind
assistance during a somewhat difficult observing run and for help with
his image reduction software, as well as Arno Riffeser, Armin Gabasch,
and Yuliana Goranova for helpful discussions. ND acknowledges support
by the Alexander von Humboldt Gesellschaft. We thank the anonymous
referee for his suggestions which helped to improve the presentation
of the paper. The Marcario Low Resolution Spectrograph is a joint
project of the Hobby-Eberly Telescope partnership and the Instituto de
Astronom\'\i a de la Universidad Nacional Aut\'onoma de M\'exico, and
was partly funded by the Deutsche Forschungsgemeinschaft, grant number
Be~1091/9--1. The Hobby-Eberly Telescope is operated by McDonald
Observatory on behalf of The University of Texas at Austin, the
Pennsylvania State University, Stanford University,
Ludwig-Maximilians-Universit\"at M\"unchen, and
Georg-August-Universit\"at G\"ottingen. This research has made use of
NASA's Astrophysics Data System (ADS) Abstract Service and the
NASA/IPAC Extragalactic Database (NED). The MUNICS project was
supported by the Deutsche Forschungsgemeinschaft, {\it
Sonderforschungsbereich 375, Astroteilchenphysik}.

\bibliographystyle{mn2e}
\bibliography{mnrasmnemonic,literature}

\label{lastpage}

\end{document}